\documentclass[aps,prb,showpacs,twocolumn,tighten,eqsecnum,amsmath,amssymb,floatfix]{revtex4-1}

\usepackage{amsmath}
\usepackage{graphicx}
\usepackage{dcolumn} 
\usepackage{bm, bbm} 
\usepackage{curves}
\usepackage{txfonts}
\usepackage{epic}
\usepackage{wasysym}
\usepackage{epsf, times}
\usepackage{color}

\begin{document}

\title{The low-frequency dielectric response of charged oblate spheroidal particles immersed in an electrolyte}

\author{Chang-Yu Hou}
\affiliation{Schlumberger-Doll Research, 1 Hampshire Street, Cambridge, MA 02139}
\author{Denise E. Freed}
\affiliation{Schlumberger-Doll Research, 1 Hampshire Street, Cambridge, MA 02139}
\author{Pabitra N. Sen}

\date{\today}

\begin{abstract}
We study the low-frequency polarization response of a surface-charged oblate spheroidal particle immersed in an electrolyte solution. Because the charged spheroid attracts counter-ions which form the electric double layer around the particle, using usual boundary conditions at the interface between the particle and electrolyte can be quite complicated and challenging. Hence, we generalize Fixman's boundary conditions, originally derived for spherical particles, to the case of the charged oblate spheroid. Given two different counter-ion distributions in the thin electric double layer limit, we obtain analytic expressions for the polarization coefficients to the first non-trivial order in frequency. We find that the polarization response normal to the symmetry axis depends on the total amount of charge carried by the oblate spheroid while that parallel to the symmetry axis is suppressed when there is less charge on the edge of the spheroid. We further study the overall dielectric response for a dilute suspension of charged spheroids. We find that the dielectric enhancement at low frequency, which is driven by the presence of a large $\zeta$-potential (surface charge), is suppressed by high ion concentrations in the electrolyte and depends on the size of the suspended particles. In addition, spheroids with higher aspect ratios will also lead to a stronger dielectric enhancement due to the combination of the electric double layer and textural effects. The characteristic frequency associated with the dielectric enhancement scales inversely with the square of the particle size, the major radius of the spheroid, and it has a weak dependence on the shape of the spheroids.
\end{abstract}

\maketitle

\newcommand{\beq}{\begin{equation}}
\newcommand{\eeq}{\end{equation}}
\newcommand{\bea}{\begin{eqnarray}}
\newcommand{\eea}{\end{eqnarray}}
%
%

\section{Introduction}
\label{sec:introduction}

Frequency-dependent dielectric signals of composite materials have been widely studied because they provide a way to probe essential properties of materials, such as the fraction of each constituent phase and the anisotropy of the mixture~\cite{Milton2002,Sihvola1999}. In addition to its practical uses, the dielectric response of mixtures is of interest because the properties of mixtures can be far from those of composite media. This also poses challenges for interpreting those measurements. Perhaps the most intriguing phenomenon is the observation of strong dielectric enhancements at low frequencies when charged or conducting colloids/particles are mixed with electrolytes~\cite{Chew1982,Qian2015}. To put this phenomenon in perspective, the measured value of the permittivity can be four orders of magnitude higher than that of the original constituents at frequencies lower than $1$ KHz.  Considering that values of the  permittivity for an electrolyte and a typical dielectric material are on the order of $10\sim 100$, the effective permittivity of the mixture can be on the order of $10^5$ or higher~\cite{Garrouch1994}.

Various mechanisms have been proposed to account for this strong dielectric enhancement. At very low frequencies, less than $1$ Hz, chemical interactions, such as adsorption and cation exchanges at the solid-liquid surface, can play an important role~\cite{Wong1979}. On one hand, it has been suggested that non-spherical particles immersed in an electrolyte, especially platy grains, can give rise to high dielectric enhancement due to the conventional Maxwell-Wagner effect occurring at interfaces between two dielectric materials. However, the actual size of the enhancement from this geometric effect alone cannot explain the observed data~\cite{RAYTHATHA1986}. On the other hand, Dukhin-Shilov~\cite{Dukhin1974}, Fixman~\cite{Fixman1980,Fixman1983} and Chew-Sen~\cite{Chew1982a,Chew1982} have provided solutions for the low frequency polarization response for a charged spherical particle. There, it has been shown that the presence of the out-of-phase diffusion currents of charged particles outside of the electric double layer is responsible for inducing strong dielectric responses at frequencies less than $10$ MHz~\cite{Chew1982}. In addition, numerical studies also support results of these analytical solutions~\cite{DeLacey1981,OBRIEN1982,Mangelsdorf1998a,Mangelsdorf1998b}. However, many colloidal particles are not spherical. Hence, it is important to explore whether a combined effect of double-layer polarization and non-spherical shape will give further enhancement.



The polarization response of a non-spherical, charged particle immersed in an electrolyte is, however, not as straightforward to evaluate. Chassagne and Bedeaux have obtained an analytic expression by assuming that the electric double layer is a shell of dielectric material bridging the spheroidal particle and the electrolyte~\cite{Chassagne2008}. In addition, an analogy to the solution of a charged sphere is used to obtain solutions for the spheroid without directly solving for the polarization responses due to the presence of an external electric field. These assumptions in principle limit the applicable regions of their solutions to spheroids with aspect ratios close to one.

In this paper, we will study the low frequency dielectric response of charged oblate spheroids immersed in an electrolyte.  We will ignore the effects of chemical interactions and electrophoretic flows. For charged spheres, the dielectric response remains qualitatively the same, with small quantitative modifications, when one takes into account the effects of electrophoretic flow~\cite{Fixman1983,Chassagne2008}. It is unclear whether such a conclusion still holds for charged spheroids. Nevertheless, we will neglect the contribution from electrophoretic flow to simplify our current study.  

Our approach is to first solve for the single-particle polarization response of a charged, oblate spheroid immersed in an electrolyte with approximate boundary conditions at the outer surface of the electric double layer, following Fixman's idea for spheres~\cite{Fixman1980,Fixman1983}.  The use of Fixman's boundary conditions allows us to partially circumvent the need to solve for the equilibrium ion distribution in the electric double layer from the Poisson-Boltzmann equation, which is in itself an interesting and difficult problem. We will then use these solutions to understand the properties of the low-frequency dielectric enhancement for a dilute suspension of charged, oblate spheroids in the electrolyte.  Because many particles in nature, such as blood cells, platy colloids, and clay grains, can be well approximated as oblate spheroids, our results can potentially be used for interpreting dielectric measurements for biological, colloidal, agriculture, and geological systems.

We will organize this paper as follows: In Sec.~\ref{sec:Governing}, we describe the setup for a charged spheroid immersed in an electrolyte and discuss the governing equations and proper boundary conditions for obtaining the single particle polarization response. In Sec.~\ref{sec:methodology}, we briefly discuss our methodology for solving for the polarization response for the setup in Sec.~\ref{sec:Governing}.  In Sec.~\ref{sec:results-polarization-coefficients}, we summarize our results for the single particle polarization coefficients for two different ion distributions, which are detailed in Appendices~\ref{app:Polarization-uniform} and~\ref{app:Polarization-non-uniform}.  In Sec.~\ref{sec:dilute}, we discuss the dielectric response for dilute suspensions of charged oblate spheroids immersed in an electrolyte. Finally, we give a brief summary in Sec.~\ref{sec:conclusions}. In addition, we define our conventions for associated Legendre's polynomials and spheroidal wave functions in Appendices~\ref{app:Legendre} and~\ref{app:spheroidal-wave-function}, respectively.

\section{Governing equations and boundary conditions}
\label{sec:Governing}

In electrodynamics, the polarization coefficient of a non-charged particle embedded in a dielectric material can be obtained from the induced dipole response (far-field electric potential) of the system subjected to an externally applied electric field~\cite{Jackson1998,Landau1984}. To follow a similar approach for extracting polarization coefficients, we need to properly establish the relevant governing equations and boundary conditions which allow us to solve for the resultant electric potential in the presence of an applied oscillating electric field.

For a charged particle immersed in an electrolyte, the governing equations that control the electric response are the Poisson equation and the continuity equations for cations and anions~\cite{Chew1982}.  At equilibrium, because the ions follow the Boltzmann distribution, the Poisson-Boltzmann equation becomes the only governing equation.  However, due to the non-linear nature of the Poisson-Boltzmann equation, the equilibrium cation and anion concentrations cannot be easily solved analytically for a non-spherical particle with arbitrary charge distributions.~\cite{Yoon1989,Hsu1997}  In addition, what we are ultimately interested in is the induced dipole response to the externally applied electric field.  Without a firm understanding of the equilibrium state, it becomes non-trivial to make concrete predictions for the perturbed state. Furthermore, because of the accumulation of counter-ions forming the electric double layer around the particle surface, directly using boundary conditions at the interface of a charged spheroid to solve for the dipole responses can be difficult. To circumvent this difficulty, we follow Fixman's idea, originally proposed for charged spheres, to derive boundary conditions outside a thin double layer for the charged oblate spheroid~\cite{Fixman1980}. These boundary conditions in principle require the knowledge of the integrated surface distributions of the cations and anions in the double layer. To proceed, we will assume two particular surface distributions of cations and anions in solving for the corresponding polarization response.

\subsection{Governing equations}

The fundamental equation that governs the electrodynamic properties of a charged particle immersed in an electrolyte is given by~\cite{Jackson1998}
\beq
\label{eq:Poisson-eq-total}
\nabla^{2} \Psi^{\rm t} (\boldsymbol{r},t) = \frac{-1}{2 N_0 \lambda_{\rm D}^2} \left(N^{\rm t}_{+}(\boldsymbol{r},t)-N^{\rm t}_{-}(\boldsymbol{r},t) \right).
\eeq
where $N^{\rm t}_{\pm}(\boldsymbol{r})$ are the total concentrations of cations and anions as a function of the position $\boldsymbol{r}$ and time $t$, $\Psi^{\rm t} = (Z e_0/k_{\rm B} T) \tilde{\Psi}^{\rm t}$ is the scaled electric potential with $\tilde{\Psi}^{\rm t}$ defined as the original electric potential, and $\lambda_{\rm D} = \sqrt{k_{\rm B} T \varepsilon_0 \varepsilon_{w}/2 (Z e_0)^2 N_0}$ is the Debye length. For simplicity, we only consider
symmetric electrolytes, i.e., the cations and anions carry a charge of $\pm Z e_0$, and their intrinsic concentrations far away from the charged particle and in the absence of an applied electric field are both given by $N_0$. Here, $T$ is the temperature in degrees Kelvin, $e_0$ is the absolute value of an electron charge, $k_{B}$ is the Boltzmann constant, $\varepsilon_0$ is the vacuum permittivity, and $\varepsilon_{w}$ is the permittivity of the electrolyte.

As we are ultimately interested in the polarization response upon the application of an external electric field, it is convenient to separate the total potential and ion concentrations into equilibrium and perturbed parts as~\cite{Chew1982} 
\beq
\label{eq:scaled-total-potential-Npm}
\Psi^{\rm t}= \Psi^{\rm eq}+ \psi, \; {\rm and} \; N^{\rm t}_{\pm}= N^{\rm eq}_{\pm} + n_{\pm}.
\eeq
Here, the superscript $^{\rm eq}$ denotes the equilibrium part that includes the response due to the presence of the charged particle, and $\psi(\boldsymbol{r})$ and $n_{\pm}(\boldsymbol{r})$ denote the perturbed part of the potential and ion concentrations due to the externally applied electric field, respectively. Because the solution to Eq.~\eqref{eq:Poisson-eq-total} obeys the superposition principle, we have the Poisson-Boltzmann equation
\beq
\label{eq:Poisson-Boltzmann-eq}
\nabla^2 \Psi^{\rm eq} (\boldsymbol{r}) = \frac{-1}{2 N_0 \lambda_{\rm D} ^2} \left(N^{\rm eq}_{+} -N^{\rm eq}_{-} \right) = \frac{1}{\lambda_{\rm D} ^2} \sinh \Psi^{\rm eq} (\boldsymbol{r}),
\eeq 
for the equilibrium part of the response, using the definition in Eq.~\eqref{eq:scaled-total-potential-Npm}. Here, we have used the Boltzmann distributions $N^{\rm eq}_{\pm}(\boldsymbol{r})=N_0 e^{\mp \Psi^{\rm eq} (\boldsymbol{r})}$ for the equilibrium ion distributions. In Eq.~\eqref{eq:Poisson-Boltzmann-eq}, the Debye length $\lambda_{\rm D} $ sets the electric screening length scale. The double layer thickness is thus on the order of a Debye length, $\delta_{\rm DL} \sim \lambda_{\rm D}$. Because the Debye length is on the order of nanometers ($\lambda_{\rm D}  \sim 2.4$nm for a $1$ ppk sodium chloride electrolyte at $25$ $^\circ$C), we will focus on the thin double layer limit, in which the double layer thickness is much smaller than the size of the particle, throughout our discussions.

Solving for the equilibrium electric potential and, hence, the ion distributions associated with the Poisson-Boltzmann equation~\eqref{eq:Poisson-Boltzmann-eq} for a charged spheroid immersed in an electrolyte is an interesting and non-trivial problem. The linearized approximation is often used to proceed in the limit of a weak surface $\zeta$-potential~\cite{Yoon1989}, while the solution for the nonlinear Poisson-Boltzmann equation involves extensive perturbative calculations~\cite{Hsu1997}. Some physical insights can still be obtained without explicitly solving for it. First, we expect $N^{\rm eq}_{\pm} \to N_0$ and $\Psi^{\rm eq} \to 0$ outside the double layer. Second, the total charge of the ions in the double (screening) layer must equal the total charge carried by the particle. Unfortunately, connecting the equilibrium ion concentrations in the electric double layer to the charge distribution on the spheroid requires solving Eq.~\eqref{eq:Poisson-Boltzmann-eq} explicitly, which is beyond the scope of this paper. As discussed below, we will circumvent this complication by assuming two particular surface ion distributions.



Upon application of an electic field, the perturbed electric potential and ion distributions obey an equation of motion similar to the one for the equilibrium potential and ion distributions:
\beq
\label{eq:EOM-psi-npm}
\nabla^2 \psi (\boldsymbol{r},t) = \frac{-1}{2 N_0 \lambda_{\rm D} ^2} \left(n_{+}(\boldsymbol{r},t)-n_{-}(\boldsymbol{r},t) \right).
\eeq 
Again, the Debye length is the domiinant length scale for the perturbed electric field and ion concentrations. However, because there is no simple relation between the perturbed ion concentrations $n_{\pm}(\boldsymbol{r})$ and the perturbed electric field $\psi$, another set of equations is required to solve for them.

In the absence of external sources or drains, cations and anions are independently conserved and obey the continuity equations
\beq
\nabla \cdot J^{\rm t}_{\pm} (\boldsymbol{r},t) = - \frac{\partial N_{\pm}^{t} (\boldsymbol{r},t) }{\partial t} = - \frac{\partial}{\partial t} n_{\pm}(\boldsymbol{r},t) 
\eeq
where $J^{\rm t}_{\pm} (\boldsymbol{r},t)$ are the total current densities for cations and anions, respectively. Because the equilibrium concentrations of ions are independent of time, we have used $\partial N^{\rm eq}_{\pm}/\partial t =0$. In the frequency domain, we then have 
\beq
\label{eq:continuity-eq}
\nabla \cdot J^{\rm t}_{\pm} (\boldsymbol{r},\omega) = i \omega  n_{\pm}(\boldsymbol{r},\omega). 
\eeq
The current density in this equation can be expressed in terms of the diffusive current density as
\bea
\label{eq:diffusive-current-1}
J^{\rm t}_{\pm} (\boldsymbol{r},t) =& - D_{\pm} N_{\pm}^{\rm t} (\boldsymbol{r},t) \nabla \mu_{\pm} (\boldsymbol{r},t),
\\
\label{eq:chmical-potential}
\mu_{\pm} (\boldsymbol{r},t) =& \ln (N_{\pm}^{\rm t}(\boldsymbol{r},t)/N_0) \pm \Psi^{\rm t} (\boldsymbol{r},t),
\eea
where $D_\pm$ are the diffusion coefficients for the cations and anions, and $\mu_{\pm}$ are the dimensionless chemical potentials for the cations and anions. For simplicity, we will also assume $D_+=D_{-} \equiv D$ in the following discussions. In equilibrium,  $J^{\rm eq}_{\pm}=0$, and, according to  Eq.~\eqref{eq:diffusive-current-1}, the ions follow the Boltzmann distribution.  By inserting Eq.~\eqref{eq:scaled-total-potential-Npm} into Eq.~\eqref{eq:diffusive-current-1}, we obtain
\begin{equation}
\nonumber
J^{\rm t}_{\pm} (\boldsymbol{r},t) \approx - D \left[ \nabla n_{\pm} (\boldsymbol{r},t)  \pm \left( N_{\pm}^{\rm eq}(\boldsymbol{r})  \nabla \psi (\boldsymbol{r},t) + n_{\pm} (\boldsymbol{r},t) \nabla \Psi^{\rm eq}(\boldsymbol{r})  \right) \right],
\end{equation}
to linear order in the applied electric field. 

As pointed out by Chew and Sen in Ref.~\onlinecite{Chew1982}, the low frequency dielectric response for charged particles immersed in an electrolyte is dominated by the neutral current outside the double layer.  Hence, we will focus on this region where the diffusion currents can be simplified to
\beq
\label{eq:J-pm-outside-DL}
J^{\rm t}_{\pm} (\boldsymbol{r},t) = - N_0 D \left[ \frac{\nabla n_{\pm} (\boldsymbol{r},t)}{N_0} \pm  \nabla \psi (\boldsymbol{r},t) \right].
\eeq
We can now combine Eqs.~\eqref{eq:continuity-eq} and~\eqref{eq:J-pm-outside-DL}, which yields
\beq
\label{eq:continuity-eq-2}
i \omega n_{\pm} (\boldsymbol{r},\omega) = -N_0 D  \left[ \frac{\nabla^2 n_{\pm} (\boldsymbol{r},\omega)}{N_0} \pm  \nabla^2 \psi (\boldsymbol{r},\omega) \right],
\eeq
Using Eq.~\eqref{eq:EOM-psi-npm} to relate $n_{\pm}(\boldsymbol{r},\omega)$ and $\psi(\boldsymbol{r},\omega)$ gives
\beq
\label{eq:sum-n}
\nabla^2 \bar{n} (\boldsymbol{r},\omega) + i \frac{\omega}{D} \bar{n} (\boldsymbol{r},\omega) = 0,  
\eeq
and 
\beq
\label{eq:sub-n}
\nabla^2 \Delta n (\boldsymbol{r},\omega) - \left( \frac{1}{\lambda_{\rm D}^2} - i \frac{\omega}{D} \right) \Delta n (\boldsymbol{r},\omega) = 0,  
\eeq
where $\bar{n}\equiv n_{+}+ n_{-}$ and $\Delta n \equiv n_{+} - n_{-}$ are the sum and difference of the perturbed cation and anion concentrations, respectively.


From Eq.~\eqref{eq:sum-n} and Eq.~\eqref{eq:sub-n}, we observe that $\Delta n$ decays much faster than $\bar{n}$ when $\omega\ll D/\lambda_{\rm D}^2$. In this limit, we can take $n_+ = n_- \equiv n$ outside the electric double layer, which yields two equations of motion as follows:
\begin{subequations}
	\label{eq:EOM-outside-DL}
	\bea
	\label{eq:EOM-outside-DL-a}
	\nabla^2 \psi(\boldsymbol{r},\omega) =& 0,
	\\
	\label{eq:EOM-outside-DL-b}
	\left(\nabla^2  +i \frac{\omega}{D} \right)  n(\boldsymbol{r},\omega) =&  0.
	\eea
\end{subequations}
To solve these differential equations, we must now derive proper boundary conditions (BC) for the electric potential and ion concentrations, both at $|\boldsymbol{r}| \to \infty$ and at the outer surface of the electric double layer.

\subsection{Boundary Conditions in spheroidal coordinates}

Let us start with the boundary condition at $|\boldsymbol{r}| \to \infty$.  The ion concentration at $|\boldsymbol{r}| \to \infty$ must satisfy the BC
\beq
\label{eq:BC-n-infinity}
n(\boldsymbol{r},\omega)\big|_{|\boldsymbol{r}| \to \infty} = 0, 
\eeq
because the perturbed ion concentration should decay to zero far from the charged particle to avoid singularities.  In the presence of an applied electric field, $(k_{\rm B} T/Z e_0) E_0 \hat{E}$ in the direction $\hat{E}$, the perturbed electric potential obeys the BC
\beq
\label{eq:BC-psi-infinity}
\psi(\boldsymbol{r},\omega) \big|_{|\boldsymbol{r}| \to \infty} = - E_0 (\hat{E} \cdot \boldsymbol{r}),
\eeq
which fits the potential profile of the electric field in the absence of the charged spheroid.

Our task now is to derive BCs at the outer surface of the electric double layer, which allows us to study the polarization response. Here, we will generalize the boundary conditions originally proposed by Fixman for a charged sphere to a charged oblate spheroid~\cite{Fixman1980}.

It is useful to introduce the oblate spheroidal coordinates with which a spheroid placed at the origin can be easily described. Let us define oblate spheroidal coordinates through their relation to Cartesian coordinates as~\cite{Flammer1957}
\beq
\label{eq:spheroidal-Cartesian}
\begin{split}
x =& h \sqrt{ (1+\xi^2)(1-\eta^2)} \cos \phi,
\\
y =& h \sqrt{ (1+\xi^2)(1-\eta^2)} \sin \phi,
\end{split}
\qquad
z = h\xi \eta,
\eeq
where $1\ge\eta\ge -1$, $\xi\ge 0$, $0\le\phi\le 2 \pi$, and $h$ is the half distance between the two foci of the ellipse. Pictorially, oblate spheroidal coordinates are a three-dimensional curvilinear coordinate system that results from rotating a two-dimensional elliptic coordinate about its minor axis. For our choice of coordinates in Eq.~\eqref{eq:spheroidal-Cartesian}, the symmetry (rotation) axis is along the $z$-axis. Here, $\xi$ can be viewed as the radial direction of the coordinate system. Because the constant $\xi_0>0$ defines an oblate spheroid with its major radius $a = h \sqrt{1+\xi_0^2}$ and its minor radius $b = h \xi_0$, the boundary conditions of an oblate spheroidal shell are easily defined. To fully define curvilinear coordinates, we need to introduce the scaling factors, $h_{\xi}$, $h_{\eta}$, and $h_{\phi}$, such that the actual distances along each orthogonal direction are $h_{q} dq$. Detailed descriptions of the scaling factors and spheroidal coordinates are provided in Appendix~\ref{app:spheroidal-wave-function}. 



To infer the proper BCs at the outer surface of the electric double layer, we start with the continuity equation in Eq.~\eqref{eq:continuity-eq}, combine it with the diffusive currents in Eq.~\eqref{eq:diffusive-current-1}, and then separate the vector derivative into the parts normal and parallel ($\parallel$) to an oblate spheroidal shell defined by holding $\xi$ fixed. Using definitions of vector derivatives in a curvilinear coordinate system, we can cast the continuity equations  into the form
\beq
\label{eq:continuity-eq-1}
\begin{split}
i \omega n_{\pm}(\boldsymbol{r},\omega) =  \frac{1}{h_{\xi} h_{\eta} h_{\phi}} &\frac{\partial }{\partial \xi}\left(h_{\eta} h_{\phi} J^{\rm t}_{\pm,\xi} (\boldsymbol{r},\omega) \right) 
\\
&- D  \nabla_{\parallel} \cdot \left( N_{\pm}^{\rm t} (\boldsymbol{r},\omega) \nabla_{\parallel} \mu_{\pm}(\boldsymbol{r},\omega) \right),
\end{split}
\eeq
where $J^{\rm t}_{\pm,\xi} (\boldsymbol{r},\omega) $ are the diffusive currents in the $\xi$-direction. We now multiply both sides of Eq.~\eqref{eq:continuity-eq-1} by the scaling factor $h_{\xi}$ and then integrate $\xi$ over the diffuse layer from $\xi=\xi_0$ to $\xi=\xi_0+\zeta$ to obtain
\beq
\label{eq:continuity-BC}
\begin{split}
i\omega \sigma_{\pm} =& J^{\rm t}_{\pm,\xi} (\boldsymbol{r},\omega)\big|_{\xi_0+\zeta} 
\\
& - D  \int_{\xi_0}^{\xi_0+\zeta} d \xi h_{\xi}\nabla_{\parallel} \cdot \left( N_{\pm}^{\rm t} (\boldsymbol{r},\omega) \nabla_{\parallel} \mu_{\pm}(\boldsymbol{r},\omega) \right) 
\\
&- \int_{\xi_0}^{\xi_0+\zeta} d \xi \frac{\partial}{\partial \xi} \left( \frac{1}{h_{\eta} h_{\phi}}\right) h_{\eta} h_{\phi} J^{\rm t}_{\pm,\xi} (\boldsymbol{r},\omega),
\end{split}
\eeq
where we have used $J^{\rm t}_{\pm,\xi}(\boldsymbol{r},\omega)\big|_{\xi_0}=0$ because currents normal to the particle surface should vanish. Here, $\zeta$ corresponds to the thickness of the double layer and can be defined by $\delta_{\rm DL} \sim h_{\xi}(\xi_0,\eta) \zeta$ for the thin double-layer approximation where $\zeta \ll \xi_0$. We also define the integrated perturbed surface ion distributions $\sigma_\pm$ as
\beq
\sigma_{\pm} = \int_{\xi_0}^{\xi_0+\zeta} h_{\xi} d \xi n_{\pm} (\boldsymbol{r},\omega).
\eeq

In the thin double-layer limit, $\zeta\ll \xi_0$, we will neglect contributions from the last term in Eq.~\eqref{eq:continuity-BC} because it is generally suppressed by $\delta_{\rm DL}/h$ when compared with the first term. The scaling factors in the transverse vector derivative can be treated as constants inside the integral. Furthermore, $\nabla_{\parallel} \mu_{\pm}$ can also be approximated as a constant over a thin double layer due to the general vector derivative property, $\nabla \times (\nabla \mu_{\pm}) =0$, together with Stoke's theorem, c.f. Refs.~\onlinecite{Fixman1980} and~\onlinecite{Dukhin1980}. With these simplifications, we have
\beq
\label{eq:BC-all-frequency}
i\omega \sigma_{\pm} = J^{\rm t}_{\pm,\xi} (\boldsymbol{r},\omega)\big|_{\xi_0+\zeta} - D   h_{\xi}\nabla_{\parallel} \cdot \left( \frac{\Gamma_{\pm}}{h_{\xi}} \nabla_{\parallel} \mu_{\pm}(\boldsymbol{r},\omega)\right)\big|_{\xi_0+\zeta},
\eeq
where the integrated surface ion distributions, $\Gamma_{\pm}$, are defined as
\beq
\label{eq:def-Gamma-pm}
\Gamma_{\pm} = \int_{\xi_0}^{\xi_0+\zeta} h_{\xi} d \xi   N^{\rm t}_{\pm} (\boldsymbol{r}) \approx  \int_{\xi_0}^{\xi_0+\zeta} h_{\xi} d \xi  N^{\rm eq}_{\pm} (\boldsymbol{r}) ,
\eeq
and are dominated by the equilibrium ion concentrations.  The BCs in Eq.~\eqref{eq:BC-all-frequency} are valid for all frequencies, provided that $\sigma_{\pm}$ and $\Gamma_{\pm}$ are known, in the thin double layer limit.


At low frequencies, $\omega < \omega_h\equiv D/h^2$, we can make a further approximation in these BCs by setting $\omega=0$ on the left-hand side of Eq.~\eqref{eq:BC-all-frequency}, which yields
\beq
\label{eq:BC-low-frequency}
0= J^{\rm t}_{\pm,\xi} (\boldsymbol{r},\omega)\big|_{\xi_0+\zeta} - D   h_{\xi}\nabla_{\parallel} \cdot \left( \frac{\Gamma_{\pm}}{h_{\xi}} \nabla_{\parallel} \mu_{\pm}(\boldsymbol{r},\omega)\right)\big|_{\xi_0+\zeta}. 
\eeq
These simplified BCs at low frequency essentially neglect the interface effect between the electrolyte and spheroid. As a result, the Maxwell-Wagner effect which plays an important role in dielectric dispersion at higher frequencies, $\omega> \omega_c$, is neglected~\cite{Freed-ex}.  The frequency $\omega_c$, defined below, is a characteristic frequency associated with dielectric enhancement. It is possible to include interface effects if we use the BCs in Eq.~\eqref{eq:BC-all-frequency}. However, additional BCs at the surface of the spheroid or explicit expressions for $\sigma_{\pm}$ are required to proceed, which further complicates the exposition. As a result, we will focus on the low-frequency response in this paper and present the more complicate results for wider frequency ranges elsewhere. 

The functional forms of the integrated ion distributions $\Gamma_{\pm}$ in principle come from solutions of the Poisson-Boltzmann equation~\eqref{eq:Poisson-Boltzmann-eq}. As discussed earlier, it is a non-trivial problem and beyond the scope of this paper. Instead, we will investigate two examples of ion distributions which mimic reasonable situations: (I) uniformly distributed cations and no anions, $\Gamma^{\rm I}_{+}= \Gamma_0 $ and $\Gamma^{\ }_{-}\approx 0$; (II) non-uniformly distributed cations and no anions, given by $\Gamma^{\rm II}_{+}= \widetilde{\Gamma}_0 h_{\xi}(\xi_0,\eta)/h$ and $\Gamma^{\ }_{-}\approx 0$. In this paper, we assume that the spheroid carries negative charge. Hence, the cations are the primary counter-ions and the amount of anions can be neglected in the double layer as long as $\Gamma_-/(N_0\lambda)\ll 1$. The surface cation densities, $\Gamma_+^{\rm I}$ and $\Gamma_{-}^{\rm II}$, can be related via surface integrals to the total charge $Q$ carried by the charged particle, as follows:
\beq
\label{eq:Q-Gamma}
|Q| = Z e_0 \int dS \Gamma_{+}^{\rm I (II)}.
\eeq 

The non-uniform ion distribution has less charge along the edge of the oblate spheroid. There are two primary reasons for investigating the particular form of the ion distribution given in case (II): First, there is less mixing of angular modes, c.f. $\ell$ indices in Eq.~\eqref{eq:general-solution-psi}, for the non-uniform charge distribution. Namely, to solve for the polarization coefficients for case (II) requires less and also cleaner approximations. Second, it provides insight on how changing the ion distribution will affect the polarization response.

\section{Brief discussion of methodology}
\label{sec:methodology}

In classical electrodynamics~\cite{Jackson1998}, the dipole response can be extracted from the $\ell =1$ component of the angular expansions of the electric potential, c.f. Eq.~\eqref{eq:general-solution-psi} below. We can hence start with the general solutions of $n(\boldsymbol{r}, \omega)$ and $\psi(\boldsymbol{r}, \omega)$ which satisfy the boundary conditions at $|\boldsymbol{r}|\to \infty$ in Eqs.~\eqref{eq:BC-n-infinity} and~\eqref{eq:BC-psi-infinity}, and then match the BCs in Eq.~\eqref{eq:BC-low-frequency}. In this section, we first summarize the formal solutions for the perturbed electric potential $\psi$ and the ion concentration $n$, and then comment on how to match the boundary conditions. The goal here is to provide a road map for obtaining the polarization response of a charged oblate spheroid immersed in an electrolyte. The detailed descriptions and derivations will be given in Appendices~\ref{app:Polarization-uniform} and~\ref{app:Polarization-non-uniform}.

Solutions to the Poisson equation~\eqref{eq:EOM-outside-DL-a} in oblate spheroidal coordinates can be written as an expansion in terms of associated Legendre polynomials~\cite{Morse1953}. With the BC in Eq.~\eqref{eq:BC-psi-infinity}, the perturbed electric field can be expanded as
\beq
\label{eq:general-solution-psi}
\begin{split}
&\psi(\xi,\eta,\phi, \omega) = - E_0 (\hat{E} \cdot \boldsymbol{r})
\\
&+ \sum_{\ell,m}  A_{\ell}^{m} \cdot Q_{\ell}^{m} (i \xi) \cdot  P_{\ell}^{m} (\eta) \cdot \left( U_m \cos m\phi + V_m \sin m\phi \right).
\end{split}
\eeq
Here, the coefficients $A_{\ell}^{m}$ are functions of $\omega$. Our conventions for the associated Legendre polynomials of the first and second kinds, $P_{\ell}^{m} (x)$ and $Q_{\ell}^{m} (x)$, are summarized in Appendix~\ref{app:Legendre}. Legendre polynomials of the second kind are dropped in the expression for the $\eta$ because $Q_{\ell}^{m} (\eta)$ diverges at $\eta=\pm 1$.  The radial terms $Q_{\ell}^{m} (i \xi) $ are always included in the solution because $Q_{\ell}^{m} (i \infty) \to 0$. On the other hand, because the $P_{\ell}^{m} (i \infty)$ either diverge or go to a constant, they are allowed only for matching the BC at $\xi \to \infty$, which appears as the first term in Eq.~\eqref{eq:general-solution-psi}.

The general solution of the diffusion equation~\eqref{eq:EOM-outside-DL-b} for the ion concentration in oblate spheroidal coordinates is, however, more complicated. Formally, it can be expanded in terms of spheroidal wave functions as
\begin{widetext}
\beq
n(\xi,\eta, \phi,\omega)= \sum_{n, m} \left( \alpha_{n}^{m} R_{mn}^{(3)} (-i e^{i s(\omega) \pi/4} h q, i \xi ) + \beta_{n}^{m} R_{mn}^{(4)} (-i e^{i s(\omega) \pi/4} h q, i \xi ) \right) \cdot S_{mn}^{(1)} (-i e^{i s(\omega) \pi/4} h q, \eta ) \cdot \left( u_m \cos m\phi + v_m \sin m\phi \right), 
\eeq
\end{widetext}
where $ R_{mn}^{(3),(4)} (c, x)$ and $ S_{mn}^{(1)} (c, x)$ are radial and angular spheroidal wave functions, respectively, $q \equiv \sqrt{|\omega|/D}$ and $s(\omega) \equiv {\rm sign} (\omega)$. Again, the expansion coefficients, $\alpha_{n}^{m}$, are functions of $\omega$. A brief discussion that defines the necessary notation for spheroidal wave functions is given in Appendix~\ref{app:spheroidal-wave-function}.

Because $R_{mn}^{(3)} \to 0 \; (\infty)$ and $R_{mn}^{(4)} \to \infty \; (0)$ for $\omega>0$ ($\omega<0$) as $\xi\to \infty$, the BC in Eq.~\eqref{eq:BC-n-infinity} for the ion concentration requires 
\beq
\label{eq:general-solution-n-omega+}
\begin{split}
n_{>}(\xi,\eta, &\phi,\omega)= \sum_{n, m} \alpha_{n}^{m}(\omega) \cdot R_{mn}^{(3)} (-i e^{i \pi/4} h q, i \xi )
\\
& \cdot S_{mn}^{(1)} (-i e^{i \pi/4} h q, \eta ) \cdot \left( u_m \cos m\phi + v_m \sin m\phi \right),
\end{split}
\eeq
for $\omega>0$, and 
\beq
\label{eq:general-solution-n-omega-}
\begin{split}
n_{<}(\xi,\eta,& \phi,\omega)= \sum_{n, m} \beta_{n}^{m} (\omega)  R_{mn}^{(4)} (-i e^{-i \pi/4} h q, i \xi )
\\
& \cdot S_{mn}^{(1)} (-i e^{-i \pi/4} h q, \eta ) \cdot \left( u_m \cos m\phi + v_m \sin m\phi \right),
\end{split}
\eeq
for $\omega<0$. Here, the subscripts $_{>}$ and $_{<}$ indicate solutions for $\omega>0$ and $\omega>0$, respectively. These solutions are not independent and should obey the relation $n_{>}^{*}(\xi,\eta, \phi,\omega) = n_{<}(\xi,\eta,\phi, -\omega)$ because the ion concentration is real in the time domain. Using the following identities for spheroidal wave functions, c.f. Eqs.~\eqref{eq:S1_mn-expansion} and~\eqref{eq:R_mn34-expansion-1},
\begin{subequations}
\label{eq:identities-R-S-star}
\beq
{S_{mn}^{(1)}}^* (-i e^{i s(\omega) \pi/4} h q, \eta )  = S_{mn}^{(1)} (- i e^{i s(-\omega) \pi/4} h q, \eta ) , 
\eeq
and
\beq
{R_{mn}^{(3)}}^* (-i e^{i s(\omega) \pi/4} h q, i \xi ) = R_{mn}^{(4)} ( -i e^{i s(-\omega) \pi/4} h q, i \xi ),
\eeq
\end{subequations}
we have $\beta_{n}^{m}(-\omega)= {\alpha_{n}^m}^*(\omega)$ for $\omega>0$. Hereafter, we will focus on the case $\omega>0$.

To solve for the perturbed electric potential to extract polarization coefficients, we need to match the boundary conditions in Eq.~\eqref{eq:BC-low-frequency} to obtain at least the $\ell=1$ components, \textit{i.e.}, $A_{1}^{m}$, in Eq.~\eqref{eq:general-solution-psi}. In principle, the BCs in Eq.~\eqref{eq:BC-low-frequency} should uniquely determine the solutions for $\psi$ and $n$. However, two issues complicate the process of matching these BCs. First, there is a mismatch between the general solutions in Eqs.~\eqref{eq:general-solution-psi} and~\eqref{eq:general-solution-n-omega+}. Although spheroidal wave functions can be expanded in terms of Legendre's polynomials, c.f.~Appendix~\ref{app:spheroidal-wave-function}, it makes the matching of the BCs more challenge. Second, for arbitrary ion distributions $\Gamma^{\pm}$, different angular components can mix with each other. Both issues can potentially lead to an infinite series expansion for the perturbed electric field and make the solutions rather formidable.

Because the concentration of anions is assumed to be small within the double layer and is set to zero, $\Gamma_- =0 $, in the distribution profile, the boundary condition for the anions in Eq.~\eqref{eq:BC-low-frequency} simply gives relations between the $\alpha_{n}^{m}$ and $A_{\ell}^m$ coefficients which allows us to cast the $\alpha_{n}^{m}$ in terms of $A_{\ell}^m$. In general, different angular components can couple with each other, i.e., $\alpha_{n}^{m}$ is related not only to the $\ell=n$ component but also to $\ell\neq n$ components of $A_{\ell}^m$. By inserting these relations into the BC for cations in Eq.~\eqref{eq:BC-low-frequency}, we obtain a set of linearly independent equations for the $A_{\ell}^m$ which can, in general, be organized into the form
\begin{equation}
\label{eq:generic-G-A-V}
\sum_{\ell',m'}  G_{\ell,m; \ell',m'} A_{\ell'}^{m'}  = V_{\ell}^m. 
\end{equation} 
Here, $G_{\ell,m; \ell',m'}$ and $V_{\ell}^m$ depend on the frequency $\omega$ and other known parameters. In principle, $A_{\ell'}^{m'}$ is then obtained by inverting the $G$ matrix. 

Because different angular components couple to each other, the $G$ matrix and $V$ vector in Eq.~\eqref{eq:generic-G-A-V} are neither diagonal nor truncated at finite order. The key is to show that the $\ell =1$ angular component is still the dominant part of the response when the system is subjected to an oscillating electric field. As a result, we developed a perturbation scheme that gives approximate analytical expressions for the $\ell=1$ components and truncates the higher order $\ell >1$ contributions to the perturbed electric field in Eq.~\eqref{eq:general-solution-psi}. Because matching the BCs and perturbative calculations are rather tedious, we will present them in Appendices~\ref{app:Polarization-uniform} and~\ref{app:Polarization-non-uniform}. We will summarize the essential results for further discussions in the following section.

\section{Results for the polarization coefficients}
\label{sec:results-polarization-coefficients}

In the conventional case where a dielectric material is embedded into a dielectric host medium, the polarization coefficients represent the induced dipole moment formed in response to the local electric field inside the embedded dielectric material~\cite{Jackson1998}. However, it has been shown that these induced dipole moments are not the dominant low-frequency polarization response of the system of interest, a charged particle immersed in the electrolyte~\cite{Chew1982}. Instead, the major contribution to the overall dipole polarization response comes from the out-of-phase ionic currents outside the electric double-layer, which the simplified BCs in Eq.~\eqref{eq:BC-low-frequency} aim to capture. As a result, the polarization coefficients described in this section should be understood as the total induced dipole moment due to the ionic response to the applied electric field in the electrolyte, averaged over the volume of the spheroid.

From Appendices~\ref{app:Polarization-uniform} and~\ref{app:Polarization-non-uniform}, the low-frequency polarization coefficients, $P_i$, for a charged oblate spheroid immersed in an electrolyte are expanded to order $\mathcal{O}(\omega)^{3/2}$ and can be cast into the generic form,
\beq
\label{eq:P_i-generic}
P_{i} = \frac{1}{3}  \frac{ \sigma_i -\sigma_{w,i} }{ L_i \sigma_i +(1-L_i) \sigma_{w,i} },
\eeq
when an oscillating electric field is applied in the $i$-direction. We use the convention that the $z$-direction is along the axis of symmetry of the oblate spheroid, while the $x$- and $y$-directions are normal to the symmetry axis. In this equation, the $L_i$ are the depolarization factors~\cite{Landau1984}, and $\sigma_i$ and $\sigma_{w,i}$, respectively, represent the effective particle and modified water conductivities in the $i$-direction. Due to the symmetry of the oblate spheroid, we have $L_x=L_y \equiv L_n$, $\sigma_x=\sigma_y \equiv \sigma_{n}$, $\sigma_{w,x}=\sigma_{w,y} \equiv \sigma_{w, n}$ and, hence, $P_x=P_y \equiv P_{n}$. In addition, we define $L_z\equiv L_{p}$, $\sigma_z \equiv \sigma_{p}$, $\sigma_{w,z}\equiv \sigma_{w,p}$ and $P_z \equiv P_{p}$. Here, the subscripts $p$ and $n$ stand for "parallel" and "normal" to the symmetry axis, respectively.

Alternatively, the generic form of the polarization coefficients in Eq.~\eqref{eq:P_i-generic} can be written as
\beq
P_{i}= \frac{1}{3}  \frac{ \left(\sigma_i \sigma_w/\sigma_{w,i}\right) -\sigma_{w} }{ L_i \left(\sigma_i \sigma_w/\sigma_{w,i}\right) +(1-L_i) \sigma_{w} }.
\eeq
With this form, the entire polarization response due to the ion flows in the electrolyte is grouped into an effective particle conductivity $\sigma_i'=\sigma_i \sigma_w/\sigma_{w,i}$. However, because this alternative form is less compatible with our approach, i.e., it is less transparent how to keep the relevant orders of the expansions, we will use Eq.~\eqref{eq:P_i-generic} as the generic form for our discussion. It is also worth noting that the solutions for the polarization coefficients obtained by Chassagne and Bedeaux in Ref.~\onlinecite{Chassagne2008} have the same generic form as in Eq.~\eqref{eq:P_i-generic} in the proper limit. However, the detailed structure of $\sigma_{p(n)}$ and $\sigma_{w,p(n)}$ in their solution are different from what we present below.

Because the depolarization factors are purely geometric quantities which only depend on the shape (or the aspect ratio) of the oblate spheroid, they are a function of $\xi_0$, the radius of the oblate spheroid in spheroidal coordinates, and are given by~\cite{Landau1984}
\beq
\label{eq:L_p}
L_{p}= - (1+\xi_0^2) Q_{1} (i \xi_0) = 1- \xi_0 (1+\xi_0^2) \frac{d Q_{1} (i \xi_0)}{d \xi_0},
\eeq
and 
\beq
\label{eq:L_n}
L_{n}= i \frac{\xi_0}{2} (1+\xi_0^2) \frac{Q_{1}^{1}(i\xi_0)}{P_{1}^{1}(i\xi_0)} = 1 + i  \frac{\xi_0}{2} (1+\xi_0^2) \frac{ \frac{d Q_{1}^{1}(i\xi_0)}{d\xi_0}}{ \frac{d P_{1}^{1}(i\xi_0)}{d \xi_0} }.
\eeq

The polarization coefficients for the uniform surface ion distribution, $\Gamma^{\rm I}_{+}= \Gamma_0 $ and $\Gamma^{\ }_{-}= 0$, are derived in Appendix~\ref{app:Polarization-uniform}. In the $z$-direction, the effective particle and modified water conductivities for $P_z=P_p^{\rm I}$ are given by  
\beq
\label{eq:sigma-p-I}
\begin{split}
	\sigma^{\rm I}_{ p} = & - \sigma_w  \frac{ \Gamma_0}{a N_0} \xi_0 \frac{b_{1,1}(\xi_0)}{2},
	\\
	\sigma_{w,p}^{\rm I} \sim& \sigma_w \left( 1 + \frac{\Gamma_0}{a N_0} \frac{b_{1,1}(\xi_0)}{2}  \mathcal{P} [ \Sigma^{(0)}_{p}(hq, \xi_0)]^{(1,2)}  \right),
\end{split}
\eeq
where $\sigma_w$ is the water (electrolyte) conductivity. The superscript $\rm I$ indicates that the results are for case (I). The two functions $b_{1,1}(\xi_0)$ and $ \mathcal{P} [ \Sigma^{(0)}_{p}(hq, \xi_0)]^{(1,2)}$ are given by
\beq
\label{eq:b_11}
b_{1,1}(\xi_0)= \frac{3}{2} \left[ \sqrt{1+\xi_0^2} -\left(1+\frac{\xi_0^2}{2}\right) \ln\left( \frac{\sqrt{1+\xi_0^2}+1 }{\sqrt{1+\xi_0^2}-1} \right) \right],
\eeq
and
\beq
\label{eq:Sigma_0_p-Pade}
\mathcal{P} [ \Sigma^{(0)}_{p}]^{(1,2)} \sim \frac{Q_{1}(i\xi_0)}{d Q_{1}(i\xi_0)/d\xi_0 } \frac{1 + \varpi_{p}^1 hq }{ 1+ \varpi_{p}^1 hq  + \varpi_{p}^2 h^2q^2} ,
\eeq
with
\bea
\varpi_{p}^2 &=& -i  \left( \frac{Q_3(i\xi_0)}{25 Q_1(i\xi_0)} - \frac{1}{6 Q_1(i\xi_0)} -  \frac{d Q_{3}(i\xi_0)/d \xi_0 }{ 25 (d Q_{1}(i\xi_0)/d \xi_0 )}  \right),
\\
\label{eq:varpi_p_1-Pade}
\varpi_{p}^1 &=& \frac{e^{-i\pi/4}}{9 \varpi_{ p}^2 } \left( \frac{P_1(i \xi_0)}{Q_1(i\xi_0)} -  \frac{d P_{1}(i\xi_0)/d \xi_0 }{ d Q_{1}(i\xi_0)/d \xi_0}  \right). 
\eea
Here, $\mathcal{P} [f]^{(1,2)}$ indicates that we have taken the Pad\' e approximation for the function $f$ to the $(1,2)$ order. We have used the approximation for the variable $hq=h \sqrt{\omega/D}$.

In the $x$- and $y$-directions, $P_x=P_y=P^{\rm I}_{n}$, the effective particle and modified water conductivities have the form
\beq
\label{eq:sigma-n-I}
\begin{split}
	\sigma_{n}^{\rm I} =&  - \sigma_w  \frac{\Gamma_0}{2 a N_0 } b^{\perp}_{1,1}(\xi_0)  \frac{P^1_1( i \xi_0)}{ d P^1_1( i \xi_0)/d\xi_0}  ,
	\\
	\sigma_{w,n}^{\rm I} (\omega) \sim& \sigma_w \left( 1 + \frac{\Gamma_0}{a N_0} \frac{b^{\perp}_{1,1}(\xi_0)}{2}   \mathcal{P}[ \Sigma^{(0)}_{n}(hq, \xi_0)]^{(1,2)} \right).
\end{split}
\eeq
In these equations, the two functions $b^{\perp}_{1,1}(\xi_0)$ and $\mathcal{P}[ \Sigma^{(0)}_{n}(hq, \xi_0)]^{(1,2)}$ are given by
\beq
\label{eq:b11-per}
\begin{split}
b_{11}^{\bot}(\xi_0) =&
- \frac{3}{4} \left[ \sqrt{\xi_0^2 +1}+ \frac{1}{\sqrt{\xi_0^2 +1}} \right.
\\
& \qquad  \qquad \left. - \frac{\xi_0^4}{2(1+\xi_0^2)} 
\ln \left( \frac{\sqrt{\xi_0^2 +1}+1}{\sqrt{\xi_0^2 +1}-1} \right) \right],
\end{split}
\eeq
and
\beq
\label{eq:Sigma_0_n-Pade}
\mathcal{P}[ \Sigma^{(0)}_{n}]^{(1,2)} \sim \frac{Q^{1}_{1}(i\xi_0)}{d Q^{1}_{1}(i\xi_0)/d\xi_0 } \frac{1 + \varpi^1_{n} hq }{ 1+ \varpi^1_{n} hq  + \varpi^2_{n} h^2q^2 } ,
\eeq
with
\bea
\varpi_{n}^2 &=& -i  \left( \frac{Q^1_3(i\xi_0)}{75 Q^1_1(i\xi_0)} - \frac{Q^1_{-1}(i\xi_0)}{3 Q^1_1(i\xi_0) } \right.
\\
\nonumber
&& \quad \left.  -  \frac{d Q^1_{3}(i\xi_0)/d \xi_0 }{ 75 (d Q^1_{1}(i\xi_0)/d \xi_0 )}  + \frac{d Q^1_{-1}(i\xi_0)/d \xi_0 }{3 (d Q^1_{1}(i\xi_0)/d \xi_0 )}  \right),
\\
\label{eq:varpi_n_1-Pade}
\varpi_{n}^1 &=& \frac{2 i e^{i\pi/4} }{9 \varpi_{n}^2  }  \left( \frac{P^1_1(i \xi_0)}{Q^1_1(i\xi_0)} -  \frac{d P^1_1(i\xi_0)/d \xi_0 }{ d Q^1_1(i\xi_0)/d \xi_0}  \right). 
\eea

As detailed in Appendices~\ref{app:Polarization-uniform} and~\ref{app:Polarization-non-uniform}, the functions $\Sigma^{(0)}_{p}$ and $\Sigma^{(0)}_{n}$ are both Taylor expanded to order $\mathcal{O}(hq)^{3/2}$. However, the Taylor expansion will cause unphysical behavior once $\omega$ becomes large enough, and it does not approach the proper limit at high frequencies~\cite{Freed-ex}. To regularize this problem, we further perform a Pad\' e approximation for the functions $\Sigma_{p}^{(0)}$ and $\Sigma_{n}^{(0)}$. Even though they are unconventional as we only perform the Pad\' e approximation in part of the full solution, these approximate Pad\' e expansions have several advantages. First, they provide low frequency approximations that are as good as the Taylor expansions. Second, they smoothly interpolate to the proper limiting solutions in the high frequency regime~\cite{Freed-ex}. Remarkably, the Pad\' e approximated solutions exactly match the polarization coefficient of a charged sphere immersed in an electrolyte in the spherical limit~\cite{Chew1982}, where $h\to0$ and $\xi_0\to \infty$ with $h \sqrt{1+\xi_0^2} = a$ held constant.

For the second example, the polarization coefficients for the non-uniform surface-ion distribution, given by  $\Gamma^{\rm II }_{+}= \widetilde{\Gamma}_0 h_{\xi}(\xi_0)/h $ and $\Gamma^{\ }_{-}= 0$, are derived in Appendix~\ref{app:Polarization-non-uniform}. This particular choice for the ion distribution  leads to less mixing of angular modes. As discussed in Appendix~\ref{app:Polarization-non-uniform}, with this choice of non-uniform charge distribution, the BC does not intrinsically cause mode mixing. Instead, the mixing of modes only occurs due to the mismatch of the mode expansions for solutions of the diffusion equation and the Laplace equation.

In the $z$-direction, $P_z=P_{p}^{\rm II}$, the effective particle and the modified water conductivities are given by
\beq
\label{eq:sigma-p-II}
\begin{split}
	\sigma^{\rm II}_{p} =& \sigma_w  \frac{ \widetilde{\Gamma}_0}{a N_0} \xi_0/\sqrt{1+\xi_0^{2}},
	\\
	\sigma_{w,p}^{\rm II} =& \sigma_w \left( 1 - \frac{\widetilde{\Gamma}_0}{a N_0}   \mathcal{P} [ \Sigma^{(0)}_{p}(hq, \xi_0)]^{(1,2)}  /\sqrt{1+\xi_0^{2}} \right),
\end{split}
\eeq
Interestingly, $\mathcal{P} [\Sigma^{(0)}_{p}(hq, \xi_0)]^{(1,2)}$ in Eq.~\eqref{eq:Sigma_0_p-Pade} also appears in the expression for the polarization response of the non-uniform surface ion distribution.

In the $x$- and $y$-directions, $P_x=P_y=P^{\rm II}_n$, the effective particle conductivity and the modified water conductivity are given by
\beq
\label{eq:sigma-n-II}
\begin{split}
\sigma^{\rm II}_{n} =&  -\sigma_w  \frac{\widetilde{\Gamma}_0}{2 a N_0 } \frac{ \beta_{1}^{1} }{\sqrt{1+\xi_0^2}} \frac{P^1_1( i \xi_0)}{ d P^1_1( i \xi_0)/d\xi_0}  \\
\sigma_{w, n}^{\rm II}  \sim& \sigma_w \left( 1+  \frac{\widetilde{\Gamma}_0}{a N_0 }  \frac{ \beta_{1}^{1}(\xi_0)  }{2 \sqrt{1+\xi_0^2}}   \mathcal{P}[\Sigma^{(0)}_{n}( h q, \xi_0)]^{(1,2)}  \right).
\end{split}
\eeq
Again, $\mathcal{P} [ \Sigma^{(0)}_{n}(hq, \xi_0)]^{(1,2)}$ is given by Eq.~\eqref{eq:Sigma_0_n-Pade} and the function $\beta^1_{1}(\xi_0) = -2+1/(1+\xi_0^2)$.

\begin{figure}
	\includegraphics[width=0.45\textwidth]{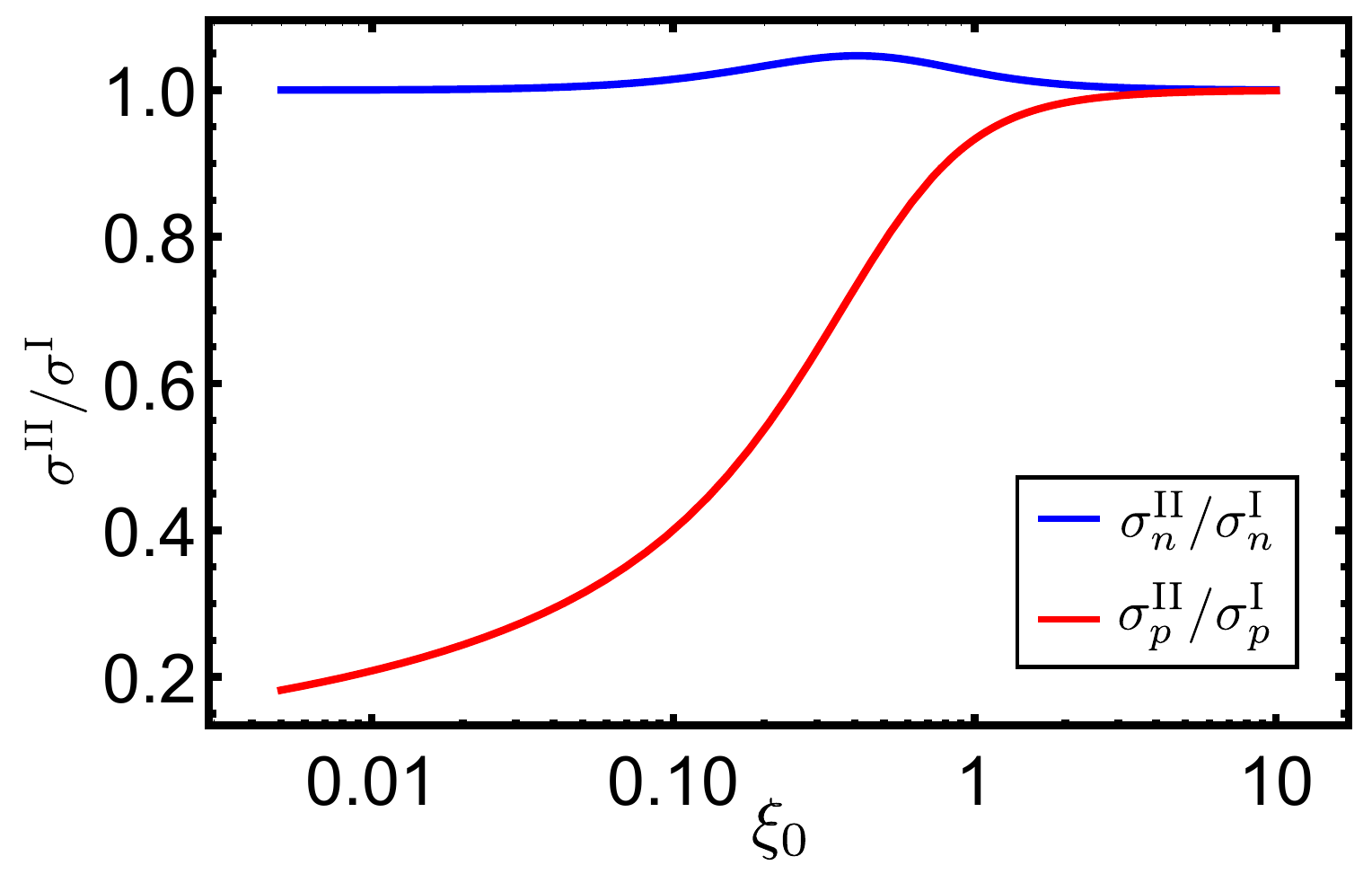}
	\caption{The ratio of $\sigma_{n}^{\rm II}/\sigma_{n}^{\rm I}$ and $\sigma_{p}^{\rm II}/\sigma_{p}^{\rm I}$ are plotted with the blue and red curves, respectively, as a function of the $\xi_0$.} 
	\label{fig:sigma_n_p_ratio}
\end{figure}

Let us compare the polarization response for the two different ion distributions. The polarization response for the two cases differs only in the form of the conductivities $\sigma_{p,n}$ and $\sigma_{w,p(n)}$. It is useful to note the following relations:
\beq
\begin{split}
\frac{\sigma_{p}^{\rm II}}{\sigma_{p}^{\rm I}}=& \frac{\sigma_{w,p}^{\rm II} -\sigma_w}{\sigma_{w,p}^{\rm I}-\sigma_w} = - \frac{2 \widetilde{\Gamma}_0}{\Gamma_0 \sqrt{1+\xi_0^2} b_{1,1}(\xi_0)},
\\
\frac{\sigma_{n}^{\rm II}}{\sigma_{n}^{\rm I}}=& \frac{\sigma_{w,n}^{\rm II} -\sigma_w}{\sigma_{w,n}^{\rm I}-\sigma_w} =  \frac{ \widetilde{\Gamma}_0 \beta_{1}^{1}(\xi_0)}{\Gamma_0 b^{\perp}_{1,1}(\xi_0)}
\end{split}
.
\eeq
If we assume that the particle has a fixed amount of total surface charge, then $\widetilde{\Gamma}_0/\Gamma_0$ can be obtained by using Eq.~\eqref{eq:Q-Gamma} and is given by
\beq
\frac{\widetilde{\Gamma}_0}{\Gamma_0} = \frac{3}{2} \frac{1+\xi_0^2}{1+3 \xi_0^2} \left( 1+ \frac{\xi_0^2}{\sqrt{1+\xi_0^2}} \tanh^{-1}\left( \frac{1}{\sqrt{1+\xi_0^2}}\right) \right).
\eeq
As shown in Fig.~\ref{fig:sigma_n_p_ratio}, the ratio $\sigma_{n}^{\rm II}/\sigma_{n}^{\rm I}$ is around $1$ for the full range of aspect ratios, but $\sigma_{p}^{\rm II}/\sigma_{p}^{\rm I}$ becomes smaller for $\xi_0 \ll 1$ and approaches $1$ for the spherical limit $\xi_0\gg 1$.
In the limit $\xi_0\ll 1$, we have
\beq
\begin{split}
\frac{\sigma_{p}^{\rm II}}{\sigma_{p}^{\rm I}} \sim& \frac{2}{2 \ln 2 -1 -2 \ln \xi_0}+ \mathcal{O}(\xi_0^2)
\\
\frac{\sigma_{n}^{\rm II}}{\sigma_{n}^{\rm I}} \sim& 1+ \mathcal{O}(\xi_0^2)
\end{split}
.
\eeq
Thus, $\sigma_n$ and $\sigma_{w,n}-\sigma_w$ are largely unaffected by the form of the ion distribution in the two cases, which means $P_n$ depends mainly on the total charge. In addition, these results show that $\sigma_{p}^{\rm II}$ and $\sigma_{w,p}^{\rm II} -\sigma_w$ are suppressed and hence imply a weaker low-frequency polarization response for $P_p^{\rm II}$ due to less charge on the side of the oblate spheroid. Because both $\sigma_{w,p}^{\rm II} -\sigma_w$ and $\sigma_{w,p}^{\rm I} -\sigma_w$ share the common factor $\mathcal{P} [ \Sigma^{(0)}_{p}(hq, \xi_0)]^{(1,2)}$, we expect that the polarization coefficients $P_p^{\rm I}$ and $P_p^{\rm II}$ have qualitatively the same frequency dependence but with different magnitudes. Hence, we will focus on case I with uniform surface ion distributions in the next section. The dielectric response will remain qualitatively the same but with weaker signals for non-uniform surface ion distributions when the electric field is applied along the axis of symmetry.

We shall close this section by commenting on the region of validity for our solutions. Even if we assume that Eq.~\eqref{eq:BC-low-frequency} is the proper BC for all frequencies, our solutions for the polarization coefficients are good approximations only in the frequency range, $\omega < \omega_h \equiv D/h^2$, because we kept only the first few orders in the expansion for small $h \omega^{1/2}$ of the spheroidal wave functions in Eq.~\eqref{eq:general-solution-n-omega+}. Although the additional Pad\' e approximation helps in joining these solutions smoothly to the high frequency solution, strictly speaking, the validity of these solutions in the region $\omega > \omega_h$ is still unknown. As discussed earlier, the full solutions for the polarization response which are valid in the frequency range $\omega>\omega_c$ require the inclusion of the Maxwell-Wagner effect. This will require the introduction of either of additional BCs at the interface between the particle and the electrolyte or the proper form of $\sigma_\pm$ in Eq.~\eqref{eq:BC-all-frequency}. Finally, because, in the spherical limit, $h\to 0$ and $a$ is held fixed and because the characteristic frequency is primarily determined by the factor $\omega_a \equiv D/a^2< \omega_h$, our solution will provide a good description over a wide dynamical range for spheroids with aspect ratios close to one. This is in turn why our solution exactly recovers the results for a spherical particle. 

\section{Dilute suspensions and dielectric enhancement}
\label{sec:dilute}

\begin{figure}
	\includegraphics[width=0.45\textwidth]{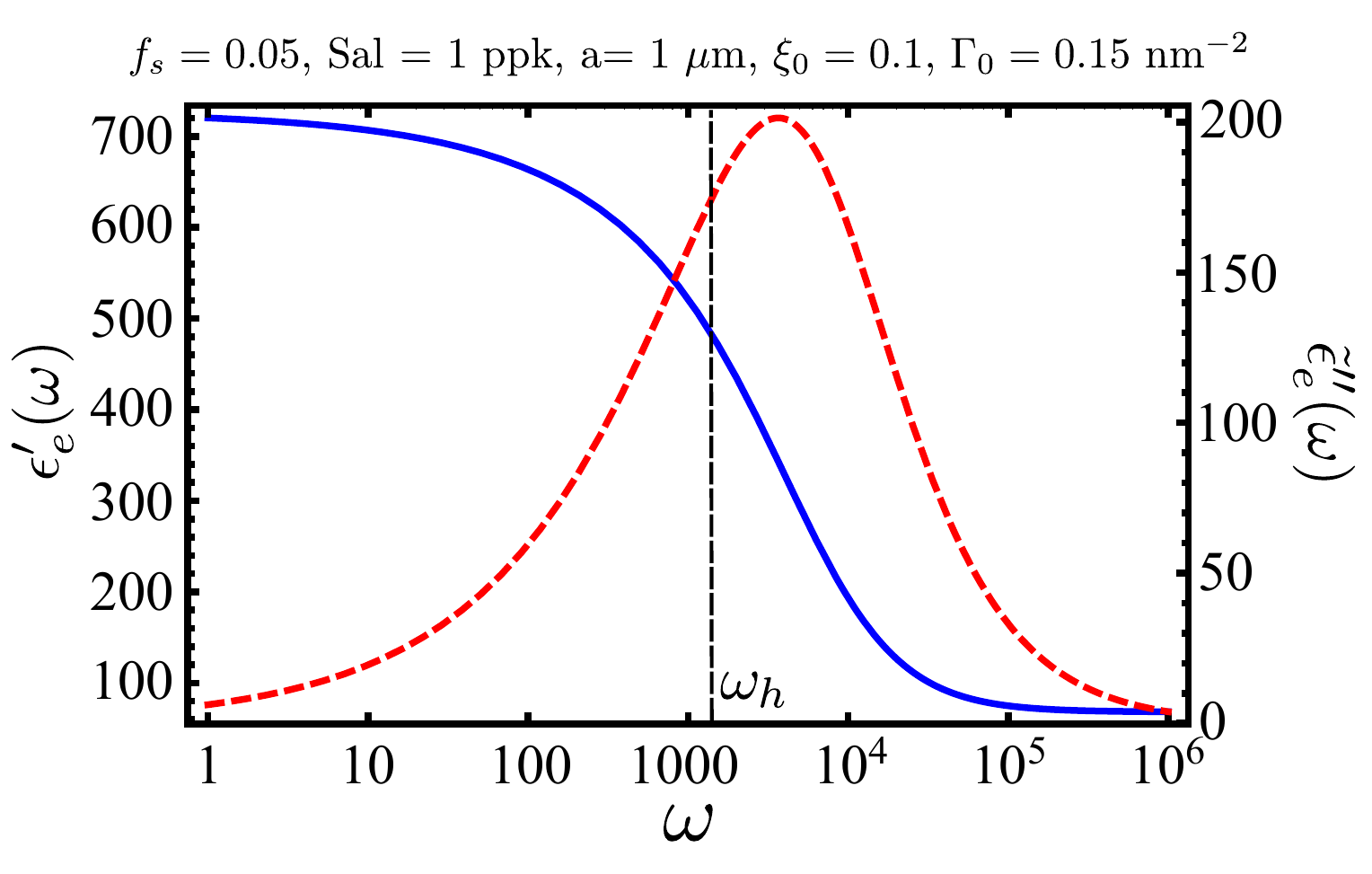}
	\caption{Frequency dependences of the relative permittivity $\epsilon'_e$ and the imaginary part of the dielectric constant after subtracting the DC contribution, $\tilde{\epsilon}''_e$, are plotted with the blue solid curve and the red dashed curve, respectively. Here, we used the following parameters: the sodium chloride electrolyte salinity, $1$ ppk, which has $\sigma_w\approx 0.2$ S/m and $\epsilon_w\approx 80$; the surface ion density, $\Gamma_0= 0.15$ nm$^{-2}$; the major radius of spheroids, $a=1$ $\mu$m; the volume fraction of spheroids, $f_s=0.05$; and $\xi_0=0.1$. We also use the diffusion coefficient $D=1.334\times 10^-9$ m$^2$/sec. An apparent enhancement of the relative permittivity is observed with a characteristic frequency around $\omega_c \sim 3500$ Hz as indicated by the peak position of the $\tilde{\epsilon}''_e$. The vertical dashed line indicates the position of $\omega_h=D/h^2$.} 
	\label{fig:example}
\end{figure}

\begin{figure}
	\includegraphics[width=0.45\textwidth]{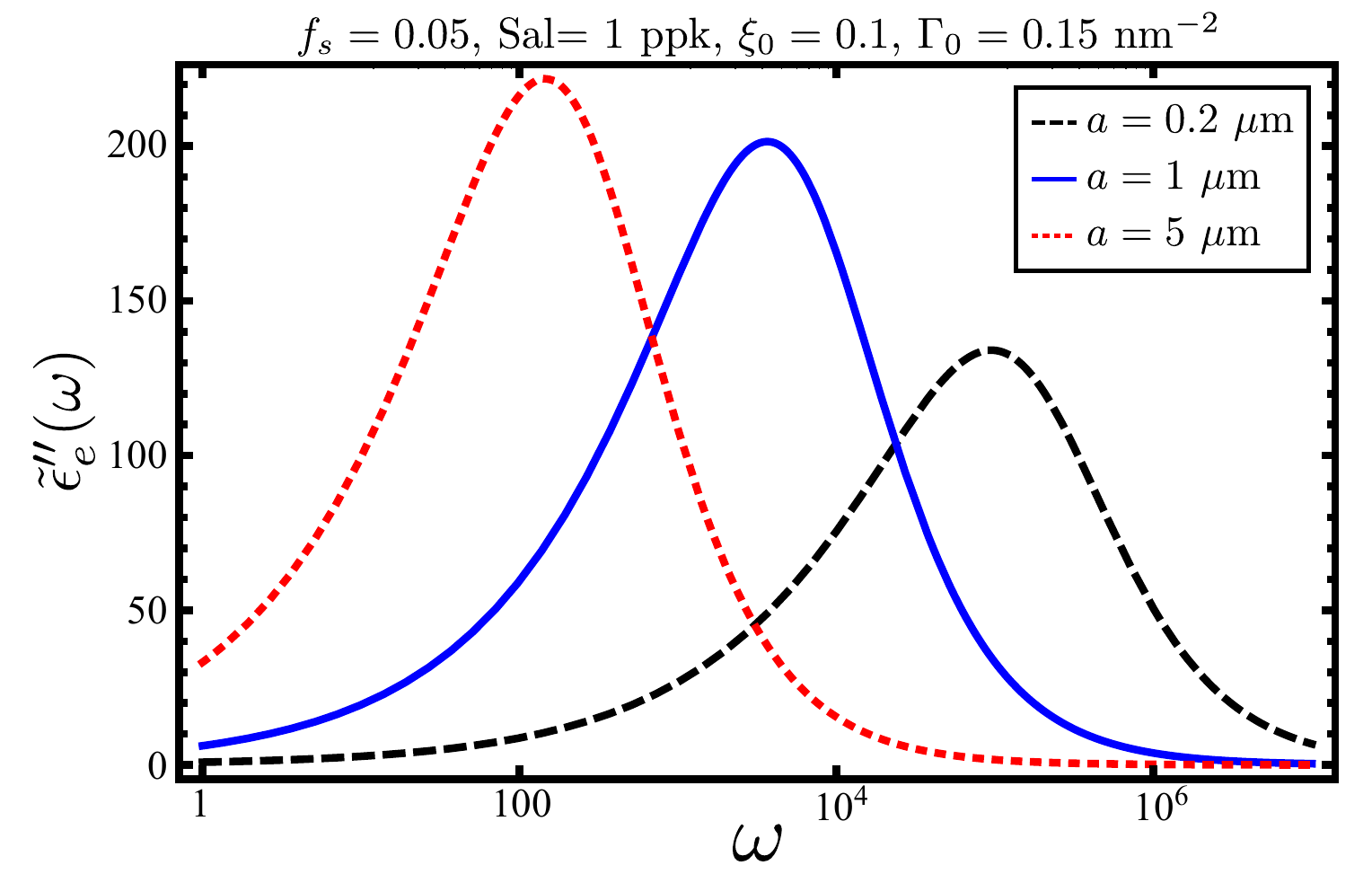}
	\caption{The $\tilde{\epsilon}''_e$ is plotted as a function of the frequency for three values of the major radius $a$. Apart from using different values of $a$, the rest of the parameters are the same as those in Fig.~\ref{fig:example}.} 
	\label{fig:peak-a_dep}
\end{figure}

\begin{figure}
	\includegraphics[width=0.45\textwidth]{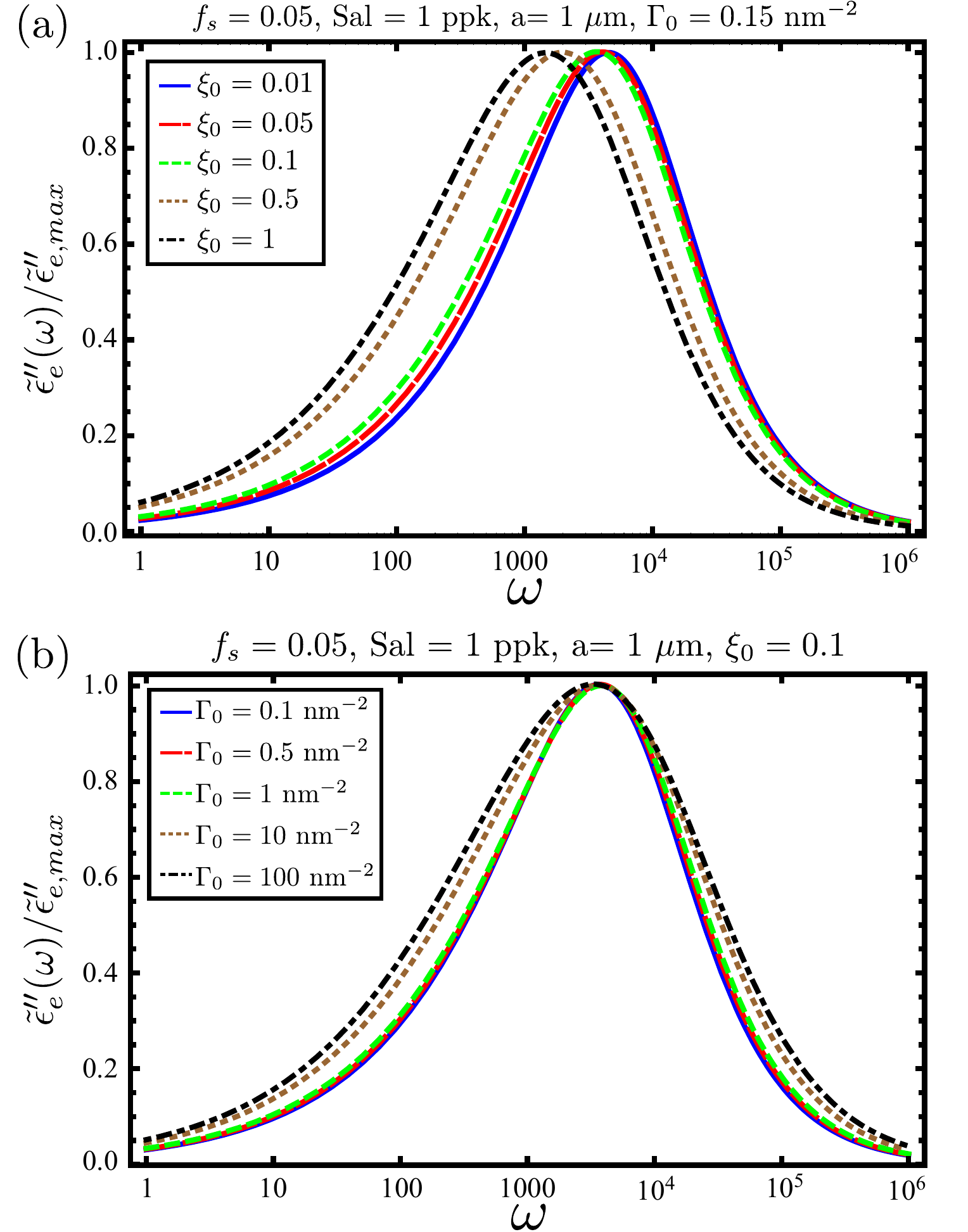}
	\caption{The $\tilde{\epsilon}''_e$ normalized by its maximum value, is plotted as a function of frequency (a) for five different values of $\xi_0$ and (b) for five different values of $\Gamma_0$. The values of $\xi_0$ and $\Gamma_0$ are indicated on plots. The rest of parameters are the same as those in Fig.~\ref{fig:example}.} 
	\label{fig:peak-xi0-Gamma0_dep}
\end{figure}

In this section, we aim to understand the consequences of the polarization coefficients derived in the previous section by considering the dielectric response of dilute suspensions of charged spheroids in an electrolyte. We derive two major results: (1) the strength of the dielectric enhancement at low frequency and (2) the characteristic frequency associated with this enhancement.  The former requires the explicit use of the solutions for the polarization coefficients summarized in Sec.~\ref{sec:results-polarization-coefficients}. Because these solutions are rigorous for small frequency and approach the exact representations when $\omega\ll \omega_h$, the strength of the dielectric enhancement derived in this section is very robust without any approximation. For the latter, we will invoke scaling arguments which still remain correct beyond the strictly applicable region of our solution and which allow us to identify the important parameters affecting the characteristic frequency. Finally, because the Pad\'e approximated solutions for the polarization coefficients ensure the proper low-frequency and high-frequency behavior~\cite{Freed-ex}, we shall argue, less rigorously, that they are still good representations in the intermediate frequency range. Hence, the peaks of imaginary part of dielectric constant after subtracting the DC contribution, i.e. $\tilde{\epsilon}''_e=(\sigma_e(\omega)-\sigma_e(\omega=0))/\varepsilon_0 \omega$, c.f. Figs.~\ref{fig:example},~\ref{fig:peak-a_dep} and~\ref{fig:peak-xi0-Gamma0_dep}, associated with the characteristic frequency represent good approximations.

In the dilute limit, \textit{i.e.}, when the volume fraction of spheroids $f_s\ll 1$, the effective dielectric constant $\varepsilon_{e}$ of an isotropic suspension can be approximated by the Maxwell-Garnett mixing formula as~\cite{Sihvola1999,Milton2002}
\begin{equation}
\label{eq:Maxwell-Garnett}
\varepsilon_{e}\equiv \left(\epsilon'_{e}+ i \frac{\sigma_{e}}{\varepsilon_0 \omega} \right)  = \left(\epsilon_w+ i \frac{\sigma_w}{\varepsilon_0 \omega} \right) \cdot \left( 1+ f_s(2 P_{n} +P_{p}) \right).
\end{equation}
Here, $\epsilon'_{e}$ and $\sigma_{e}$ are the relative permittivity (or the real part of the dielectric constant) and the conductivity of the suspension, respectively. As usual, $\varepsilon_0$ is the vacuum permittivity and $\omega$ is the frequency. 

To be concrete, we will consider the case of a uniform surface ion distribution with $P_{n}=P_{n}^{\rm I}$ and $P_{p}=P_{p}^{\rm I}$ in the following discussion. By partitioning the polarization coefficients into their real and imaginary parts, $P_{n}^{\rm I}= P_{n}^{\rm I'}+ i P_{n}^{\rm I''}$ and $P_{p}^{\rm I}= P_{p}^{\rm I'}+ i P_{p}^{\rm I''}$, we obtain
\beq
\label{eq:epsilon_e}
\epsilon'_e= \epsilon_w \left(1+ f_s \left(2 P_{n}^{\rm I'}+P_{p}^{\rm I'} \right) \right) - f_s \frac{\sigma_w}{\varepsilon_0 \omega} \left(2 P_{n}^{\rm I''}+P_p^{\rm I''} \right),
\eeq
and
\beq
\label{eq:sigma_e}
\frac{\sigma_{e}}{\varepsilon_0 \omega}= \frac{\sigma_w}{\varepsilon_0 \omega} \left(1+ f_s \left(2 P_{n}^{\rm I'}+P_{p}^{\rm I'} \right) \right) + f_s \epsilon_w \left(2 P_{n}^{\rm I''}+P_{p}^{\rm I''} \right). 
\eeq
Because the real parts of the polarization coefficients, $P_{n}^{\rm I'}$ and $P_{p}^{\rm I'}$, are on the order of $1$, the first term of each expression does not move the values of the effective permittivity and conductivity substantially away from those of the electrolyte when $ f_s\ll 1$. On the other hand, if the imaginary parts of the polarization coefficients have terms that are linearly proportional to $\omega$, a strong enhancement of the relative permittivity at low frequency becomes possible because the vacuum permittivity, $\varepsilon_0=8.85\times 10^{-12}$ F$\cdot$m$^{-1}$, is a very small value~\cite{Chew1982}.

Before we explore our analytical expressions in more detail, it is worthwhile to first give an illustration of a typical dielectric response when the enhancement of the relative permittivity becomes apparent. In Fig.~\ref{fig:example}, we plot the relative permittivity $\epsilon'_e$ with the blue solid curve as a function of the frequency. In addition, $\tilde{\epsilon}''_e$ is depicted with the dashed red curve. In this example, the electrolyte is a $1$ ppk sodium chloride solution, which has $\sigma_w\approx0.2$ S/m and $\epsilon_w=80$. The diffusion constant of the sodium ion is $D=1.334\times 10^{-9}$~m$^2/$sec.~\cite{CRC} If the suspended particles are glass beads, the surface charge density is approximately $\Gamma_0=0.15$ nm$^{-2}$ for a $1$ ppk electrolyte with neutral acidity~\cite{Behrens2001}. We also use the volume fraction $f_s=0.05$, $\xi_0=0.1$, and the major radius $a=1$~$\mu$m.

As depicted in Fig.~\ref{fig:example}, the relative permittivity is enhanced because its value is almost an order of magnitude larger than that of the electrolyte, $\epsilon_w \approx 80$, for low frequency. We will use $\epsilon'_e(\omega=0)$ to quantify the dielectric enhancement strength. We expect that the relative permittivity of a dilute suspension will reach $\epsilon_e \sim \epsilon_w$ for frequencies $\omega\gg \omega_c$ because we ignore the Maxwell-Wagner effect  in our solution. Finally, $\tilde{\epsilon}''_e$ has a prominent peak which indicates that there is a characteristic frequency, $\omega_c$, associated with the dielectric enhancement. 

Let us first focus on the relative permittivity of the suspension in the DC limit, which gives us the enhancement strength. After expanding the polarization coefficients to linear order in $\omega$ and performing some algebra, we have
\beq
\label{eq:P_p-linear-expansion}
P_{p}^{\rm I} \approx P_{p,\omega=0}^{\rm I} + i \frac{1}{3}  \frac{ \frac{Q_1(i \xi_0)}{d Q_1(i \xi_0)/d\xi_0 } \frac{i \varpi_{p}^2}{\xi_0 (1+\xi_0^2)} \frac{a^2 \omega}{D} }{ \left(L_{p} \sigma^{\rm I}_{p}  +(1-L_{p} ) \sigma^{\rm I}_{w,p}(\omega=0) \right)^2} (\sigma^{\rm I}_{p})^2,
\eeq
and
\beq
\label{eq:P_n-linear-expansion}
P_{n}^{\rm I} \approx P_{n,\omega=0}^{\rm I} + i \frac{1}{3}  \frac{ \frac{Q^1_1(i \xi_0)}{d Q^1_1(i \xi_0)/d\xi_0 } \frac{d P^1_1(i \xi_0)/d\xi_0}{P^1_1(i \xi_0) } \frac{i \varpi_{n}^2}{ (1+\xi_0^2)} \frac{a^2 \omega}{D} }{ \left(L_{n} \sigma^{\rm I}_{n} +(1-L_{n} )\sigma^{\rm I}_{w,n}(\omega=0) \right)^2} (\sigma^{\rm I}_{n})^2 .
\eeq
It is straightforward to show that the first term in each expansion is a real function. Because $i \varpi_{p}^2$, $i \varpi_{n}^2$, $\sigma^{\rm I}_{w,p}(\omega=0)$ and $\sigma^{\rm I}_{w,n}(\omega=0)$ are all real functions, the second term in each expansion is a pure imaginary function.

\begin{figure}
	\includegraphics[width=0.45\textwidth]{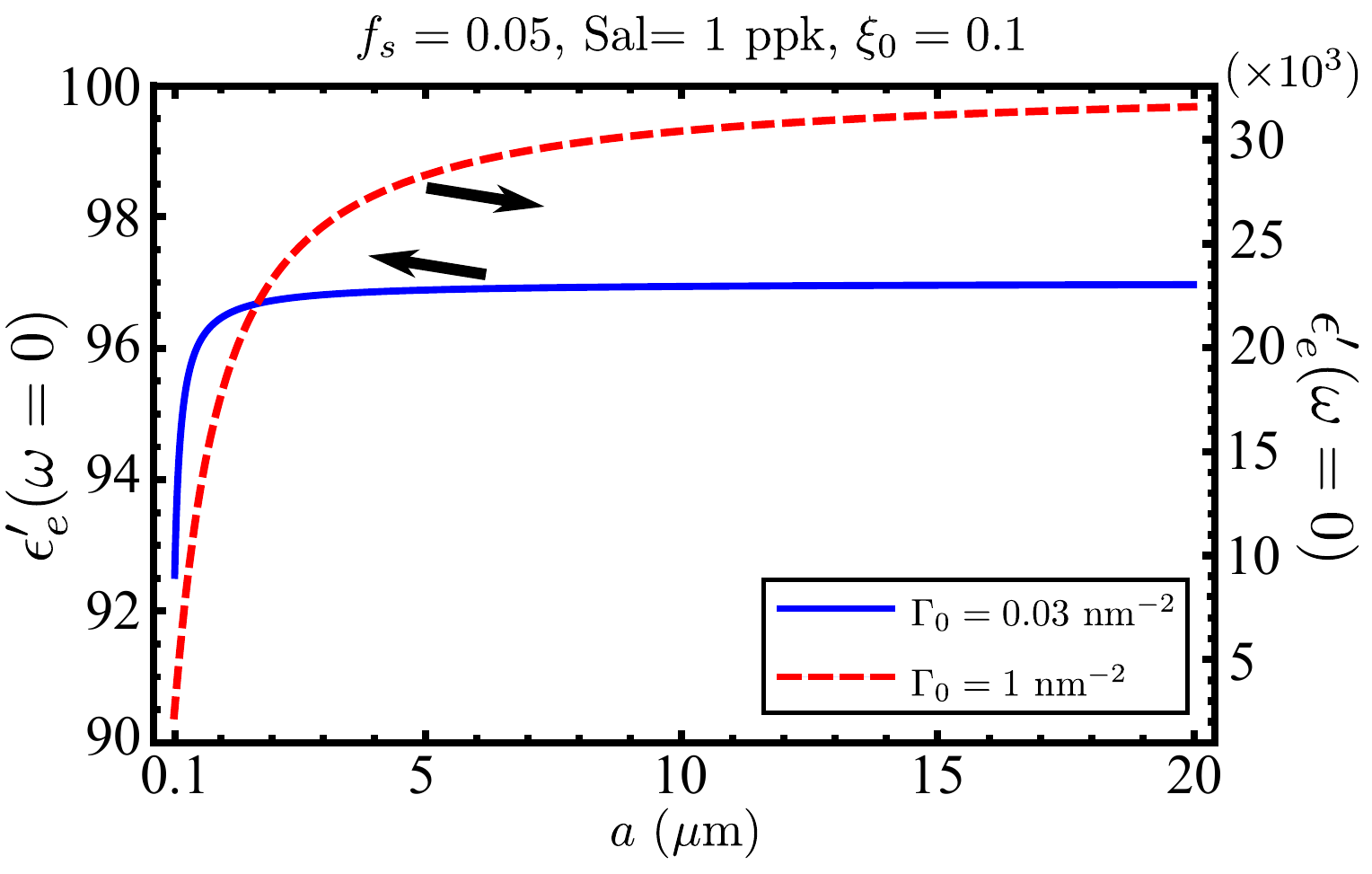}
	\caption{The zero-frequency relative permittivity is plotted as a function of the major radius of the spheroid, $a$, for two values of surface ion density, $\Gamma_0 = 0.03$ and $1$ nm$^{-2}$. Apart from taking different values of $\Gamma_0$, varying $a$ and taking the limit $\omega=0$, the rest of the parameters are the same as those in Fig.~\ref{fig:example}.} 
	\label{fig:epsilon_e-a-dep}
\end{figure}

With the expansions in Eqs.~\eqref{eq:P_p-linear-expansion} and~\eqref{eq:P_n-linear-expansion}, the zero-frequency relative permittivity reads
\beq
\label{eq:epsilon_e_DC}
\begin{split}
\epsilon'_e (\omega=0)=&\epsilon_w \left(1+ f_s \left(2 P_{n,\omega=0}^{\rm I}+P_{p,\omega=0}^{\rm I} \right) \right) 
\\
& \qquad \qquad  + f_s \frac{\sigma_w a^2}{\varepsilon_0 D} \mathcal{I} \left(\xi_0,\frac{\Gamma_0}{a N_0}\right),
\end{split}
\eeq
where
\beq
\begin{split}
\label{eq:I}
\mathcal{I} \left(\xi_0,\frac{\Gamma_0}{a N_0}\right) =&  - \frac{2}{3}  \frac{ \frac{Q^1_1(i \xi_0)}{d Q^1_1(i \xi_0)/d\xi_0 } \frac{d P^1_1(i \xi_0)/d\xi_0}{P^1_1(i \xi_0) } \frac{i \varpi_{n}^2}{ (1+\xi_0^2)} }{ \left(L_{n} \sigma^{\rm I}_{n} +(1-L_{n} )\sigma^{\rm I}_{w,n}(\omega=0) \right)^2} (\sigma^{\rm I}_{n})^2
\\
& - \frac{1}{3} \frac{ \frac{Q_1(i \xi_0)}{d Q_1(i \xi_0)/d\xi_0 } \frac{i \varpi_{p}^2}{\xi_0 (1+\xi_0^2)}}{ \left(L_{p} \sigma^{\rm I}_{p}  +(1-L_{p} ) \sigma^{\rm I}_{w,p}(\omega=0) \right)^2} (\sigma^{\rm I}_{p})^2 . 
\end{split}
\eeq
We first observe that $\mathcal{I} (\xi_0, \Gamma_0/a N_0)$ vanishes in the limit $\Gamma_0/a N_0\to 0$ because $\sigma^{\rm I}_{n}$ and $\sigma^{\rm I}_{p}$ are proportional to this quantity.  This implies that there is little or no enhancement of the relative permittivity in the absence of charge on the spheroid or in a high salinity environment.

Although it is tempting to argue that the low frequency dielectric enhancement scales as $a^2$ from Eq.~\eqref{eq:epsilon_e_DC}, it is important to notice that $\mathcal{I} (\xi_0,\Gamma_0/a N_0)$ can add some size dependence as well. If the amount of charge carried by the spheroid is linearly proportional to the volume of the particle, $\Gamma_0/a$ will be a function of $\xi_0$ only.  In that case, the dielectric enhancement will scale as $a^2$. In contrast, if the amount of charge carried by the spheroid is linearly proportional to the surface area of the particle, $\mathcal{I} (\xi_0,\Gamma_0/a N_0)$ will depend on $a$ explicitly. In this paper, we will focus on the latter case, where the scaling with respect to $a$ is more complicated. In the limit where $\Gamma_0/a N_0\ll 1$, one can show that $\mathcal{I} \propto 1/a^2$. Thus, the DC relative permittivity becomes almost independent of size. As shown in Fig.~\ref{fig:epsilon_e-a-dep}, the relative permittivity has a weak dependence on $a$ when $\Gamma_0/a N_0\ll 1$, c.f. the blue solid curve for $\Gamma_0=0.03$ nm$^{-2}$ and the region $a\gg1$ of the red dashed curve for $\Gamma_0=1$ nm$^{-2}$. On the other hand, an appreciable size dependence for the dielectric enhancement is observed when $\Gamma_0/a N_0 \sim 1$, as depicted by the red dashed curve in Fig.~\ref{fig:epsilon_e-a-dep}.

\begin{figure}
	\includegraphics[width=0.45\textwidth]{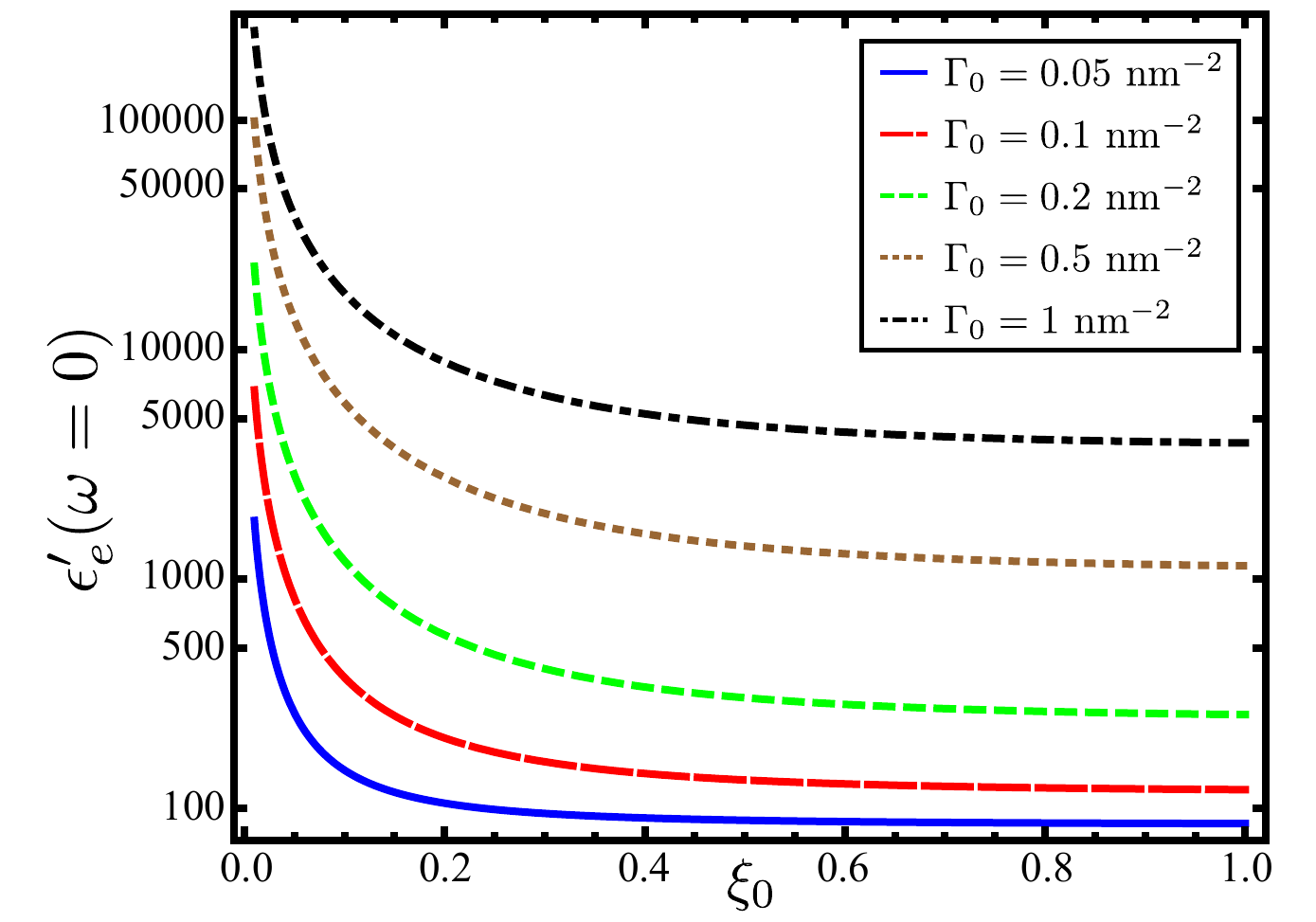}
	\caption{The zero-frequency relative permittivity is plotted as a function of $\xi_0$ for five values of surface ion density, $\Gamma_0 = 0.05$, $0.1$, $0.2$, $0.5$ and $1$ nm$^{-2}$. Apart from using different values of $\Gamma_0$, varying $\xi_0$, and taking the limit $\omega=0$, the rest of the parameters are the same as those in Fig.~\ref{fig:example}.} 
	\label{fig:epsilon_xi-dep}
\end{figure}

Because of the functional form of $\mathcal{I} (\xi_0,\Gamma_0/a N_0)$ and $\epsilon'_e(\omega=0)$ in Eq.~\eqref{eq:epsilon_e_DC}, the enhancement roughly scale as $\propto \Gamma_0^2$ and $\propto1/N_0$ in the limit $\Gamma_0/a N_0\ll 1$. Here, we have used $\sigma_w \propto N_0$. In the region $\Gamma_0/a N_0 \sim 1$, the enhancement can, however, have non-monotonic behavior with respect to the variable $\Gamma_0/a N_0$. The explicit $\xi_0$ dependence is not easy to obtain. However, as shown in Fig.~\ref{fig:epsilon_xi-dep}, the enhancement, in general, appears stronger at the smaller $\xi_0\ll 1$ region, \textit{i.e.}, the larger aspect ratio, while it becomes insensitive to $\xi_0$ for $\xi_0> 1$, \textit{i.e.}, approaching the spherical limit.

One important quantity to identify in Eqs.~\eqref{eq:epsilon_e} and~\eqref{eq:sigma_e} is the characteristic frequency, $\omega_c$, which is associated with the enhancement of the relative permittivity. As shown in Fig.~\ref{fig:example}, $\tilde{\epsilon}''_e$ has a peak centered around the frequency where the relative permittivity rises with decreasing frequency. Due to the symmetry of oblate spheroids, we will in principle have two distinct characteristic frequencies, $\omega_c^{p}$ and $\omega_c^{n}$, corresponding to the directions parallel and normal to the symmetry axis, respectively. However, because the widths of the peaks are of the same order as $\omega_c^{n,p}$, and $\Delta \omega_c = |\omega_c^{p}-\omega_c^{n}|$ is much smaller than $\omega_c^{n,p}$, we are unable to distinguish the two characteristic frequencies. Hence, for practical purposes, we will assume that there is an averaged characteristic frequency. 

From Eq.~\eqref{eq:sigma_e} and the equations for the polarization coefficient, Eq.~\eqref{eq:P_i-generic} through Eq.~\eqref{eq:varpi_n_1-Pade}, we can show that the peak position and the shape of $\tilde{\epsilon}''_e$ are mainly determined by the contribution from
\beq
\begin{split}
\tilde{\epsilon}''_e(\omega) =&\frac{\sigma_{e}(\omega)-\sigma_{e}(0) }{\varepsilon_0 \omega} 
\\
\sim&  f_s \frac{\sigma_w}{\varepsilon_0 \omega}  \left(2 (P_{n}^{\rm I'}-P_{n,\omega=0}^{\rm I})+ (P_{p}^{\rm I'}-P_{p,\omega=0}^{\rm I}) \right)
\\
=& f_s \frac{\sigma_w a^2}{\varepsilon_0 D} \mathcal{E}\left(a\sqrt{\frac{\omega}{D}}, \frac{\Gamma_0}{a N_0}, \xi_0 \right) ,
\end{split}
\eeq
where $\mathcal{E}$ is a function of $a\sqrt{\omega/D}$, $\Gamma_0/a N_0$, and $\xi_0$. Hence, the characteristic frequency can be, in general, expressed as $\omega_c= g(\xi_0,\Gamma_0/a N_0) \omega_a$, where $\omega_a \equiv D/a^2$ and $g(\xi_0,\Gamma_0/a N_0)$ is a function of $\xi_0$ and $\Gamma_0/a N_0$. The height of the peak will directly depend on the volume fraction of spheroids $f_s$ and shows scaling behavior with the other parameters as indicated by $\mathcal{E}$. Here, the reasons for using the length of the radius $a=h\sqrt{1+\xi_0^2}$, instead of $h$, as the characteristic length scale is twofold. First, the motion of the ions in the electrolyte should ultimately be affected by the explicit physical length scale, either $a$ or $b$, of the spheroid. Second, using the length-scale $a$ makes it is easier to connect our solution to the one in the spherical limit, given by $h\to 0$ and $\xi_0\to\infty$ with $h \sqrt{1+\xi_0^2} = a$ held  constant. As show in Fig.~\ref{fig:peak-a_dep}, $\tilde{\epsilon}''_e(\omega)$ is plotted for three different values of $a=0.2$, $1$, $5$ $\mu$m. The shift of peak's position is appreciable and satisfies the scaling behavior of the characteristic frequency, given by $\omega_c \propto 1/a^2$. For the examples shown in Fig.~\ref{fig:peak-a_dep}, one can see that the characteristic frequencies of adjacent peaks differ by roughly a factor of $25$, which is consistent with a $1/a^2$ scaling behavior.

By expanding $\tilde{\epsilon}''_e(\omega)$ for small values of $\Gamma_0/a N_0$, one can show that the characteristic frequency $\omega_c$ has a very weak dependence on $\Gamma_0/a N_0$ and a moderate dependence in $\xi_0$. We expect that $\omega_c \approx \omega_a$ in spherical limit $\xi_0\to \infty$.~\cite{Chew1982} In contrast, platier spheroids, with $\xi_0\ll1$, will have higher characteristic frequencies because the true length scale of the particle should effectively be reduced from $a$ to smaller values. This trend is observed in Fig.~\ref{fig:peak-xi0-Gamma0_dep}(a), where the peak position shifts from higher to lower frequencies and approaches $\omega_a$ when $\xi_0$ varies from $0.01$ to $1$. We notice that the $\omega_c$ does not have a very strong dependence on $\xi_0$ because it changes by about a factor of 4 when $\xi_0$ is varied by two orders of magnitude. Thus, it has the expected response with respect to $\xi_0$. Next, Fig.~\ref{fig:peak-xi0-Gamma0_dep}(b) shows how the peak position of $\tilde{\epsilon}''_{e}(\omega)$ varies as a function of $\Gamma_0$. We observe that the characteristic frequency depends very weakly on $\Gamma_0$ because it changes by only $15$ \% when $\Gamma_0$ varies by three orders of magnitude. In Figs.~\ref{fig:peak-xi0-Gamma0_dep}(a) and (b), we have normalized $\tilde{\epsilon}''_{e}(\omega)$ with respect to its  maximum for each value of $\xi_0$ and $\Gamma_0$ to clearly show how the peak position, or characteristic frequency, shifts. The actual peak height can have a strong dependence on $\xi_0$ and $\Gamma_0$. In general, it becomes larger for smaller values of $\xi_0$ as well as for larger values of $\Gamma_0$.

\section{Conclusions}
\label{sec:conclusions}

In this paper, we studied the low-frequency polarization response of a charged, oblate spheroid immersed in an electrolyte and obtained analytical solutions for the polarization coefficients for two different surface-ion distributions. When there are fewer cations on the edge of the oblate spheroid, the polarization response becomes weaker along the symmetry axis but remains roughly constant in the direction normal to the symmetry axis. Our approach is based on approximate BCs in the thin electric double layer limit, and on  expansions of spheroidal wave functions for the solutions of the diffusion equation.  We make two low-frequency approximations.  First, in the BCs, we do not take into account the conventional Maxwell-Wagner effect for the dielectric enhancement, and, second, we keep only the first few leading terms in the expansions of the spheroidal wave functions.   Because both approximations become rigorous in the low frequency regime, $\omega\ll \omega_h$, the derived solutions are, strictly speaking, valid only at low frequency. However, using a Pad\'e approximation, we are able to smoothly match our solution onto the high-frequency polarization coefficients~\cite{Freed-ex}. Remarkably, in the spherical limit, the Pad\' e approximated polarization coefficients for the charged spheroid approach those for the charged sphere when the Maxwell-Wagner effect is ignored~\cite{Chew1982,Chassagne2008}. This indicates that our solution will be applicable for a wider range of frequencies for particles with aspect ratios close to one.

We then incorporate our solutions into the Maxwell-Garnett mixing law to understand the dielectric response of a dilute suspension of charged spheroids in an electrolyte. We find that the enhancement of the relative permittivity depends on the surface charge density, the size (the major radius), the concentration of cations, and the shape (aspect ratio) of the charged spheroids. The functional dependence of the enhancement on the parameters is rather complicated. In the limit $\Gamma_0/a N_0\ll 1$, it generally scales as $\Gamma_0^2$ and $1/N_0$, and it becomes independent of the size $a$ of the spheroid. Interestingly, spheroids with higher aspect ratios ($\xi_0\ll 1$), which lead to a stronger enhancement due to the Maxwell-Wagner effect, also cause stronger enhancements due to the double layer effect.  Finally, using a scaling argument, we found that the characteristic frequency associated with the enhancement is largely determined by the size of the spheroids and has a generic functional form, $\omega_c= g(\xi_0,\Gamma_0/a N_0) D/a^2$, which depends moderately on $\xi_0$ (or aspect ratio) and very weakly on $\Gamma_0/a N_0$.

Several outstanding issues could be of interest for future research. First, throughout our discussion, we ignored the direct interfacial effect between the particle and the electrolyte, which becomes important in the frequency range $\omega > \omega_c$. Including this effect requires introducing additional BCs right at the interface or using the BCs in Eq.~\eqref{eq:BC-all-frequency} instead of those in Eq.~\eqref{eq:BC-low-frequency}, along with self-consistent perturbed ion surface densities, $\sigma_{\pm}$. In addition, we ignored the effect of convective polarization. To take this into account, one must solve for the velocity profile of the fluid in the double layer~\cite{Fixman1983}. Third, numerical studies, similar to ones for the charged sphere and prolate spheroid~\cite{Mangelsdorf1998a,Mangelsdorf1998b,Chassagne2013}, could be beneficial for understanding the polarization response of a charged spheroid immersed in an electrolyte. Finally, systematic experiments with controlled parameters would be essential for qualitatively and quantitatively confirming the validity of the current theory.

\section*{Acknowledgment}

The authors would like to thank M. Hurlimann, D. L. Johnson, J. Qian and N. Seleznev for useful discussions. P. N. Sen also thanks R. Kan for early collaborations on related subjects. This work is supported by the Petrophysics program at Schlumberger-Doll Research.

\begin{appendix}

\section{Legendre's functions}
\label{app:Legendre}

In this Appendix, the basic properties and our conventions for the associated Legendre function are reviewed. The Legendre functions are the solutions of the Poisson equations in spheroidal coordinates. They will also be used as the basis for the expansions of the spheroidal wave functions in Appendix~\ref{app:spheroidal-wave-function}. For a more extensive summary, we refer the reader to the Mathematical Handbook by Abramowitz and Stegun~\cite{Abramowitz1964}. The associated Legendre functions are a class of solutions to the following differential equation
\beq
\label{eq:Legendre-equation}
(1-z^2) \frac{d^{2} w}{d z} - 2 z \frac{d w}{dz}+ \left[ \nu (\nu+1) -\frac{\mu^2}{1-z^2} \right]w=0
\eeq  
where $z$ can be a complex variable, and $\nu$ and $\mu$ can be arbitrary complex parameters. For our purposes, we only need those with integer indices, $\nu\equiv\ell$ and $\mu \equiv m$. The solutions of Eq.~\eqref{eq:Legendre-equation} are then denoted by $P_{\ell}^{m}(z)$ and $Q_{\ell}^{m}(z)$ for the associated Legendre polynomials of the first and second kinds, respectively. Also, the solutions for $m=0$ are called the Legendre polynomials and are often denoted by $P_\ell(z)$ and $Q_\ell (z)$ without specifying $m=0$.

The Legendre polynomials are well documented in the literature. Here, we list the ones that are useful for our application~\cite{Abramowitz1964}. The first three Legendre polynomials of the first kind with arbitrary complex variable read 
\beq
\label{eq:P-0-2}
P_{0} (z) = 1,\qquad  P_{1} (z)=z,\qquad  P_{2} (z)=\frac{1}{2}(3z^2-1). 
\eeq
The first four Legendre polynomials of the second kind with a complex variable $z$ or a real variable $|z|>1$ are given by
\beq
\label{eq:Q-0-2}
\begin{split}
Q_{0} (z) =& \frac{1}{2} \ln \left(\frac{z+1}{z-1}\right), \qquad Q_{1} (z)=\frac{z}{2} \ln \left(\frac{z+1}{z-1}\right)-1,
\\
Q_{2} (z)=& \frac{3z^2-1}{4} \ln \left(\frac{z+1}{z-1}\right)-\frac{3z}{2},
\\
Q_{3} = & \frac{z(5z^2-3)}{4}   \ln \left(\frac{z+1}{z-1}\right) +\frac{2}{3} -\frac{5z^2}{2}.
\end{split}
\eeq
For Legendre polynomials of the second kind with real variable, $z \to x$ and $|x|<1$, one needs to change the functional form of $\ln \left(\frac{z+1}{z-1}\right)$ to $\ln \left(\frac{1+x}{1-x}\right)$ in Eq.~\eqref{eq:Q-0-2} for a proper branch cut. 

For complex variables, the associated Legendre polynomials with $m\neq 0$ can be defined by the  following relations  
\beq
\label{eq:def-P-l-m}
P_{\ell}^{m} (z)= (z^2-1)^{m/2} \frac{d^m}{dz^m} P_{\ell}(z), \; Q_{\ell}^{m} (z)=  (z^2-1)^{m/2} \frac{d^m  }{dz^m} Q_{\ell}(z).
\eeq
When $z\to x$, for $x$ a real variable with $|x|<1$, the associated Legendre's polynomials follow the relation
\beq
\label{eq:def-P-l-m-real}
P_{\ell}^{m} (x)= (1-x^2)^{m/2} \frac{d^m}{dx^m} P_{\ell}(x), \; Q_{\ell}^{m} (x)=  (1-x^2)^{m/2} \frac{d^m}{dx^m} Q_{\ell}(x).
\eeq
We note that definitions in Eq.~\eqref{eq:def-P-l-m-real} are different from that in Ref.~\onlinecite{Abramowitz1964} by a factor of $(-1)^m$, but are consistent with that in Ref.~\onlinecite{Flammer1957}, which discusses the spheroidal wave functions. 

In our discussion, explicit forms of the following associate Legendre's functions are used for the angular solutions with $-1<\eta<1$, 
\beq
P_{1} (\eta)= \eta,\qquad P_{1}^{1} (\eta) = \sqrt{1-\eta^2}.
\eeq
In addition, the radial sector for the spheroidal wave functions, discussed in Appendix~\ref{app:spheroidal-wave-function} for $ 0 < \xi_0 < \xi$, are expanded in terms of associated Legendre polynomials with pure imaginary variables. The following Legendre polynomials,
\beq
\begin{split}
P_{1}(i \xi) =& i \xi, \qquad Q_{1}(i \xi) = \xi \tan^{-1}(1/\xi) - 1,
\\
Q_{3}(i \xi) =& -  \frac{\xi}{2} (3+5 \xi^2) \tan^{-1}(1/\xi) +\frac{2}{3}+\frac{5 \xi^2}{2},
\end{split}
\eeq
and associated Legendre polynomials,
\beq
\begin{split}
P^{1}_{1}(i \xi) =& i \sqrt{\xi^2+1}, \; Q^{1}_{-1}(i \xi) =i \xi Q^{1}_{0}(i \xi) = - \frac{\xi}{\sqrt{\xi^2+1}},
\\
Q^{1}_{1}(i \xi) =&  \sqrt{\xi^2+1} \tan^{-1}(1/\xi)- \frac{\xi}{\sqrt{\xi^2+1}},
\\
Q^{1}_{3}(i \xi) =& \frac{13 \xi+15 \xi^3}{2 \sqrt{\xi^2+1}} -\frac{3  (1+5\xi^{2}) \sqrt{\xi^2+1}}{2} \tan^{-1}(1/\xi),
\end{split}
\eeq
are used for our discussions.

\section{Review of spheroidal wave functions}
\label{app:spheroidal-wave-function}

In this Appendix, we review the spheroidal wave functions that can be used to express general solutions of the diffusion equation for the perturbed ion density in Eq.~\eqref{eq:EOM-outside-DL-b}. Here, we follow closely the conventions for spheroidal wave functions in Ref.~\onlinecite{Flammer1957}. Although we are mainly interested in the oblate spheroid, we will also review the prolate case with the understanding that our approach can be applied to study the prolate spheroid as well.

Let us start by defining two families of spheroidal coordinates by their relation to Cartesian coordinates. For the prolate case, we have~\cite{Flammer1957} 
\beq
\begin{split}
x=& h \sqrt{(\xi^2-1)(1-\eta^2)} \cos \phi,
\\
y=& h \sqrt{(\xi^2-1)(1-\eta^2)} \sin \phi,
\\
z=& h \xi \eta,
\end{split}
\eeq
where $h$ is the half distance between the two focal points along the $z$-axis. Here, $1 \leq \xi < \infty$ can be understood as the radial direction, $-1 \leq \eta \leq 1$ plays the role of the polar angle and $0 \leq \phi \leq 2\pi$ is the azimuthal angle. The scale factors for the prolate spheroidal coordinates are given by
\beq
\begin{split}
h_{\xi} =& h \sqrt{ \frac{\xi^2-\eta^2}{\xi^2-1} }, \qquad h_{\eta} = h \sqrt{ \frac{\xi^2-\eta^2}{1- \eta^2} },
\\
h_{\phi} =&  h \sqrt{(\xi^2-1)(1-\eta^2)}.
\end{split}
\eeq
The relations between oblate spheroidal and Cartesian coordinates are given in Eq.~\eqref{eq:spheroidal-Cartesian} with their scale factors given by
\beq
\begin{split}
h_{\xi} =& h \sqrt{ \frac{\xi^2+\eta^2}{\xi^2+1} }, \qquad h_{\eta} = h \sqrt{ \frac{\xi^2+\eta^2}{1- \eta^2} },
\\
h_{\phi} =& h \sqrt{(\xi^2+1)(1-\eta^2)}.
\end{split}
\eeq

We are interested in solutions to the scalar Helmholtz differential equation 
\beq
\label{eq:Helmholtz-eq}
\left(\nabla^2 + \kappa^2  \right) f(\boldsymbol{r}) =0,
\eeq
where $\kappa$ can be a complex parameter. This equation can describe many physical phenomena. For instance, lossless wave propagation in the frequency domain is governed by Eq.~\eqref{eq:Helmholtz-eq} when $\kappa$ is a real number. In addition, the diffusion equation in the frequency domain, such as Eq.~\eqref{eq:EOM-outside-DL-b}, is also described by Eq.~\eqref{eq:Helmholtz-eq} with $\kappa^2$ being purely imaginary. The Helmholtz differential equation can be written explicitly as
\begin{equation}
\label{eq:H-eq-prolate}
\begin{split}
&\left[\frac{\partial}{\partial \xi} (\xi^2-1) \frac{\partial}{\partial \xi} + \frac{\partial}{\partial \eta} (1-\eta^2) \frac{\partial}{\partial \eta} \right.
\\
 & \qquad \left.  + \frac{\xi^2-\eta^2}{(\xi^2-1)(1-\eta^2)}\frac{\partial^2}{\partial \phi^2} + c^2 (\xi^2-\eta^2)  \right] f=0,
\end{split}
\end{equation}
for prolate spheroidal coordinates and 
\beq
\label{eq:H-eq-oblate}
\begin{split}
&\left[\frac{\partial}{\partial \xi} (\xi^2+1) \frac{\partial}{\partial \xi} + \frac{\partial}{\partial \eta} (1-\eta^2) \frac{\partial}{\partial \eta} \right.
\\
&  \qquad \left.+ \frac{\xi^2+\eta^2}{(\xi^2+1)(1-\eta^2)}\frac{\partial^2}{\partial \phi^2} +c^2 (\xi^2+\eta^2)  \right] f=0,
\end{split}
\eeq
for oblate spheroidal coordinates. Here $c \equiv h \kappa$. Because Eq.~\eqref{eq:H-eq-prolate} can be transformed to Eq.~\eqref{eq:H-eq-oblate} with $\xi \to \pm i \xi$ and $c \to \mp i c$, we can use these transformations to relate solutions in the prolate spheroidal coordinates to those in oblate spheroidal coordinates.

With the standard separation-of-variable technique, solutions to the Helmholtz equation Eq.~\eqref{eq:H-eq-prolate} in prolate spheroidal coordinates can be expressed as~\cite{Flammer1957}
\begin{align}
\label{eq:H-eq-solution-prolate}
f_{mn}= R_{mn} (c, \xi) \cdot S_{mn}(c, \eta) \cdot
\begin{split}
&\cos
\\
&\sin 
\end{split}
m \phi
\end{align}
where $R_{mn} (c, \xi)$ and $S_{mn}(c, \eta)$ satisfy the ordinary differential equations
\beq
\label{eq:ODE-xi-eta-prolate}
\begin{split}
\frac{d}{d \xi} \left[(\xi^2-1)\frac{d}{d \xi} R_{mn} \right] - \left[ \lambda_{mn} -c^2 \xi^2 + \frac{m^2}{\xi^2-1}\right] R_{mn}&=0,
\\
\frac{d}{d \eta} \left[(1-\eta^2)\frac{d}{d \eta} S_{mn}  \right] +  \left[ \lambda_{mn} -c^2 \eta^2 - \frac{m^2}{1-\eta^2}\right] S_{mn} &=0.
\end{split}
\eeq
Here, the separation constants $m$ and $\lambda_{mn}$ are the same in both of the equations in Eq.~\eqref{eq:ODE-xi-eta-prolate}. According to the transformation relating Eqs.~\eqref{eq:H-eq-prolate} and~\eqref{eq:H-eq-oblate}, the solutions of Eq.~\eqref{eq:H-eq-oblate} in oblate spheroidal coordinates are
\begin{align}
\label{eq:H-eq-solution-oblate}
f_{mn}= R_{mn} (-i c, i \xi) \cdot S_{mn}(-i c, \eta) \cdot
\begin{split}
&\cos
\\
&\sin 
\end{split}
m \phi
\end{align}
where $R_{mn} (-i c, i \xi)$ and $S_{mn}(-i c, \eta)$ satisfy the ordinary differential equations in Eq.~\eqref{eq:ODE-xi-eta-prolate} with the transformations, $\xi \to  i \xi$ and $c \to - i c$.
We note that all the equations in~\eqref{eq:ODE-xi-eta-prolate} reduce to Legendre equations when $c=0$. Hence, we have $\lambda_{mn} (c=0) = n (n+1)$ with $n$ an integer. As a result, it is useful to expand both $R_{mn}$ and $S_{mn}$ in terms of Legendre polynomials, especially in the region $c \ll 1$.

In what follows, we will focus on the solutions in oblate spheroidal coordinates.

\subsection{Solutions for the angular sector}

The two orthogonal solutions $S_{mn}(- i c, \eta )$ for the angular coordinate $\eta$, given a set of eigenvalue $\lambda_{mn}$ and azimuthal index $m$, can be formally expressed as expansions over Legendre polynomials as
\begin{subequations}
	\label{eq:S_mn-expansion}
	\bea
	\label{eq:S1_mn-expansion}
	S_{mn}^{(1)}(- i c, \eta ) = \sideset{}{'}\sum_{r=0,1}^{\infty} d_{r}^{mn}(- i c)  P_{m+r}^m(\eta),
    \\
    \label{eq:S2_mn-expansion}
    S_{mn}^{(2)}(- i c, \eta ) = \sideset{}{'}\sum_{r=-\infty}^{\infty}d_{r}^{mn}(- i c)  Q_{m+r}^m(\eta).
	\eea
\end{subequations}
Here, the $\sum\nolimits'$ indicates that the summation is over even integer $r$ for even $n-m$  and is over odd integer $r$ for odd $n-m \in$. The expansion coefficients, $d_{r}^{mn}(- i c)$, are functions of $- i c$. Because $Q^{m}_{\ell}(\eta)$ is divergent at $\eta=\pm 1$, we will focus on the first kind type of solution for the angular sector. When $c=h\kappa =0$, $d_{n-m}^{mn}(0)=1$ and $d_{r\neq n-m}^{mn}(0)=0$ because Eq.~\eqref{eq:ODE-xi-eta-prolate} reduces to Legendre's equation. As a result, we expect that $r=n-m$ is the dominant term for the expansion in the limit $c \ll 1$. 

Using the recursion relations for the Legendre polynomials and the expansion in Eq.~\eqref{eq:S1_mn-expansion}, one can derive a recursion formula for the coefficients $d_{r}^{mn}$, c.f.~\onlinecite{Flammer1957}. This allows us to express the eigenvalues $\lambda_{mn}$ in terms of expansions of $c\equiv h\kappa$ as
\beq
\label{eq:lambda_mn-expansion}
\lambda_{mn} (- i c) = n(n+1) + \sum_{k=1} (-1)^k \cdot l_{2k}^{mn} \cdot c^{2k}, 
\eeq
where $l_{2k}^{mn}$ are coefficients depending on $n$ and $m$ which are well documented in Refs.~\onlinecite{Flammer1957} and~\onlinecite{Abramowitz1964}. This expansion satisfies $\lambda_{mn}\to n(n+1)$ for $c \to 0$. The expansion coefficients, $d_{r}^{mn}(- i c)$, are completely determined by the recursion formula given a proper choice of normalization. We will follow the convention in Ref.~\onlinecite{Flammer1957} for normalization and require 
\begin{equation}
\begin{split}
&S_{mn}(-i c, \eta=0) = P_{n}^{m}(\eta=0)
\\
=& \left\{
\begin{split}
\frac{(-1)^{\frac{n-m}{2}} (n+m)! }{2^n \left(\frac{n-m}{2}\right)!  \left(\frac{n+m}{2} \right)!}, \quad \forall \quad (n-m) \; even 
\\
\frac{(-1)^{\frac{n-m-1}{2}} (n+m+1)! }{2^n \left(\frac{n-m-1}{2}\right)!  \left(\frac{n+m+1}{2} \right)!}, \quad \forall \quad (n-m) \; odd 
\end{split}
  \right.
  \end{split}
\end{equation}
Ultimately, the angular functions, $S_{mn}(-i c, \eta)$, form an orthogonal set of functions for $-1 \leq \eta \leq 1$. As a result, we have
\beq
\int_{-1}^{1} S_{mn} (-i c, \eta) S_{m n'} (-i c, \eta) d \eta = \delta_{n n'} N_{mn}, 
\eeq
where
\begin{equation}
N_{mn} = 2 \sideset{}{'}\sum_{r=0,1}^{\infty} \frac{(r+2m)!(d_{r}^{mn})^2}{(2r +2 m +1) r!}.
\end{equation}

Although the second type of solution, $S^{(2)}(-i c, z)$, is seldom used for the angular sector, it will become useful when $|z|>1$ and $z$ is complex. In this case, we have to use the proper definition of Legendre polynomials of complex variables, c.f. Eqs.~\eqref{eq:Q-0-2} and~\eqref{eq:def-P-l-m}. Furthermore, the summation of $S^{(2)}(-i c, z)$ in Eq.~\eqref{eq:S2_mn-expansion} extends to negative values of $r$, which requires extra care because $Q_{n}^{m}(z)$ diverges for $n<-m$. By combining the fact that $d_{r}^{mn}=0$ for all $r<-2m$ and using the identity for Legendre Polynomials with negative valued  indices, c.f. Ref.~\onlinecite{Flammer1957} and Eq. (8.2.2) in Ref.~\onlinecite{Abramowitz1964}, a non-divergent expansion of $S^{(2)}(-i c, z)$ can be obtained, which reads
\begin{equation}
\label{eq:S2_mn-expansion-1}
\begin{split}
S_{mn}^{(2)}(- i c, z ) =& \sideset{}{'}\sum_{r=-2m,-2m+1}^{\infty} d_{r}^{mn}(- i c)  Q_{m+r}^m(z) 
\\
& + \sideset{}{'}\sum_{r=2m+2,2m+1}^{\infty} d_{\rho| r}^{mn}(- i c) P_{r-m-1}^m (z),
\end{split}
\end{equation}
with coefficients $d_{\rho|r}^{mn} (-i c)$ functions of $-ic$.

\subsection{Solutions of radial sector}


Conventionally, the two orthogonal solutions for the radial sector for a given set of $\lambda_{mn}$ and $m$ are expanded in terms of spherical Bessel and Neumann functions, $j_\ell(x)$ and $n_\ell(x)$, as~\cite{Abramowitz1964}
\begin{widetext}
\begin{subequations}
\label{eq:R_mn-expansion-1}
\bea
\label{eq:R_mn1-expansion-1}
R_{mn}^{(1)} (-i c, i \xi) = \left(\sideset{}{'}\sum_{r=0,1}^{\infty} d_{r}^{mn}(i c) \frac{(2m+r)!}{r!} \right)^{-1}\cdot \left(\frac{\xi^2+1}{\xi^2}\right)^{m/2} \cdot \sideset{}{'} \sum_{r=0,1}^{\infty} i^{r+m-n} \frac{(2m+r)!}{r!} d_{r}^{mn} (-i c)  j_{m+r} (c \xi),
\\
\label{eq:R_mn2-expansion-1}
R_{mn}^{(2)} (-i c, i \xi) = \left(\sideset{}{'}\sum_{r=0,1}^{\infty} d_{r}^{mn}(i c) \frac{(2m+r)!}{r!} \right)^{-1}\cdot \left(\frac{\xi^2+1}{\xi^2}\right)^{m/2} \cdot \sideset{}{'} \sum_{r=0,1}^{\infty} i^{r+m-n} \frac{(2m+r)!}{r!} d_{r}^{mn} (-i c)  n_{m+r} (c \xi).
\eea
\end{subequations} 
\end{widetext}
By adding and subtracting $R_{mn}^{(1)}$ and $R_{mn}^{(2)}$, we can rewrite these solutions as
\begin{widetext}
\beq
\label{eq:R_mn34-expansion-1}
R_{mn}^{(3), (4)} (-i c, i \xi) \equiv R_{mn}^{(1)} \pm i R_{mn}^{(2)} = \left(\sideset{}{'}\sum_{r=0,1}^{\infty} d_{r}^{mn}(i c) \frac{(2m+r)!}{r!} \right)^{-1}\cdot \left(\frac{\xi^2+1}{\xi^2}\right)^{m/2} \cdot \sideset{}{'} \sum_{r=0,1}^{\infty} i^{r+m-n} \frac{(2m+r)!}{r!} d_{r}^{mn} (-i c)  h^{(1), (2)}_{m+r} (c \xi).
\eeq
\end{widetext}
Here, $h^{(1), (2)}_\ell(c \xi)$ are Hankel's functions. For the diffusion equation, \textit{i.e.}, when $c$ is a complex number, one of $R_{mn}^{(3),(4)}$ converges and the other diverges at large distance, $\xi \to \infty$, which allows us to easily select the solution with the proper boundary condition. However, although the expansion in Eq.~\eqref{eq:R_mn34-expansion-1} converge rapidly for long distances, $\xi\gg 1$, the convergence of these series for short distances can be very slow and problematic at times~\cite{Li1998}.

Because we are interested in matching BCs at the outer surface of the double layer, alternative expansions of $R_{mn}^{(3),(4)}$ which converge well at short distances are desired. Let us observe that the radial and angular equations in Eq.~\eqref{eq:ODE-xi-eta-prolate} are essentially the same. Hence, solutions to them must be proportional to each other. As a result, it is possible to identify solutions for radial wave functions of the form~\cite{Flammer1957}
\begin{equation}
\label{eq:R_mn-expansion-2}
R_{mn}^{(1), (2)} (-i c, i \xi) = \frac{S_{mn}^{(1), (2)} (-i c, i \xi) }{\kappa_{mn}^{(1),(2)} (-i c)}, 
\end{equation} 
with joining factors $\kappa_{mn}^{(1),(2)} (-i c)$ given by
\begin{widetext}
\beq
\label{eq:joining-factor-kappa1}
\kappa_{mn}^{(1)} (-i c) =
\left\{
\begin{split}
& \frac{ (2m+1)\cdot (n+m)! }{2^{n+m} \cdot d_{0}^{mn}(-ic) \cdot (-i c)^m \cdot m! \cdot \left(\frac{n-m}{2}\right)! \cdot \left(\frac{n+m}{2}\right)! } \cdot  \sideset{}{'}\sum_{r=0}^{\infty} d_{r}^{mn} (-i c) \frac{(2 m+r)!}{r!} , \qquad \qquad \forall \quad {\rm (n-m) \; even}
\\
& \frac{ (2m+3)\cdot (n+m+1)!  }{2^{n+m} \cdot d_{1}^{mn}(-ic)  \cdot (-i c)^{m+1} \cdot m!\cdot \left(\frac{n-m-1}{2}\right)!\cdot \left(\frac{n+m+1}{2}\right)! } \cdot  \sideset{}{'}\sum_{r=1}^{\infty} d_{r}^{mn} (-i c) \frac{(2 m+r)!}{r!}  , \; \; \,\quad \forall \quad {\rm (n-m) \; odd}
\end{split}
\right. ,
\eeq
and
\beq
\label{eq:joining-factor-kappa2}
\kappa_{mn}^{(2)} (-i c) =
\left\{
\begin{split}
	& \frac{ 2^{n-m} \cdot (2m)!\cdot \left(\frac{n-m}{2}\right)! \cdot \left(\frac{n+m}{2}\right)! \cdot d_{-2m}^{mn}(-ic) }{(2m -1) \cdot m! \cdot (n+m)! \cdot (-ic)^{m-1}  } \cdot  \sideset{}{'}\sum_{r=0}^{\infty} d_{r}^{mn} (-i c) \frac{(2 m+r)!}{r!} , \qquad \qquad \forall \quad {\rm (n-m) \; even}
	\\
	& -\frac{ 2^{n-m} \cdot (2m)!\cdot \left(\frac{n-m-1}{2}\right)! \cdot \left(\frac{n+m+1}{2}\right)! \cdot d_{-2m+1}^{mn}(-ic) }{(2m -3) (2m -1) \cdot m! \cdot (n+m+1)! \cdot (-ic)^{m-2}  } \cdot  \sideset{}{'}\sum_{r=1}^{\infty} d_{r}^{mn} (-i c) \frac{(2 m+r)!}{r!} , \quad \forall \quad {\rm (n-m) \; odd}
\end{split}
\right. . 
\eeq
\end{widetext}

Because Eq.~\eqref{eq:R_mn-expansion-2} is simply an alternative representation of Eq.~\eqref{eq:R_mn-expansion-1}, we have
\beq
\label{eq:Rmn_final}
R_{mn}^{(3),(4)}(-i c, i \xi) = \frac{S_{mn}^{(1)} (-i c, i \xi) }{\kappa_{mn}^{(1)} (-i c)} \pm i \frac{S_{mn}^{(2)} (-i c, i \xi) }{\kappa_{mn}^{(2)} (-i c)}.
\eeq
This is the expansion we will use for the radial sector because it converges well at short distances and goes to the correct limit for large $\xi_0$.

Finally, to use the solutions reviewed in this Appendix for the charge density which satisfies the diffusion equation~\eqref{eq:EOM-outside-DL-b}, we can simply take $c= e^{-is(\omega)\pi/4} h q$ with $q\equiv \sqrt{|\omega|/D}$.

\section{Polarization response of a charged oblate spheroid with a uniform surface ion distribution}
\label{app:Polarization-uniform}

In this appendix, we provide a detailed derivation to obtain the polarization coefficients of a charged oblate spheroid that is immersed in an electrolyte and has a uniform surface ion distribution given by $\Gamma^{\ }_{+}= \Gamma_0 $ and $\Gamma^{\ }_{-}= 0$. The case for a non-uniform ion distribution will be studied in the next Appendix. As mentioned in the main text, the angular components do not naturally decouple in our problem. Hence, a perturbation scheme is developed to solve for an approximate analytic expression for the polarization coefficients in the low frequency limit.

Formally, we need to match the BCs in Eq.~\eqref{eq:BC-low-frequency}
\beq
\label{eq:BC-DL}
J^{\rm t}_{\pm,\xi} (\boldsymbol{r},\omega)\big|_{\xi_0+\zeta} =  D h_{\xi}\nabla_{\parallel} \cdot \left( \frac{\Gamma_{\pm}}{h_{\xi}} \nabla_{\parallel} \mu_{\pm}(\boldsymbol{r},\omega)\right)\big|_{\xi_0+\zeta},
\eeq
at the outer surface of the double layer. In the thin double layer limit, we will further set $\zeta\to 0$ to simplify our discussion. Using definitions of the diffusive currents in Eq.~\eqref{eq:diffusive-current-1} and chemical potentials in Eq.~\eqref{eq:chmical-potential}, we obtain explicit expressions for the two coupled BCs, given by
\beq
\label{eq:BC-anion-uniform}
\left[ \frac{\partial}{\partial \xi} n_{>}(\xi,\eta, \phi,\omega)  -  N_0 \frac{\partial}{\partial \xi} \psi(\xi,\eta, \phi,\omega)  \right]_{\xi=\xi_0} =0,
\eeq
and 
\begin{widetext}
\begin{eqnarray}
\label{eq:BC-cation-uniform}
&&- \frac{1}{h_{\xi} (\xi_0)} \left[ \frac{\partial}{\partial \xi} n_{>}(\xi,\eta, \phi,\omega)  +  N_0 \frac{\partial}{\partial \xi} \psi(\xi,\eta, \phi,\omega)  \right]_{\xi=\xi_0}
\\
\nonumber
&=& \frac{\Gamma_0}{N_0} \left\{ \frac{1}{h_{\eta} h_{\phi}}  \frac{\partial}{\partial \eta} \left[ \frac{h_{\phi}}{h_\eta} \left( \frac{\partial}{\partial \eta} n_{>}(\xi,\eta,\phi,\omega) + N_0 \frac{\partial}{\partial \eta} \psi(\xi,\eta,\phi,\omega)  \right) \right] +  \frac{1}{ h_{\phi}^2 } \left( \frac{\partial^2}{\partial \phi^2} n_>(\xi,\eta,\phi,\omega) +  N_0 \frac{\partial^2}{\partial \phi^2} \psi(\xi,\eta,\phi,\omega) \right) \right\}_{\xi=\xi_0}. 
\end{eqnarray}
\end{widetext}

Because the $\ell = 1$ component of the perturbed electric field in Eq.~\eqref{eq:general-solution-psi} gives the strength of the dipole response, our primary goal is to obtain $A_1^m U_m$ or $A_1^m V_m$ to extract the polarization coefficients by matching the BCs, given in Eqs.~\eqref{eq:BC-anion-uniform} and~\eqref{eq:BC-cation-uniform}, when the electric field is applied along the two major axes. According to our derivation, we can also see that the $\ell = 1$ component is the dominant part of the response to the externally applied electric field. We will now summarize the key steps and essential results for evaluating the polarization coefficients in the $z$- and $x$-directions.

\subsection{Polarization coefficient $P_z$ in the $z$-direction}
\label{sec:P-z-uniform}

When an electric field $\boldsymbol{E}=E_0 \hat{z}$ is applied in the $z$-direction, the solutions for $\psi$ and $n$ are independent of the angular variable $\phi$ due to the azimuthal symmetry. Thus, the expansions given in Eqs.~\eqref{eq:general-solution-psi} and \eqref{eq:general-solution-n-omega+} for the electric potential and ion concentration when $\omega>0$ become
\beq
\label{eq:psi-z}
\psi^{z}= i E_0 h P_1 (i \xi) P_1(\eta)  + \sum_{\ell}  A_{\ell}(\omega)  \cdot Q_{\ell} (i \xi) \cdot  P_{\ell} (\eta),
\eeq
and
\beq
\label{eq:n-z}
n^{z}_{>} = \sum_{\ell} \alpha_{\ell}(\omega)  \cdot R_{0\ell}^{(3)} (-i e^{i \pi/4} h q, i \xi )  \cdot S_{0\ell}^{(1)} (-i e^{i \pi/4} h q, \eta ) .
\eeq
Here, the definition of spheroidal coordinates is used to write $z= h \xi \eta= -i h P_{1}(i \xi) P_1(\eta)$.  Because $m=0$, we also suppress the index $m$ in the coefficients $A_{\ell}$ and the $\alpha_{\ell}$. 

By inserting Eqs.~\eqref{eq:psi-z} and~\eqref{eq:n-z} into the anion BC in Eq.~\eqref{eq:BC-anion-uniform} and using the expansions for the angular spheroidal wave functions in Eq.~\eqref{eq:S1_mn-expansion}, we obtain
\begin{widetext}
\beq
\sum_{\ell} \sideset{}{'}\sum_{r=0,1}^{\infty} \alpha_{\ell} (\omega) \frac{ d R_{0\ell}^{(3)} (-i e^{i \pi/4} h q, i \xi_0 )}{d \xi_0}  d_{r}^{0\ell}(-i e^{i \pi/4} h q)  P_{r}(\eta) + N_0 \left( E_0 h P_1(\eta)  - \sum_{\ell}  A_{\ell}(\omega) \frac{d Q_{\ell} (i \xi_0)}{d \xi_0}   P_{\ell} (\eta) \right) =0,
\eeq
\end{widetext}
where the symbol $\sum\nolimits'$ indicates that the summation is over even $r$ for $\ell$ even and odd $r$ for $\ell$ odd. Using the orthogonal property of the Legendre polynomials, $\int_{-1}^{1} d \eta P_{\ell}(\eta) P_{\ell'}(\eta)=c_{\ell} \delta_{\ell \ell'}$ with $c_\ell= 2/(2\ell +1)$, we have
\beq
\label{eq:BC-anion-z-1}
\sideset{}{'}\sum_{\ell=0,1}^{\infty} K^z_{\ell',\ell} \alpha_{\ell} (\omega) = - N_0 \left( E_0 h \delta_{1 \ell'}  -  A_{\ell'}(\omega) \frac{d Q_{\ell} (i \xi_0)}{d \xi_0} \right).
\eeq
where $\delta_{\ell \ell'}$ is the Kronecker delta function, and elements of the matrix $K^{z}$ are given by
\beq
\label{eq:def-K_l_l}
K^z_{\ell',\ell} =D^{z}_{\ell', \ell}  \frac{ d R_{0, \ell}^{(3)} (-i e^{i \pi/4} h q, i \xi_0 )}{d \xi_0},
\;
D^{z}_{\ell', \ell} \equiv  d_{\ell'}^{0, \ell} (-i e^{i \pi/4} h q) .
\eeq
Here, we have defined the matrix $D^z$ matrix for convenience.

Formally, we can now write $\alpha_\ell$ in terms of the $A_{\ell}$ by using the inverse matrix $\bar{K}^z$ as follows
\beq
\label{eq:alpha_l-A_l_inverse}
\begin{split}
	\alpha_{\ell} =& N_0 \sum_{\ell'=0}^{\infty} \bar{K}^{z}_{\ell,\ell'}  \left( - E_0 h \delta_{\ell'1} + \frac{d Q_{\ell'} (i \xi_0)}{d \xi_0}  A_{\ell'} \right),
	\\
	\bar{K}^z_{\ell,\ell'} =& \left( \frac{ d R_{0, \ell}^{(3)} (-i e^{i \pi/4} h q, i \xi_0 )}{d \xi_0}\right)^{-1} \cdot \bar{D}^z_{\ell,\ell'} .
\end{split}
\eeq
where $\bar{D}^z$ is the  matrix inverse of $D^{z}$. Some useful observations are in order. From the properties of $d_{\ell'}^{0, \ell}$, it follows that $D^z_{\ell',\ell} =0$ when $\ell+\ell'$ is odd. So, sectors with even and odd indices in Eq.~\eqref{eq:BC-anion-z-1} are decoupled. As a result, $\bar{D}^z_{\ell,\ell'}$ and $\bar{K}^{z}_{\ell,\ell'}$ are also zero when $\ell+\ell'$ is odd, \textit{i.e.}, sectors with even or odd indices are decoupled for the inverse matrix. This decoupling structure will simplify the solution for the perturbed electric potential dramatically.



Let us now turn to the BC for the cations in Eq.~\eqref{eq:BC-cation-uniform}. Again, due to  azimuthal symmetry, the $\phi$ dependence on the BC disappears. After plugging in the explicit form of the scaling factors and  $\psi^{z}$ and $n^{z}_{>}$ given in Eqs.~\eqref{eq:psi-z} and~\eqref{eq:n-z}, we can replace $\alpha_\ell$ with the relation in given  Eq.~\eqref{eq:alpha_l-A_l_inverse} which yields 
\begin{widetext}
\begin{equation}
\label{eq:BC-cation-z-uniform-1}
\begin{split}
&- \sum_{\ell,\ell'=0}^{\infty}  \bar{D}^z_{\ell',\ell} \cdot \left( - E_0 h \delta_{\ell 1} + \frac{d Q_{\ell} (i \xi_0)}{d \xi_0}  A_{\ell} \right) S_{0\ell'}^{(1)} (-i e^{i \pi/4} h q, \eta ) + \left( E_0 h P_1(\eta)  - \sum_{\ell}  A_{\ell} \frac{d Q_{\ell} (i \xi_0)}{d \xi_0}   P_{\ell} (\eta) \right) 
\\
=&
\frac{\Gamma_0}{a N_0}  \frac{d}{d \eta} \left[ \frac{(1-\eta^2)}{\sqrt{\xi_0^2+\eta^2}} \left( \sum_{\ell, \ell'=0}^{\infty}  \bar{D}^z_{\ell',\ell} \cdot   \left( - E_0 h \delta_{\ell 1} + \frac{d Q_{\ell} (i \xi_0)}{d \xi_0}  A_{\ell} \right)   \frac{ R_{0\ell'}^{(3)} (-i e^{i \pi/4} h q, i \xi_0 )}{ d R_{0, \ell'}^{(3)} (-i e^{i \pi/4} h q, i \xi_0 )/d \xi_0} \frac{ d S_{0\ell'}^{(1)} (-i e^{i \pi/4} h q, \eta )}{d \eta} \right. \right.
\\
&\left. \left. \qquad \qquad \qquad \qquad \qquad \qquad \qquad \qquad \qquad \qquad \qquad \qquad +   \left(i E_0 h  P_{1} (i \xi_0)   + \sum_{\ell}  A_{\ell} Q_{\ell} (i \xi_0)  \frac{d  P_{\ell} (\eta)}{d \eta} \right)  \right) \right] .
\end{split}
\end{equation}
Next, using the expansion of $S_{0\ell'}^{(1)}$ in Eq.~\eqref{eq:S1_mn-expansion} and the definition of $D^{z}_{\ell,\ell'}$ in Eq.~\eqref{eq:def-K_l_l} to replace $d_{\ell}^{0\ell'}$, Eq.~\eqref{eq:BC-cation-z-uniform-1} can be reorganized as
\begin{equation}
\label{eq:BC-cation-z-uniform-2}
\begin{split}
& 2 E_0 h P_{1}(\eta) - 2 \sum_{\ell=0}^{\infty}     P_{\ell} (\eta)    \frac{d Q_{\ell} (i \xi_0)}{d \xi_0}  A_{\ell}
\\
= &
(-E_0 h) \frac{\Gamma_0}{a N_0 }  \frac{d}{d \eta} \left[\frac{(1-\eta^2)}{\sqrt{\xi_0^2+\eta^2}} \left( \xi_0 \frac{ d P_{1}(\eta)}{d \eta} +  \sum_{ \ell'=0}^{\infty} \sideset{}{'}\sum_{r=1}^{\infty}  D^{z}_{r,\ell'} \bar{D}^z_{\ell',1} \frac{ R_{0\ell'}^{(3)} (-i e^{i \pi/4} h q, i \xi_0 )}{ d R_{0, \ell'}^{(3)} (-i e^{i \pi/4} h q, i \xi_0 )/d \xi_0}  \frac{d P_{r}(\eta)}{d \eta} \right)  \right]
\\
&+
\frac{\Gamma_0}{a N_0 }  \frac{d}{d \eta} \left[ \frac{(1-\eta^2)}{\sqrt{\xi_0^2+\eta^2}} \sum_{\ell}  A_{\ell} \left( \sum_{ \ell'=0}^{\infty}\sideset{}{'}\sum_{r=0,1}^{\infty} D^{z}_{r,\ell'} \bar{D}^z_{\ell',\ell}   \frac{   R_{0\ell'}^{(3)} (-i e^{i \pi/4} h q, i \xi_0 )}{ d R_{0, \ell'}^{(3)} (-i e^{i \pi/4} h q, i \xi_0 )/d \xi_0} \frac{d Q_{\ell} (i \xi_0)}{d \xi_0}  \frac{ d P_{r}(\eta)}{d \eta} +  Q_{\ell} (i \xi_0)  \frac{d  P_{\ell} (\eta)}{d \eta}  \right) \right] .
\end{split}
\end{equation}
Here, we have used the fact that $\bar{D}^z$ is the inverse matrix of $D^z$.

We now employ the orthogonality properties of the Legendre polynomials by first multiplying both sides of Eq.~\eqref{eq:BC-cation-z-uniform-2} by $P_{\ell''}(\eta)$ and then integrating over the interval $-1<\eta<1$.  After an intgration by parts and some algebra, we obtain
\begin{align}
\label{eq:BC-cation-z-uniform-3}
& E_0 h\left( 2 \delta_{1 \ell''} + \frac{\Gamma_0}{a N_0 }  \left( \xi_0 b_{\ell'' 1} (\xi_0)   +   \sum_{ \ell'=0} \sideset{}{'}\sum_{ r=1}^{\infty}  b_{\ell'', r}(\xi_0)  D^z_{r,\ell'} \bar{D}^z_{\ell',1} \frac{ R_{0\ell'}^{(3)} (-i e^{i \pi/4} h q, i \xi_0 )}{ d R_{0, \ell'}^{(3)} (-i e^{i \pi/4} h q, i \xi_0 )/d \xi_0}    \right) \right) 
\\
\nonumber
= & 2 \sum_{\ell=0}^{\infty}  A_{\ell}  \frac{d Q_{\ell} (i \xi_0)}{d \xi_0} \delta_{\ell \ell''}
+\frac{\Gamma_0}{a N_0 }  \sum_{\ell}  A_{\ell} \left( b_{\ell'',\ell}(\xi_0)   Q_{\ell} (i \xi_0) +  \sum_{ \ell'=0}^{\infty} \sideset{}{'}\sum_{r=0,1}^{\infty}  b_{\ell'',r}(\xi_0)  D^z_{r,\ell'} \bar{D}^z_{\ell',\ell}  \frac{R_{0\ell'}^{(3)} (-i e^{i \pi/4} h q, i \xi_0 )}{ d R_{0, \ell'}^{(3)} (-i e^{i \pi/4} h q, i \xi_0 )/d \xi_0} \frac{d Q_{\ell} (i \xi_0)}{d \xi_0}    \right), 
\end{align}
\end{widetext}
where $a=h\sqrt{1+\xi_0^2}$ and the functions $b_{\ell,\ell'} (\xi_0)$ are defined as
\beq
b_{\ell,\ell'} (\xi_0)= - \frac{1}{c_{\ell}}\int_{-1}^{1} d \eta \frac{1}{\sqrt{\xi_0^2+\eta^2}}P^{1}_{\ell}(\eta) P^{1}_{\ell'}(\eta) .
\eeq
Because $b_{\ell,\ell''}=0$ for $\ell +\ell'$ odd, sectors with even and odd indices are decoupled as well in Eq.~\eqref{eq:BC-cation-z-uniform-3}. Upon closer inspection, one can show that all the $A_{\ell}=0$ for $\ell$ even. For the sector with odd indices, a matrix equation for $A_{2n'+1}$ can be written as $\sum_{n'} G^z_{2n+1, 2n'+1} A_{2n'+1} = V^z_{2n+1}$ with elements given by
\begin{widetext}
\begin{align}
\label{eq:G-z-uniform}
G^{z}_{2n+1, 2n'+1} =&   2 \frac{d Q_{2n'+1} (i \xi_0)}{d \xi_0} \delta_{n n'} + \frac{\Gamma_0}{a N_0}   b_{2n+1,2n'+1}(\xi_0) Q_{2n'+1} (i \xi_0)
\\
\nonumber
&  + \frac{\Gamma_0}{a N_0}   \frac{d Q_{2n'+1} (i \xi_0)}{d \xi_0}  \sum_{\mu,\nu=0}^{\infty}  b_{2n+1,2\mu+1} D^z_{2\mu+1,2\nu+1} \bar{D}^z_{2\nu+1,2n'+1}  \frac{R_{0,2\nu+1}^{(3)} (-i e^{i \pi/4} h q, i \xi_0 )}{ d R_{0, 2\nu+1}^{(3)} (-i e^{i \pi/4} h q, i \xi_0 )/d \xi_0} ,
\end{align}
and
\begin{equation}
\label{eq:V-z-uniform}
V^{z}_{2n+1} =  E_0 h \left[ 2\delta_{n0}  + \frac{\Gamma_0}{a N_0}  \left(\xi_0 b_{2n+1,1}+ \sum_{\mu,\nu=0}^{\infty} b_{2n+1,2\mu+1}  D^z_{2\mu+1,2\nu+1} \bar{D}^z_{2\nu+1,1} \frac{ R_{0,2\nu+1}^{(3)} (-i e^{i \pi/4} h q, i \xi_0 )}{ d R_{0, 2\nu+1}^{(3)} (-i e^{i \pi/4} h q, i \xi_0 )/d \xi_0} \right) \right].
\end{equation}

It is useful to separate both the $G^z$ matrix and $V^z$ vector into two parts, $G^z=G^{z(0)}+\delta G^z$ and $V^z=V^{z(0)}+\delta V^z$.  The elements of the leading order terms, $G^{z(0)}$ and $V^{z(0)}$, are given by
\begin{equation}
\label{eq:G-z0-uniform}
G^{z(0)}_{2n+1, 2n'+1} =   2 \frac{d Q_{2n'+1} (i \xi_0)}{d \xi_0} \delta_{n n'} + \frac{\Gamma_0}{a N_0}   b_{2n+1,2n'+1} \left( Q_{2n'+1} (i \xi_0) +  D^z_{2n'+1,2n'+1} \bar{D}^z_{2n'+1,2n'+1}  \frac{d Q_{2n'+1} (i \xi_0)}{d \xi_0}  \frac{R_{0,2n'+1}^{(3)} (-i e^{i \pi/4} h q, i \xi_0 )}{ d R_{0, 2n'+1}^{(3)} (-i e^{i \pi/4} h q, i \xi_0 )/d \xi_0}\right),
\end{equation}
and
\begin{equation}
\label{eq:V-z0-uniform}
V^{z(0)}_{2n+1} =  E_0 h \left[ 2\delta_{n0}  + \frac{\Gamma_0}{a N_0}  b_{2n+1,1}(\xi_0) \left(\xi_0 +  D^z_{1,1} \bar{D}^z_{1,1} \frac{ R_{0,1}^{(3)} (-i e^{i \pi/4} h q, i \xi_0 )}{ d R_{0, 1}^{(3)} (-i e^{i \pi/4} h q, i \xi_0 )/d \xi_0} \right) \right].
\end{equation}
With respect to $hq$, these elements are of the order $\mathcal{O}(hq)^0$ with the $\mathcal{O}(hq)^2$ with higher oder contributions coming solely from the last term of each expression.  The elements of the higher-order terms, $\delta G^{z}$ and $\delta V^{z}$, are given by
\begin{align}
\label{eq:delta_G-z-uniform}
\delta G^{z}_{2n+1, 2n'+1} =  \frac{\Gamma_0}{a N_0}   \frac{d Q_{2n'+1} (i \xi_0)}{d \xi_0}  \sum_{\substack{\mu,\nu=0 \\ \neg \mu=\nu=n'  } }^{\infty}  b_{2n+1,2\mu+1} D^z_{2\mu+1,2\nu+1} \bar{D}^z_{2\nu+1,2n'+1}  \frac{R_{0,2\nu+1}^{(3)} (-i e^{i \pi/4} h q, i \xi_0 )}{ d R_{0, 2\nu+1}^{(3)} (-i e^{i \pi/4} h q, i \xi_0 )/d \xi_0} ,
\end{align}
and
\begin{equation}
\label{eq:delta_V-z-uniform}
\delta V^{z}_{2n+1} =  E_0 h  \frac{\Gamma_0}{a N_0}   \sum_{\substack{\mu,\nu=0 \\ \neg \mu=\nu=0  }}^{\infty} b_{2n+1,2\mu+1}  D^z_{2\mu+1,2\nu+1} \bar{D}^z_{2\nu+1,1} \frac{ R_{0,2\nu+1}^{(3)} (-i e^{i \pi/4} h q, i \xi_0 )}{ d R_{0, 2\nu+1}^{(3)} (-i e^{i \pi/4} h q, i \xi_0 )/d \xi_0}  .
\end{equation}
\end{widetext}
Here, the $\neg A$ symbol indicates that $A$ is not included in the summation. In these equations, $\delta G^{z}_{2n+1, 2n'+1}$ and $\delta V^{z}_{2n+1}$ are on the order of $\mathcal{O}(hq)^2$, with higher-order corrections starting at $\mathcal{O}(hq)^4$.

We are now interested in solving for an approximate expression for the component $A_1$. Let us first consider the contributions from the leading-order terms of the matrix $G^z$ and vector $V^{z}$.   From the structure of  $G^{z(0)}$ and  $V^{z(0)}$, we can show that the leading component $A^{(0)}_{1}$  dominates the response of the perturbed electric field, \textit{i.e.}, $A^{(0)}_1\gg A^{(0)}_{2n+1}$ for $n \ge 1$, in the following two limits~\cite{Dukhin1980}: (i) When $|\Gamma_0/ N_0 a| \ll |(dQ_{\ell}(i \xi_0)/d\xi_0)/Q_{\ell}(i \xi_0)|$, in which case the matrix $G^{z(0)}$  is nearly diagonal, and (ii)  When $|\Gamma_0/ N_0 a| \gg |(dQ_{\ell}(i \xi_0)/d\xi_0)/Q_{\ell}(i \xi_0)|$, in which case the first column of the matrix is proportional to the source vector, $G^{z(0)}_{2n+1, 1}\propto V^{z(0)}_{2n+1}$. Depending on the salinity (ion concentration) of the electrolyte and the amount of charges carried by the particle, our system can be in limit (i),  in limit (ii), or in the intermediate regime. However, because cases (i) and (ii) are opposite limits, we expect that $A^{(0)}_1$ will most likely still be the most dominant component of the perturbed electric field even in the intermediate regime. Hence, we will first approximate the perturbed electric potential as
\beq
\label{eq:psi-z-uniform-approx}
\psi^{z} \approx -E_0 h \xi \eta + A^{(0)}_1 Q_1(i \xi) P_{1}(\eta),
\eeq
with the $A_1^{(0)}$ component approximated by $A^{(0)}_{1} \approx V^{z(0)}_1/G^{z(0)}_{11}$
\beq
\frac{A^{(0)}_{1}}{ E_0 h } \approx    \frac{  \left( 1   + \frac{\Gamma_0 }{ 2 a N_0 }  b_{1,1}  \left( \xi_0   + \Sigma^{(0)}_{\rm p} \right) \right) }{ \frac{d Q_{1} (i \xi_0)}{d \xi_0}  + \frac{\Gamma_0 }{ 2 a N_0 }  b_{1,1} \left( Q_{1} (i \xi_0) +   \Sigma^{(0)}_{\rm p} \frac{d Q_{1} (i \xi_0)}{d \xi_0}   \right)   }.
\eeq
where
\beq
\label{eq:Sigma_z0_11-uniform}
\Sigma^{(0)}_{\rm p}( h q,  \xi_0) = D^z_{1,1} \bar{D}^z_{1,1} \frac{ R_{0,1}^{(3)} (-i e^{i \pi/4} h q, i \xi_0 )}{ d R_{0,1}^{(3)} (-i e^{i \pi/4} h q, i \xi_0 )/d \xi_0}. 
\eeq

Because the higher-order terms, $\delta G^{z}$ and $\delta V^{z}$, also contribute to the component $A_1$  at order $\mathcal{O}(hq)^2$, the actual $\mathcal{O}(hq)^2$ dependence of  $A_1$  can, in principle, differ from $A_1^{(0)}$. However, by explicitly examining the structure of the inversion of $G^z$ matrix and performing perturbative expansions in $hq$, we can show that the $\mathcal{O}(hq)^2$ order contribution from $\delta G^{z}$ and $\delta V^{z}$ is suppressed by a factor of $\Gamma_0 / (a N_0)$ in limit (i) and by a factor of $a N_0/\Gamma_0$ in  limit (ii), compared to the $\mathcal{O}(hq)^2$ order contributions in $A^{(0)}_{1}$. It is also important to note that the second order leading contribution from the addition of the $\delta G^{z}$ and $\delta V^{z}$ terms is of order $\mathcal{O}(hq)^4$. We will hence keep only the $A^{(0)}_{1}$ part of the contribution to the perturbed electric field to order $\mathcal{O}(hq)^3$ for the following discussions.

In the limit as $\xi \to \infty$, the second term in Eq.~\eqref{eq:psi-z-uniform-approx} exaclty matches the dipole polarization response $a^2 b P_{z} \boldsymbol{E} \cdot \boldsymbol{r}/r^3$, with $P_{z}$ defined as the polarization coefficient in the z-direction. Using $Q_{1} (i \xi) \to - 1/(3 \xi^2)$ for $\xi \to \infty$ and $P_1(\eta)=\eta$, we have
\beq
\label{eq:Pz-uniform-1}
P_{z}=  -\frac{1}{3}  \frac{ \left( 1+ \frac{\Gamma_0 }{ 2 a N_0 } b_{1,1} \left( \xi_0   + \Sigma^{(0)}_{\rm p}\right)\right)/ \xi_0 (1+\xi_0^2)  }{  \frac{d Q_{1} (i \xi_0)}{d \xi_0}  +  \frac{\Gamma_0 }{ 2 a N_0 } b_{1,1} \left( Q_{1} (i \xi_0) +   \Sigma^{(0)}_{\rm p} \frac{d Q_{1} (i \xi_0)}{d \xi_0}   \right)   }.
\eeq
Upon multiplying both the denominator and numerator in Eq.~\eqref{eq:Pz-uniform-1} by the water conductivity $\sigma_w$  and using the depolarization factor $L_{p}$ given in Eq.~\eqref{eq:L_p}, we obtain
\beq
\label{eq:Pz-uniform-2}
P_{z}=P_{p}^{\rm I} = \frac{1}{3}  \frac{ \sigma_{p}^{\rm I} -\sigma_{w,p}^{\rm I} }{ L_{p} \sigma_{p}^{\rm I} +(1-L_{p}) \sigma_{w,p}^{\rm I} },
\eeq
where the superscript $\rm I$ denotes  case (I) for uniform surface ion distributions, and the subscript $p$ means that the electric field is parallel to the symmetry axis of the oblate spheroid. In this equation, the effective particle and  modified water conductivities along the z-direction are given by
\beq
\label{eq:sigma-p-uniform}
\begin{split}
\sigma_p^{\rm I} =& - \sigma_w  \frac{ \Gamma_0}{a N_0} \xi_0 \frac{b_{1,1}(\xi_0)}{2},
\\
\sigma_{w,p}^{\rm I} =& \sigma_w \left( 1 + \frac{\Gamma_0}{ a N_0}\frac{b_{1,1}(\xi_0)}{2} \Sigma^{(0)}_{p}( h q,  \xi_0) \right).
\end{split}
\eeq
We note that the effective particle conductivity $\sigma_p^{\rm I}$ has no frequency dependence and remains the same throughout all frequencies.  The expression for $\sigma_p^{\rm I}$ found here is the same as that found for the high-frequency response of a charged spheroid in Reference \onlinecite{Freed-ex}. 

By expanding the expression for $\Sigma^{(0)}_{p}(h q, \xi_0)$, given in Eq.~\eqref{eq:Sigma_z0_11-uniform}, to order $(hq)^3$, we obtain 
\begin{widetext}
 \beq
 \label{eq:Sigma_z0_11-uniform-expand}
 \begin{split}
 	&\Sigma^{(0)}_{p}( h q,  \xi_0)\sim  \frac{ R_{0,1}^{(3)} (-i e^{i \pi/4} h q, i \xi_0 )}{ d R_{0, 1}^{(3)} (-i e^{i \pi/4} h q, i \xi_0 )/d \xi_0}
 	\\
 	\sim& \frac{Q_{1}(i\xi_0)}{d Q_{1}(i\xi_0)/d\xi_0 } \left[ 1 +  i \frac{h^2\omega}{D} \left( \frac{Q_3(i\xi_0)}{25 Q_1(i\xi_0)} - \frac{1}{6 Q_1(i\xi_0)} -  \frac{d Q_{3}(i\xi_0)/d \xi_0 }{ 25 (d Q_{1}(i\xi_0)/d \xi_0 )}  \right) +  e^{-i\pi/4} \frac{h^3\omega^{3/2}}{9 D^{3/2}} \left( \frac{P_1(i \xi_0)}{Q_1(i\xi_0)} -  \frac{d P_{1}(i\xi_0)/d \xi_0 }{ d Q_{1}(i\xi_0)/d \xi_0}  \right)   \right] ,
 \end{split}
\eeq
\end{widetext}
where we have used the expression in Eq.~\eqref{eq:Rmn_final} for $R_{0,1}^{(3)}$ and only kept the terms with expansion coefficients, $d_{1}^{01}(-i e^{i \pi/4} h q) \approx 1+ 3i(hq)^2/50$, $d_{3}^{01}(-i e^{i \pi/4} h q)\approx i(hq)^2/25$, and $d_{\rho|1}^{01} \approx -i (hq)^2/6$, with $q=\sqrt{\omega/D}$. All the other expansion coefficients are higher order in $hq$. To obtain Eq.~\eqref{eq:Sigma_z0_11-uniform-expand}, we have also made use of the expansion $D^z_{1,1} \bar{D}^z_{1,1}\sim 1 + \mathcal{O}(hq)^4 $ and the definitions of the joining factors. Although the expression for the joining factors in Eqs.~\eqref{eq:joining-factor-kappa1} and~\eqref{eq:joining-factor-kappa2} are quite tedious, the infinite summation always cancel out because $\Sigma^{(0)}_{p}( h q,  \xi_0)$ is approximated as a ratio between $ R_{0,1}^{(3)} (-i e^{i \pi/4} h q, i \xi_0 )$ and $d R_{0, 1}^{(3)} (-i e^{i \pi/4} h q, i \xi_0 )/d \xi_0$.
 
Because the expansions of the spheroidal wave functions in terms of $d_{\ell}^{0\ell'}(hq)$ are quite accurate for $hq \leq 1$, the Taylor expansion of $\Sigma^{(0)}_{p}$ to order $(hq)^3$ in Eq.~\eqref{eq:Sigma_z0_11-uniform-expand} should provide a good approximation for the polarization response at low frequencies when $\omega \leq \omega_h= D/h^2$. However, the polynomial expansion with a finite cutoff is bound to give unreasonable predictions for $\omega \gg \omega_h$. We hence perform an additional Pad\' e approximation to order $(1,2)$,  which reorganizes the expansion in Eq.~\eqref{eq:Sigma_z0_11-uniform-expand} into the form of a rational function.  To obtain the Pad\'e approximation for $\Sigma^{(0)}_{p}$, denoted by $\mathcal{P} [ \Sigma^{(0)}_{p}]^{(1,2)}$, we make use of the leading-order behavior as a function of $\omega^{-1}$ of the high frequency solution, given in Reference~\onlinecite{Freed-ex}, as well as the small $\omega$ solution given in Eq.~\eqref{eq:Sigma_z0_11-uniform-expand}.  We find
\beq
\label{eq:Sigma_z0_11-uniform-Pade}
\mathcal{P} [ \Sigma^{(0)}_{p}]^{(1,2)} \sim \frac{Q_{1}(i\xi_0)}{d Q_{1}(i\xi_0)/d\xi_0 } \frac{1 + \varpi_{p}^1 \frac{h\omega^{1/2}}{D^{1/2}} }{ 1+ \varpi^1_{p} \frac{h\omega^{1/2}}{D^{1/2}} + \varpi_{p}^2 \frac{h^2\omega}{D} } ,
\eeq
with
\bea
\varpi^2_{p} =& -i  \left( \frac{Q_3(i\xi_0)}{25 Q_1(i\xi_0)} - \frac{1}{6 Q_1(i\xi_0)} -  \frac{d Q_{3}(i\xi_0)/d \xi_0 }{ 25 (d Q_{1}(i\xi_0)/d \xi_0 )}  \right),
\\
\varpi_{p}^1 =& \frac{e^{-i\pi/4}}{9 \varpi_{p}^2 } \left( \frac{P_1(i \xi_0)}{Q_1(i\xi_0)} -  \frac{d P_{1}(i\xi_0)/d \xi_0 }{ d Q_{1}(i\xi_0)/d \xi_0}  \right). 
\eea
There are several advantages to using the Pad\' e approximation in Eq.~\eqref{eq:Sigma_z0_11-uniform-Pade}. First, it has the same low-frequency response given by the Taylor expansion in Eq.~\eqref{eq:Sigma_z0_11-uniform-expand}.  Second, it gives the desired high-frequency behavior, $\mathcal{P} [ \Sigma^{(0)}_{p}(\omega\to \infty)]^{(1,2)}  \to 0$, which is consistent with the high frequency solution~\cite{Freed-ex}. This at least provides a smooth transition from the low to high frequency regions without unphysical behavior inbetween. Remarkably, in the spherical limit, the Pad\'e approximation $\mathcal{P} [ \Sigma^{(0)}_{p}]^{(1,2)}$ reproduces the exact polarization response of a charged sphere immersed in an electrolyte~\cite{Chew1982}. As a result, in this paper we will use the Pad\'e approximation in the epxression for the modified water conductivity, which yields
\beq
\label{eq:modified-sigma_w-z-uniform}
\sigma_{w,p}^{\rm I} (\omega) \sim \sigma_w \left( 1 + \frac{\Gamma_0}{a N_0} \frac{b_{1,1}(\xi_0)}{2}  \mathcal{P} [ \Sigma^{(0)}_{p}(hq, \xi_0)]^{(1,2)}  \right).
\eeq

\subsection{Polarization coefficient $P_x$ in $x$-direction}
\label{sec:P-x-uniform}

In this section, we consider the case when the electric field is applied in the $x$-direction, so that it is normal to the spheroid's axis of symmetry.  For the most part, the derivation is very similar to the one given in the preceding section, except that now the perturbed electric field and the ion concentration depend on the coordinate $\phi$, and $m =1$.

When the electric field is applied in the $x$-direction, the perturbed electric field and the ion concentration are proportional to $\cos \phi$.  Thus, the general solutions for them become 
\bea
\label{eq:psi-x}
\psi^{x}&=& \left(i E_0 h P_{1}^{1} (i \xi) P_{1}^{1} (\eta) + \sum_{\ell=1}^{\infty} A_{\ell}^{1} Q_{\ell}^{1} (i \xi) P_{\ell}^{1}(\eta)\right) \cos \phi,
\\
\label{eq:n-x}
n^{x}_{>}&=&  \sum_{\ell=1}^{\infty} \alpha_{\ell}^{1} R_{1\ell}^{(3)} (-i e^{i \pi/4} h q, i \xi )  S_{1\ell}^{(1)} (-i e^{i \pi/4} h q, \eta ) \cos \phi,
\eea
Here, we have used the relation $x= -i h P_{1}^{1}(i \xi_0) P_{1}^{1}(\eta) \cos\phi $.  

Again, we start by substituting the above expressions for the perturbed electric field $\psi^{x}$ and ion concentration $n^{x}_{>}$ into the BC for the anions. Using the expansion for the angular spheroidal wave function given in Eq.~\eqref{eq:S1_mn-expansion} and employing the orthogonality properties of the associated Legendre polynomials, we obtain the relations
\beq
\label{eq:BC-anion-x-uniform-1}
\begin{split}
\sideset{}{'}\sum_{\ell=1,2}^{\infty} K^{x}_{\ell',\ell} \alpha_{\ell}^{1}  =&  N_0  \left( i E_0 h \frac{d P_{1}^{1} (i \xi_0)}{d \xi_0}  \delta_{1 \ell'}  +  A_{\ell'}^{1}\frac{d Q_{\ell'}^{1} (i \xi_0)}{d \xi_0}  \right)
\\
 K^{x}_{\ell',\ell} =& D^x_{\ell',\ell} \frac{d R_{1\ell}^{(3)} (-i e^{i \pi/4} h q, i \xi_0 )}{d \xi_0}.
 \end{split}
\eeq
In this equation, we define a matrix $D^x$  with elements given by $D^x_{\ell',\ell}\equiv d_{\ell'-1}^{1\ell}(-i e^{i \pi/4} h q) $, and the symbol $\sum\nolimits'$ indicates that the summation is over even $\ell$ for $\ell'$ even and odd $\ell$ for $\ell'$ odd.  Next, the coefficients $\alpha_{\ell}^1$ can be expressed in terms of the $A_{\ell'}^1$ as
\bea
\nonumber
\alpha_{\ell}^{1}  &=&  N_0 \sideset{}{'}\sum_{\ell'=1,2}^{\infty}  \bar{K}^{x}_{\ell,\ell'}  \left( i E_0 h \frac{d P_{1}^{1} (i \xi_0)}{d \xi_0}  \delta_{1 \ell'}  +  A_{\ell'}^{1} \frac{d Q_{\ell'}^{1} (i \xi_0)}{d \xi_0}  \right) ,
\\
\label{eq:alpha_l1-A_l1-uniform}
\bar{K}^{x}_{\ell,\ell'} &=& \left(\frac{d R_{1\ell}^{(3)} (-i e^{i \pi/4} h q, i \xi_0 )}{d \xi_0}\right)^{-1} \cdot \bar{D}^x_{\ell,\ell'} .
\eea
Here, $\bar{K}^x$ is the matrix  inverse of $K^x$, and $\bar{D}^x$ is the matrix inverse of $D^x$. Because $D^x_{\ell', \ell}= d_{\ell'-1}^{1, \ell}(-i e^{i \pi/4} h q) =0$ for $\ell+\ell'$ odd, it follow that $\bar{D}^x_{\ell, \ell'}=0$ for $\ell+\ell'$ odd. As a result, even and odd index sectors of the inverse matrix $\bar{K}^x$ are decoupled.


Now, using the expansion of $S_{1\ell}^{(1)}(-i e^{i \pi/4} h q, \eta )$ in Eq.~\eqref{eq:S1_mn-expansion} and the relation between $\alpha_\ell^1$ and $A_{\ell}^1$ in Eq.~\eqref{eq:alpha_l1-A_l1-uniform}, we can write the BC for the cations in Eq.~\eqref{eq:BC-cation-uniform} as
\begin{widetext}
\bea
 \label{eq:BC-cation-x-uniform}
 &-& i E_0 h \left[ 2 \frac{d P_{1}^{1} (i \xi_0)}{d \xi_0} P_{1}^{1} (\eta) +  \frac{\Gamma_0}{a N_0}   \left( P_{1}^{1}(i\xi_0) \mathcal{O}_\eta (\xi_0) P_{1}^1(\eta) + \sum_{\ell',r=1}^{\infty}  D^x_{r,\ell'} \bar{D}^x_{\ell',1}    \frac{d P_{1}^{1} (i \xi_0)}{d \xi_0}   \frac{R_{1\ell'}^{(3)} (-i e^{i \pi/4} h q, i \xi_0 )}{d R_{1\ell'}^{(3)} (-i e^{i \pi/4} h q, i \xi_0 )/d \xi_0}   \mathcal{O}_\eta (\xi_0) P_{r}^1(\eta) \right) \right]
 \\
 \nonumber
 &=&
 \sum_{\ell=1}^{\infty}  A_{\ell}^{1} \left[ 2\frac{d Q_{\ell}^{1} (i \xi_0)}{d \xi_0}   P_{\ell}^1(\eta)  
 +
 \frac{\Gamma_0}{a N_0}   \left(Q_{\ell}^{1} (i \xi_0) \mathcal{O}_{\eta}(\xi_0) P_{\ell}^{1}(\eta) +   \frac{d Q_{\ell}^{1} (i \xi_0)}{d \xi_0} \sum_{\ell',r=1}^{\infty}  D^x_{r,\ell'} \bar{D}^x_{\ell',\ell}     \frac{R_{1\ell'}^{(3)} (-i e^{i \pi/4} h q, i \xi_0 )}{d R_{1\ell'}^{(3)} (-i e^{i \pi/4} h q, i \xi_0 )/d \xi_0}    \mathcal{O}_{\eta}(\xi_0) P_{r}^1(\eta) \right) \right].
 \eea
\end{widetext}
Here, we have defined a differential operator 
\beq
\mathcal{O}_\eta (\xi_0) \equiv   \frac{d }{d \eta}\left(\frac{1-\eta^2}{\sqrt{\xi_0^2+\eta^2}}\frac{d }{d \eta} \right) - \frac{\sqrt{\xi_0^2+\eta^2}}{ (1+\xi_0^2)(1-\eta^2)} .
\eeq

Similar to the calculation for the z-direction, we now employ the orthogonality properties of the Legendre polynomials by multiplying both sizes of Eq.~\eqref{eq:BC-cation-x-uniform} by $P_{\ell''}^{1}(\eta)$ and integrating over the interval $-1<\eta<1$.  Again, the even and odd index sectors decouple with all the $A^{1}_{\ell}=0$ for $\ell$ even. A matrix equation for the odd index sector reads
\beq
\sum_{n'=0} G^x_{2n+1, 2n'+1} A^{1}_{2n'+1} = V^x_{2n+1},
\eeq
where
\begin{widetext}
\beq
\label{eq:G-x-uniform}
\begin{split}
	G^x_{2n+1, 2n'+1} =&  2 \frac{d Q_{2n'+1}^{1} (i \xi_0)}{d \xi_0} \delta_{n n'}    +
	\frac{\Gamma_0}{a N_0 }  b^{\perp}_{2n+1,2n'+1}  Q_{2n'+1}^{1} (i \xi_0) 
	\\
	+& \frac{\Gamma_0}{a N_0}     \sum_{\mu,\nu=0}^{\infty}   b^{\perp}_{2n+1,2\mu+1}  D^x_{2\mu+1,2\nu+1} \bar{D}^x_{2\nu+1,2n'+1} \cdot     \frac{R_{1,2\nu+1}^{(3)} (-i e^{i \pi/4} h q, i \xi_0 )}{d R_{1,2\nu+1}^{(3)} (-i e^{i \pi/4} h q, i \xi_0 )/d \xi_0}  \frac{d Q_{2n'+1}^{1} (i \xi_0)}{d \xi_0}   
\end{split}
\eeq
and 
\beq
\label{eq:V-x-uniform}
\begin{split}
	V^x_{2n+1} =&  -  i E_0 h \left[ 2 \frac{d P_{1}^{1} (i \xi_0)}{d \xi_0}  \delta_{n0} +  \frac{\Gamma_0}{a N_0 }  P_{1}^{1} (i \xi_0) b^{\perp}_{2n+1,1} (\xi_0) \right.
	\\
	&\qquad \qquad + \left. \frac{\Gamma_0}{a N_0} \sum_{\mu,\nu=0}^{\infty}   b^{\perp}_{2n+1,2\mu+1}  D^x_{2\mu+1,2\nu+1} \bar{D}^x_{2\nu+1, 1}      \frac{R_{1,2\nu+1}^{(3)} (-i e^{i \pi/4} h q, i \xi_0 )}{d R_{1,2\nu+1}^{(3)} (-i e^{i \pi/4} h q, i \xi_0 )/d \xi_0} \frac{d P_{1}^{1} (i \xi_0)}{d \xi_0} \right].
\end{split}
\eeq
Here, the functions $b^{\perp}_{\ell \ell'}(\xi_0)$ are defined by the integral 
\beq
b^{\perp}_{\ell \ell'}(\xi_0)= \frac{1}{c_{\ell}^{1}}\int_{-1}^{1} d \eta P_{\ell}^{1}(\eta) \mathcal{O}_{\eta}(\xi_0) P_{\ell'}^{1}(\eta) = -\frac{1}{c_{\ell}^{1}} \left( \int_{-1}^{1}  d \eta \frac{1-\eta^2}{\sqrt{\xi_0^2+\eta^2}}  \frac{d P_{\ell}^{1}(\eta)}{d \eta} \frac{d P_{\ell'}^{1}(\eta)}{d \eta} + \frac{1}{1+\xi_0^2} \int_{-1}^{1}  d \eta \frac{\sqrt{\xi_0^2+\eta^2}}{ 1-\eta^2}  P_{\ell}^{1}(\eta)  P_{\ell'}^{1}(\eta) \right),
\eeq
\end{widetext}
and we have used the orthogonality properties of the associated Legendre polynomials, given by $\int_{-1}^{1} d \eta P^{m}_{\ell}(\eta) P^{m}_{\ell'}(\eta) = c^{m}_{\ell'} \delta_{ \ell \ell'}$, with $c^m_\ell= c_{\ell} (\ell+m)!/(\ell-m)!$.  We note that Eqs.~\eqref{eq:G-x-uniform} and \eqref{eq:V-x-uniform} have a very similar form to Eqs.~\eqref{eq:G-z-uniform} and \eqref{eq:V-z-uniform}, but with $m=0$ replaced by $m=1$ and $b_{\ell,\ell'}$ replaced with $b^{\perp}_{\ell \ell'}$.

Again, it is useful to separate both the $G^x$ matrix and $V^x$ vector into two parts, $G^x=G^{x(0)}+\delta G^x$ and $V^x=V^{x(0)}+\delta V^x$. The elements of the leading order parts, $G^{x(0)}$ and $V^{x(0)}$, are given by
\begin{widetext}
\beq
\label{eq:G-x0-uniform}
G^{x(0)}_{2n+1, 2n'+1} =  2 \frac{d Q_{2n'+1}^{1} (i \xi_0)}{d \xi_0} \delta_{n n'}    +
\frac{\Gamma_0}{a N_0 }  b^{\perp}_{2n+1,2n'+1} \left( Q_{2n'+1}^{1} (i \xi_0)   +  D^x_{2n'+1,2n'+1} \bar{D}^x_{2n'+1,2n'+1} \frac{R_{1,2n'+1}^{(3)} (-i e^{i \pi/4} h q, i \xi_0 )}{d R_{1,2n'+1}^{(3)} (-i e^{i \pi/4} h q, i \xi_0 )/d \xi_0}  \frac{d Q_{2n'+1}^{1} (i \xi_0)}{d \xi_0} \right) , 
\eeq
and 
\beq
\label{eq:V-x0-uniform}
\begin{split}
	V^{x(0)}_{2n+1} =&  -  i E_0 h \left[ 2 \frac{d P_{1}^{1} (i \xi_0)}{d \xi_0} \delta_{n0} +  \frac{\Gamma_0}{a N_0 }  b^{\perp}_{2n+1,1} (\xi_0) \left( P_{1}^{1} (i \xi_0)  +  D^x_{1,1} \bar{D}^x_{1,1} \frac{R_{1,1}^{(3)} (-i e^{i \pi/4} h q, i \xi_0 )}{d R_{1,1}^{(3)} (-i e^{i \pi/4} h q, i \xi_0 )/d \xi_0} \frac{d P_{1}^{1} (i \xi_0)}{d \xi_0} \right) \right].
\end{split}
\eeq
\end{widetext}
The leading contribution to these elements is of the order of $\mathcal{O}(hq)^0$, and the $\mathcal{O}(hq)^2$ and higher order contributions come solely from the last term of each expression. Here, we will not give the explicit form of the higher order parts, $\delta G^x$ and $\delta V^x$, but only state that, as in the case for the z-direction, their elements are of order $\mathcal{O}(hq)^2$, with the second leading order contribution of order $\mathcal{O}(hq)^4$.


For reasons similar to those in the case of the $z$-direction, $A_{1}^{1(0)}$ is the dominant response to the applied electric field and can be approximated by $A^{1(0)}_{1} \approx V_{1}^{x(0)} /G_{11}^{x(0)}$ with the result that
\beq
\label{eq:A_11-x0-uniform}
\frac{A^{1(0)}_{1}}{E_0 h} \approx - i  \frac{ \frac{d P_{1}^{1} (i \xi_0)}{d \xi_0} +  \frac{\Gamma_0}{2 a N_0 } b^{\perp}_{1,1} \left(P_{1}^{1} (i \xi_0) +\Sigma^{(0)}_{ n} \frac{d P_{1}^{1} (i \xi_0)}{d \xi_0} \right) }{   \frac{d Q_{1}^{1} (i \xi_0)}{d \xi_0}  +	\frac{\Gamma_0}{2 a N_0} b^{\perp}_{1,1} \left( Q_{1}^{1} (i \xi_0) +  \Sigma^{(0)}_{ n}  \frac{d Q_{1}^{1} (i \xi_0)}{d \xi_0}  \right) }, 
\eeq
where
\beq
\label{eq:Sigma-x0_11-uniform}
\Sigma^{(0)}_{ n}( h q,  \xi_0) =  D^x_{1,1} \bar{D}^x_{1,1} \frac{R_{1,1}^{(3)} (-i e^{i \pi/4} h q, i \xi_0 )}{d R_{1,1}^{(3)} (-i e^{i \pi/4} h q, i \xi_0 )/d \xi_0}. 
\eeq
The approximate perturbed electric field then becomes
\beq
\label{eq:psi-x-approx-uniform}
\psi^{x} \approx i E_0 h P_{1}^{1} (i \xi) P_{1}^{1} (\eta) \cos \phi +  A_{1}^{1(0)} Q_{1}^{1} (i \xi) P_{1}^{1}(\eta) \cos \phi.
\eeq 


Because $Q_{1}^{1}(\xi) \to - 2\sqrt{1+\xi^2}/(3 \xi^3)$ for $\xi\to \infty$, the second term in Eq.~\eqref{eq:psi-x-approx-uniform} is exactly the dipole polarization response $a^2b P_x^ x/r^3$ in the $x$-direction. The polarization coefficients in the $x$- and $y$-directions are hence given by
\beq
\label{eq:Px-uniform-1}
P_{n}=-  i  \frac{2 }{3}  \frac{  \frac{d P_{1}^{1} (i \xi_0)}{d \xi_0} +  \frac{\Gamma_0 b^{\perp}_{1,1} }{2 a N_0 }   \left(  P_{1}^{1} (i \xi_0) +\Sigma^{(0)}_{ n} \frac{d P_{1}^{1} (i \xi_0)}{d \xi_0} \right) }{ \xi_0  (1+\xi_0^2) \left(  \frac{d Q_{1}^{1} (i \xi_0)}{d \xi_0}  +	\frac{\Gamma_0 b^{\perp}_{1,1} }{2 a N_0}\left(  Q_{1}^{1} (i \xi_0) +  \Sigma^{(0)}_{ n} \frac{d Q_{1}^{1} (i \xi_0)}{d \xi_0}  \right) \right) }. 
\eeq
Using the depolarization factor defined in Eq.~\eqref{eq:L_n}, we have
\beq
\label{eq:Px-uniform-2}
P_{n}^{\rm I} =  \frac{1}{3} \frac{ \sigma_{n}^{\rm I} -\sigma_{ w,n}^{\rm I} }{ L_{n} \sigma_{n}^{\rm I} +(1-L_{n}) \sigma_{w,n}^{\rm I} }.
\eeq
Here, the subscript $n$ denotes that the applied electric field is normal to the symmetry axis of the oblate spheroid. The effective particle and modified water conductivities are given respectively by 
\beq
\label{eq:sigma-n-uniform}
\begin{split}
\sigma_{n}^{\rm I} =&  - \sigma_w  \frac{\Gamma_0}{2 a N_0 } b^{\perp}_{1,1}(\xi_0)  \frac{P^1_1( i \xi_0)}{ d P^1_1( i \xi_0)/d\xi_0}  ,
\\
\sigma_{w,n}^{\rm I} =& \sigma_w \left( 1+  \frac{\Gamma_0}{ a N_0 } \frac{b^{\perp}_{1,1}(\xi_0)}{2}  \Sigma^{(0)}_{n}(h q, \xi_0)   \right).
\end{split}
\eeq
Again, the effective particle conductivity $\sigma^{\rm I}_{n}$ remains the same for all frequencies~\cite{Freed-ex}. 

Next, we need to find an approximation for $\Sigma^{(0)}_{n}(h q, \xi_0)$ which gives a well-behaved polarization response over a wide range of frequencies. As in the case for the $z$-direction, we first perform the Taylor expansion of $\Sigma^{(0)}_{n}(h q, \xi_0)$ to the $(hq)^3$ order and then reorganize it by the Pad\' e approximation to the $(1,2)$ order, using the leading order behavior as $\omega \to\infty$. This procedure gives us
\beq
\label{eq:Sigma_x0_11-uniform-Pade}
\mathcal{P}[ \Sigma^{(0)}_{n}]^{(1,2)} \sim \frac{Q^{1}_{1}(i\xi_0)}{d Q^{1}_{1}(i\xi_0)/d\xi_0 } \frac{1 + \varpi^1_{n} \frac{h\omega^{1/2}}{D^{1/2}} }{ 1+ \varpi^1_{n} \frac{h\omega^{1/2}}{D^{1/2}} + \varpi^2_{n} \frac{h^2\omega}{D} } ,
\eeq
with
\bea
\varpi_{n}^2 &=& -i  \left( \frac{Q^1_3(i\xi_0)}{75 Q^1_1(i\xi_0)} - \frac{Q^1_{-1}(i\xi_0)}{3 Q^1_1(i\xi_0) } \right.
\\
\nonumber
&& \quad \left.  -  \frac{d Q^1_{3}(i\xi_0)/d \xi_0 }{ 75 (d Q^1_{1}(i\xi_0)/d \xi_0 )}  + \frac{d Q^1_{-1}(i\xi_0)/d \xi_0 }{3 (d Q^1_{1}(i\xi_0)/d \xi_0 )}  \right),
\\
\varpi_{n}^1 &=& \frac{2 i e^{i\pi/4} }{9 \varpi_{n}^2 }  \left( \frac{P^1_1(i \xi_0)}{Q^1_1(i\xi_0)} -  \frac{d P^1_1(i\xi_0)/d \xi_0 }{ d Q^1_1(i\xi_0)/d \xi_0}  \right). 
\eea
To obtain this expression, we have used the following expansion for the coefficients for $R_{1,1}^{(3)} (-i e^{i \pi/4} h q, i \xi_0 )$: $d_{0}^{11} (-i e^{i \pi/4} h q) \sim 1 + i(hq)^2/50$, $d_{-2}^{11} (-i e^{i \pi/4} h q) \sim -i(hq)^2/3$, and $d_{2}^{11} (-i e^{i \pi/4} h q) \sim i(hq)^2/75$. We also used the fact that $D^x_{1,1} \bar{D}^x_{1,1}\approx 1+ \mathcal{O}(hq)^4$, with $q=\sqrt{\omega/D}$, and the definitions of the joining factors in Eqs.~\eqref{eq:joining-factor-kappa1} and~\eqref{eq:joining-factor-kappa2}.

Similar to the case of $\mathcal{P}\left[\Sigma^{(0)}_{p} \right]^{(1,2)}$, the Pad\' e approximation for
$\Sigma^{(0)}_{n}$, $\mathcal{P}\left[ \Sigma^{(0)}_{n}\right]^{(1,2)}$, provides a good approximation for the polarization response in the frequency range $\omega \leq \omega_h= D/h^2$, and it eliminates the unphysical behavior at $\omega > \omega_h$ which occurs for the Taylor expansion. Also, $\mathcal{P}\left[ \Sigma^{(0)}_{n}(\omega\to \infty)\right]^{(1,2)} \to 0$, which gives a smooth transition to the high frequency region. Finally, in the spherical limit, the Pad\'e approximation $\mathcal{P}\left[ \Sigma^{(0)}_{n}\right]^{(1,2)} = \mathcal{P} [ \Sigma^{(0)}_{p}]^{(1,2)}$ reproduces the exact polarization response of a charged sphere immersed in an electrolyte~\cite{Chew1982}. Hence, we will use the modified water conductivity with the Pad\' e approximation, given by
\beq
\label{eq:modified-sigma_w-x-uniform}
\sigma_{w,n}^{\rm I} (\omega) \sim \sigma_w \left( 1 + \frac{\Gamma_0}{a N_0} \frac{b^{\perp}_{1,1}(\xi_0)}{2}   \mathcal{P}[ \Sigma^{(0)}_{n}(hq, \xi_0)]^{(1,2)} \right),
\eeq
for our discussion.

\section{Polarization response of a charged oblate spheroid with a non-uniform surface ion distribution}
\label{app:Polarization-non-uniform}

In this appendix, we repeat the same analysis as in Appendix~\ref{app:Polarization-uniform}, but for a non-uniform surface cation distribution of the form $\Gamma^{II}_{+}= \widetilde{\Gamma}_0 h_{\xi}/h$. The derivations will be more concise because they are largely parallel to the ones in the previous appendix. The main difference between the calculation for our particular choice of non-uniform surface cation distribution and for the uniform cation distribution is that there is less mixing of modes for the non-uniform charge distribution. There are two reasons that the modes become mixed. First, the BCs depend both on the perturbed electric potential, which is a solution of the Laplace equation, and the perturbed charge density, which is a solution of the diffusion equation, whose eigenfunctions can be expressed as infinite sums of the eigenfunctions of the Laplace equation. This mismatch of the two general solutions will cause mode mixing for any choice of charge distribution. Second, the form of the BC can intrinsically cause mode mixing. This does not occur for our choice of non-uniform charge distribution, but will occur for most other charge distributions, including the uniform one.

In this case, the BC for anions is identical to Eq.~\eqref{eq:BC-anion-uniform} while the BC for cations, due to the non-uniform distribution, becomes
\begin{widetext}
\begin{eqnarray}
\label{eq:BC-cation}
&&- \frac{1}{h_{\xi} (\xi_0)} \left[ \frac{\partial}{\partial \xi} n_{>}(\xi,\eta, \phi,\omega)  +  N_0 \frac{\partial}{\partial \xi} \psi(\xi,\eta, \phi,\omega)  \right]_{\xi=\xi_0}
\\
\nonumber
&=& \frac{\widetilde\Gamma_0}{h N_0} \left\{ \frac{1}{h_{\eta} h_{\phi}}  \frac{\partial}{\partial \eta} \left[ \frac{h_{\xi} h_{\phi}}{h_\eta} \left( \frac{\partial}{\partial \eta} n_{>}(\xi,\eta,\phi,\omega) + N_0 \frac{\partial}{\partial \eta} \psi(\xi,\eta,\phi,\omega)  \right) \right] +  \frac{h_{\xi}}{ h_{\phi}^2 } \left( \frac{\partial^2}{\partial \phi^2} n_>(\xi,\eta,\phi,\omega) +  N_0 \frac{\partial^2}{\partial \phi^2} \psi(\xi,\eta,\phi,\omega) \right) \right\}_{\xi=\xi_0}. 
\end{eqnarray}
\end{widetext}
Once we have the general solutions in Eqs.~\eqref{eq:general-solution-psi} and~\eqref{eq:general-solution-n-omega+}, the electric potential and the ion concentration are uniquely determined by the BCs. Again, we shall focus on solving for the components $A_{1}^{m}$ which  allow us to extract the polarization coefficients.

\subsection{Polarization coefficient $P_z$ in the $z$-direction}
\label{sec:P-z}

When the electric field $\boldsymbol{E}=E_0 \hat{z}$ is applied in the $z$-direction, the general solutions for $\psi^{z}$ and $n^{z}_{>}$ still take the forms in Eqs.~\eqref{eq:psi-z} and~\eqref{eq:n-z}, respectively. Because the BC for the anions is given by Eq.~\eqref{eq:BC-anion-uniform}, the $\alpha_\ell$ are related to the $A_{\ell}$ through Eq.~\eqref{eq:alpha_l-A_l_inverse} with all the $K^z$, $\bar{K}^{z}$, $D^z$ and $\bar{D}^z$ matrices defined in Appendix~\ref{app:Polarization-uniform}.

Let us now focus on the BC for cations in Eq.~\eqref{eq:BC-cation}. Using the expansions for $\psi^{z}$ and $n^{z}_{>}$ in Eqs.~\eqref{eq:psi-z} and~\eqref{eq:n-z}, replacing $\alpha_\ell$ with the relation in Eq.~\eqref{eq:alpha_l-A_l_inverse}, and finally inserting the expansion of $S_{0\ell'}^{(1)}$ in terms of the Legendre polynomials, we obtain
\begin{widetext}
\begin{align}
\label{eq:BC-cation-z}
& 2 E_0 h P_{1}(\eta) - 2 \sum_{\ell=0}^{\infty}     P_{\ell} (\eta)    \frac{d Q_{\ell} (i \xi_0)}{d \xi_0}  A_{\ell}
\\
\nonumber
= &
(-E_0 h) \frac{\widetilde{\Gamma}_0}{h N_0 (1+\xi_0^2)}  \frac{d}{d \eta} \left[(1-\eta^2) \left( \xi_0 \frac{ d P_{1}(\eta)}{d \eta} +  \sum_{ \ell'=0}^{\infty} \sideset{}{'}\sum_{r=1}^{\infty}  D^{z}_{r,\ell'} \bar{D}^z_{\ell',1} \frac{ R_{0\ell'}^{(3)} (-i e^{i \pi/4} h q, i \xi_0 )}{ d R_{0, \ell'}^{(3)} (-i e^{i \pi/4} h q, i \xi_0 )/d \xi_0}  \frac{d P_{r}(\eta)}{d \eta} \right)  \right]
\\
\nonumber
&+
\frac{\widetilde{\Gamma}_0}{h N_0 (1+\xi_0^2)}  \frac{d}{d \eta} \left[ (1-\eta^2) \sum_{\ell}  A_{\ell} \left( \sum_{ \ell'=0}^{\infty}\sideset{}{'}\sum_{r=0,1}^{\infty} D^{z}_{r,\ell'} \bar{D}^z_{\ell',\ell}   \frac{   R_{0\ell'}^{(3)} (-i e^{i \pi/4} h q, i \xi_0 )}{ d R_{0, \ell'}^{(3)} (-i e^{i \pi/4} h q, i \xi_0 )/d \xi_0} \frac{d Q_{\ell} (i \xi_0)}{d \xi_0}  \frac{ d P_{r}(\eta)}{d \eta} +  Q_{\ell} (i \xi_0)  \frac{d  P_{\ell} (\eta)}{d \eta}  \right) \right] .
\end{align}
Here, we have used the fact that $\bar{D}^z$ is the inverse matrix of $D^z$ with elements $D^{z}_{\ell,\ell'} = d_{\ell}^{0\ell'}$.

As for the uniform case, we now employ the orthogonality properties of the Legendre polynomials by multiplying both sizes of Eq.~\eqref{eq:BC-cation-z} by $P_{\ell''}(\eta)$ and integrating over the interval $-1<\eta<1$. Again, the even and odd index sectors are decoupled and $A_{\ell}=0$ for $\ell$ even. After some algebra, the matrix equation of the odd index sector reads, $\sum_{n'=0} \widetilde{G}^z_{2n+1, 2n'+1} A_{2n'+1} = \widetilde{V}^z_{2n+1}$,
where
\begin{align}
\nonumber
\widetilde{G}^{z}_{2n+1, 2n'+1} =&   \left( 2 c_{2n+1} \frac{d Q_{2n+1} (i \xi_0)}{d \xi_0}  - \frac{\widetilde{\Gamma}_0}{h N_0 (1+\xi_0^2)}  c_{2n+1}^{1} Q_{2n+1} (i \xi_0)\right)  \delta_{n n'} 
\\
\label{eq:def-G-matrix-elements}
& - \frac{\widetilde{\Gamma}_0}{h N_0 (1+\xi_0^2)}  c_{2n+1}^{1}\left( \sum_{\nu=0}^{\infty}   D^{z}_{2n+1, 2\nu+1} \bar{D}^z_{2\nu+1,2n'+1}  \frac{   R_{0,2\nu+1}^{(3)} (-i e^{i \pi/4} h q, i \xi_0 )}{ d R_{0, 2\nu+1}^{(3)} (-i e^{i \pi/4} h q, i \xi_0 )/d \xi_0} \frac{d Q_{2n'+1} (i \xi_0)}{d \xi_0}    \right),
\end{align}
and
\begin{equation}
\label{eq:def-V-z-elements}
\widetilde{V}^{z}_{2n+1} =  E_0 h \left[ 2  c_{1} \delta_{n0} - \frac{\widetilde{\Gamma}_0}{h N_0 (1+\xi_0^2)}c_{2n+1}^{1}  \left( \xi_0  \delta_{n0}  + \sum_{\nu=0}^{\infty}  D^z_{2n+1,2\nu+1} \bar{D}^z_{2\nu+1,1} \frac{ R_{0,2\nu+1}^{(3)} (-i e^{i \pi/4} h q, i \xi_0 )}{ d R_{0, 2\nu+1}^{(3)} (-i e^{i \pi/4} h q, i \xi_0 )/d \xi_0} \right) \right].
\end{equation}
\end{widetext}
For this choice of surface ion distribution, the matrix $b_{n,n'}$ from the uniform case is replaced with a
diagonal matrix, which reduces the mixing of the modes.

With arguments similar to those right before Eq.~\eqref{eq:psi-z-uniform-approx}, we can argue that the dominant response upon the application of an electric field is in the $A_1$ component. However, unlike in the case for uniform ion distributions, we do not need to introduce the leading and higher order parts, c.f. Eq.~\eqref{eq:G-z0-uniform} to Eq.~\eqref{eq:delta_V-z-uniform}, to separate $\widetilde{G}^z$ and $\widetilde{V}^z$. Instead, their scaling structure as a function of $hq$ in the low frequency limit, $hq\to 0$, provides a nice way to separate the different orders of contributions to $A_{\ell}$. 

We start by observing that the first line of Eq.~\eqref{eq:def-G-matrix-elements} contributes only to diagonal terms, $n=n'$, and is of the order $\mathcal{O}(hq)^0$, while the second line of Eq.~\eqref{eq:def-G-matrix-elements} contributes to all combinations of $n$ and $n'$ with leading contributions scaling as $\mathcal{O}(hq)^{2|n-n'|}$ according to the structure of the elements $D^z_{\ell,\ell'}$ (or, equivalently, the $d^{0\ell'}_{\ell}$). We can then conclude that the elements $\widetilde{G}^z_{2n+1,2n'+1}$ are of the order $\mathcal{O}(hq)^{2|n-n'|}$. As a result, the elements of its inverse matrix, $\widetilde{\bar{G}}^z_{2n'+1, 2n+1}$, also scale as $\mathcal{O}(hq)^{2|n-n'|} = \mathcal{O}(h^2\omega/D)^{|n-n'|}$. Because $V_{2n+1}$ scales as $\mathcal{O}(hq)^{2|n|}=\mathcal{O}(h^2\omega/D)^{|n|}$ according to the structure of the elements $D^z_{\ell,\ell'}$, the leading order of $A_{2n'+1}$ is $\mathcal{O}(hq)^{2|n'|}=\mathcal{O}(h^2\omega/D)^{|n'|}$. To collect all the terms that contribute to the leading order of $A_{2n'+1}$, the elements in the lower triangle (including the diagonal part) of the inverse matrix $\widetilde{\bar{G}}^z$ are needed. One can then argue that the upper triangle of $\widetilde{\bar{G}}^z$ contributes to $A_{2n'+1}$ starting at order $\mathcal{O}(hq)^{2|n'|+4}$. In contrast to the case discussed in Sec.~\ref{sec:P-z-uniform}, the off-diagonal terms of the matrix $\widetilde{G}^z$ contribute to the component $A_1$  starting at $\mathcal{O}(hq)^4$. 

With these observations, the $A_1$ component of the perturbed electric potential can be approximated as $A_{1}\approx \widetilde{V}^z_1/\widetilde{G}^z_{11}$
\beq
\frac{A_{1}}{ E_0 h } \approx   \frac{ 1   - \frac{\widetilde{\Gamma}_0}{ a N_0 \sqrt{1+\xi_0^2}} \left( \xi_0   + \widetilde{\Sigma}_{p}( h q,  \xi_0) \right) }{ \frac{d Q_{1} (i \xi_0)}{d \xi_0}  - \frac{\widetilde{\Gamma}_0}{ a N_0 \sqrt{1+\xi_0^2} }  \left( Q_{1} (i \xi_0) + \widetilde{\Sigma}_{p}(h q, \xi_0)  \right)   },
\eeq
where we have used $a= h \sqrt{1+\xi_0^2} $ and $c^{1}_{1}= 2 c_1= 4/3$. In addition, we have defined
\beq
\nonumber
\widetilde{\Sigma}_{p}( h q, \xi_0) = \sum_{\nu=0}^{\infty}  D^z_{1,2\nu+1} \bar{D}^z_{2\nu+1,1} \frac{ R_{0,2\nu+1}^{(3)} (-i e^{i \pi/4} h q, i \xi_0 )}{ d R_{0, 2\nu+1}^{(3)} (-i e^{i \pi/4} h q, i \xi_0 )/d \xi_0}. 
\eeq
The polarization coefficient in the z-direction can now be extracted from the $A_1$, and is given by 
\beq
\label{eq:Pz-1}
\widetilde{P}_{z} = \frac{1}{3}  \frac{ \left[\frac{\widetilde{\Gamma}_0 }{ a N_0 \sqrt{1+\xi_0^2} } \left( \xi_0   + \widetilde{\Sigma}_{p} \right) -1 \right] / \xi_0 (1+\xi_0^2)  }{\left( \frac{d Q_{1} (i \xi_0)}{d \xi_0}  - \frac{\widetilde{\Gamma}_0 }{ a N_0 \sqrt{1+\xi_0^2} }  \left( Q_{1} (i \xi_0) + \widetilde{\Sigma}_{p} \frac{d Q_{1} (i \xi_0)}{d \xi_0}   \right) \right)  }, 
\eeq


Using the depolarization factor in Eq.~\eqref{eq:L_p}, we can case the polarization coefficient into a form similar to Eq.~\eqref{eq:Pz-uniform-2}, given by  
\beq
\label{eq:Pz-2}
P^{\rm II}_{p} =\widetilde{P}_{z} = \frac{1}{3}  \frac{ \sigma^{\rm II}_{p} - \sigma_{w,p}^{\rm II} }{ L_{p} \sigma^{\rm II}_{p} +(1-L_{p}) \sigma_{w,p}^{\rm II}  },
\eeq
where the effective particle and the modified water conductivities in the $z$-direction are given by
\beq
\label{eq:sigma-p}
\begin{split}
\sigma^{\rm II}_{p} =& \sigma_w  \frac{ \widetilde{\Gamma}_0}{a N_0}  P_1(\xi_0)/\sqrt{1+\xi_0^{2}},
\\
\sigma_{w,p}^{\rm II} =& \sigma_w \left( 1 - \frac{\widetilde{\Gamma}_0}{a N_0}  \widetilde{\Sigma}_{p}( h q, \xi_0) /\sqrt{1+\xi_0^{2}} \right).
\end{split}
\eeq
The superscript $\rm II$ indicates that those quantities are related to case (II) with the non-uniform surface ion distribution. Again, the effective conductivity $\sigma^{\rm II}_{p} $ has no frequency dependence and agrees with the valus found for high frequencies.~\cite{Freed-ex}.

To order $(hq)^3$, the Taylor expansion of $\widetilde{\Sigma}_{p}(h q,  \xi_0)$ is the same as for $\Sigma^{(0)}_{p}( h q,  \xi_0) $ in Eq.~\eqref{eq:Sigma_z0_11-uniform-expand}. Thus, the Pad\' e approximation, $\mathcal{P}[\widetilde{\Sigma}_{p}(h q, \xi_0)]^{(1,2)}$, is equal to $\mathcal{P}[\Sigma^{(0)}_{p}(h q, \xi_0)]^{(1,2)}$ in Eq.~\eqref{eq:Sigma_z0_11-uniform-Pade}. The Pad\' e- approximated modified water conductivity reads
\beq
\label{eq:modified-sigma_w-z-nonuniform}
\sigma_{\rm w,p}^{\rm II}  \sim \sigma_w \left( 1 - \frac{\widetilde{\Gamma}_0}{a N_0 } \mathcal{P} [ \Sigma^{(0)}_{p} ( h q, \xi_0) ]^{(1,2)}/\sqrt{1+\xi_0^{2}}  \right).
\eeq
Again, using this Pad\' e approximation for the modified water conductivity gives a good-low frequency approximation with a smooth transition to the high frequency polarization response.

\subsection{Polarization coefficient $P_x$ in $x$-direction}
\label{sec:P-x}

Lastly, we consider the case when the electric field is perpendicular to the axis of symmetry for the non-uniform charge distribution.  When an electric field $\boldsymbol{E}=E_0 \hat{x}$ is applied, the general solutions for $\psi^{x}$ and $n^{x}_{>}$ take the forms given in Eqs.~\eqref{eq:psi-x} and ~\eqref{eq:n-x}, respectively. Because the BC for the anions is given by Eq.~\eqref{eq:BC-anion-uniform}, the $\alpha^{1}_{\ell}$ can be expressed in terms of the $A^{1}_{\ell}$ through the relation in Eq.~\eqref{eq:alpha_l1-A_l1-uniform} with all the matrices $K^x$, $\bar{K}^x$, $D^x$, and $\bar{D}^x$ defined in Appendix~\ref{app:Polarization-uniform}. For the BC for the cations in Eq.~\eqref{eq:BC-cation}, we first use the explicit form of $\psi^x$ and $n_>^x$ in Eqs.~\eqref{eq:psi-x} and~\eqref{eq:n-x}, then replace $\alpha_\ell^1$ by $A_{\ell}^1$ with the relation in Eq.~\eqref{eq:alpha_l1-A_l1-uniform}, and finally use the expansion of $S_{1\ell}^{(1)}(-i e^{i \pi/4} h q, \eta )$ in terms of the associated Legendre polynomials, in Eq.~\eqref{eq:S1_mn-expansion}, which gives
\begin{widetext}
\bea
\label{eq:BC-cation-x}
&-& \left( 2 i E_0 h \frac{d P_{1}^{1} (i \xi_0)}{d \xi_0} P_{1}^{1} (\eta)  + 2 \sum_{\ell=1}^{\infty}  A_{\ell}^{1} \frac{d Q_{\ell}^{1} (i \xi_0)}{d \xi_0}   P_{\ell}^1(\eta)   \right)
\\
\nonumber
&=&
\frac{\widetilde{\Gamma}_0}{a N_0 \sqrt{1+\xi_0^2} }   \left[ \sum_{\ell,\ell',r=1}^{\infty} \beta_{r}^{1}(\xi_0)  D^x_{r,\ell} \bar{D}^x_{\ell,\ell'} \cdot \left( i E_0 h \frac{d P_{1}^{1} (i \xi_0)}{d \xi_0}  \delta_{1 \ell'}  +  A_{\ell'}^{1} \frac{d Q_{\ell'}^{1} (i \xi_0)}{d \xi_0}  \right) \frac{R_{1\ell}^{(3)} (-i e^{i \pi/4} h q, i \xi_0 )}{d R_{1\ell}^{(3)} (-i e^{i \pi/4} h q, i \xi_0 )/d \xi_0}    P_{r}^1(\eta) \right. 
\\
\nonumber
& &  \qquad \qquad \qquad  \qquad \qquad \qquad  \qquad \qquad \qquad \qquad \qquad \left.  +  \left(\beta_{1}^{1}(\xi_0)  i E_0 h P_{1}^{1} (i \xi_0)  P_{1}^{1} (\eta)  + \sum_{\ell=1}^{\infty} \beta_{\ell}^{1} A_{\ell}^{1} Q_{\ell}^{1} (i \xi_0)  P_{\ell}^{1}(\eta) \right)  \right] .
\eea
\end{widetext}
Here, we have introduced $\beta^1_{\ell}(\xi_0) = -\ell(\ell+1)+1/(1+\xi_0^2)$ and used the following identity to simplify the expression:
\beq
\nonumber
\frac{d }{d \eta} \left[ (1-\eta^2) \frac{d }{d \eta} P_{\ell}^{1}(\eta)\right] -\frac{1}{1-\eta^2}  P_{\ell}^{1}(\eta) = -\ell(\ell+1) P_{\ell}^{1}(\eta).
\eeq

By employing the orthogonality properties of the Legendre polynomials, one can show that sectors with even and odd indices are decoupled and all the $A^1_{\ell}=0$ for $\ell$ even. Then, the matrix equation for the odd index sector reads $\sum_{n'=0}\widetilde{G}^x_{2n+1, 2n'+1} A_{2n'+1} = \widetilde{V}^x_{2n+1}$
with
\begin{widetext}
\beq
\label{eq:G-x}
\begin{split}
	\widetilde{G}^x_{2n+1, 2n'+1} =&  \left( 2 \frac{d Q_{2n+1}^{1} (i \xi_0)}{d \xi_0}  +
	\frac{\widetilde{\Gamma}_0}{a N_0 \sqrt{1+\xi_0^2}}  \beta_{2n+1}^{1}  Q_{2n+1}^{1} (i \xi_0) \right)  \delta_{n n'}  
	\\
	+& \frac{\widetilde{\Gamma}_0}{a N_0 \sqrt{1+\xi_0^2}}  \beta_{2n+1}^{1}    \sum_{\nu=0}^{\infty}   D^x_{2n+1,2\nu+1} \bar{D}^x_{2\nu+1,2n'+1} \cdot     \frac{R_{1,2\nu+1}^{(3)} (-i e^{i \pi/4} h q, i \xi_0 )}{d R_{1,2\nu+1}^{(3)} (-i e^{i \pi/4} h q, i \xi_0 )/d \xi_0}  \frac{d Q_{2n'+1}^{1} (i \xi_0)}{d \xi_0}   
\end{split}
\eeq
and 
\beq
\label{eq:V-x}
\begin{split}
	\widetilde{V}^x_{2n+1} =&  -  i E_0 h \left[ \left( 2 \frac{d P_{1}^{1} (i \xi_0)}{d \xi_0} +  \frac{\widetilde{\Gamma}_0}{a N_0 \sqrt{1+\xi_0^2}}  P_{1}^{1} (i \xi_0) \beta_{2n+1}^{1} \right) \delta_{n0} \right.
	\\
	&\qquad \qquad + \left. \frac{\widetilde{\Gamma}_0}{a N_0 \sqrt{1+\xi_0^2}} \beta_{2n+1}^{1} \sum_{\nu=0}^{\infty}   D^x_{2n+1,2\nu+1} \bar{D}^x_{2\nu+1,1}    \frac{R_{1,2\nu+1}^{(3)} (-i e^{i \pi/4} h q, i \xi_0 )}{d R_{1,2\nu+1}^{(3)} (-i e^{i \pi/4} h q, i \xi_0 )/d \xi_0} \frac{d P_{1}^{1} (i \xi_0)}{d \xi_0} \right].
\end{split}
\eeq
\end{widetext}

According to an argument similar to the one following Eq.~\eqref{eq:def-V-z-elements}, an approximate expression for $A_1^1$ is given by $A^{1}_{1} \approx  \widetilde{V}_{1}^{x}/\widetilde{G}_{11}^x$ such that
\beq
\frac{A^{1}_{1}}{E_0 h} \approx   -  i  \frac{   \frac{d P_{1}^{1} (i \xi_0)}{d \xi_0} +  \frac{\widetilde{\Gamma}_0  \beta_{1}^{1}(\xi_0)}{2a N_0 \sqrt{1+\xi_0^2}}    \left( P_{1}^{1} (i \xi_0) +\widetilde{\Sigma}_{n}\frac{d P_{1}^{1} (i \xi_0)}{d \xi_0} \right) }{   \frac{d Q_{1}^{1} (i \xi_0)}{d \xi_0}  +	\frac{\widetilde{\Gamma}_0 \beta_{1}^{1}(\xi_0)}{2 a N_0 \sqrt{1+\xi_0^2}} \left( Q_{1}^{1} (i \xi_0) +  \widetilde{\Sigma}_{n} \frac{d Q_{1}^{1} (i \xi_0)}{d \xi_0}  \right) }, 
\eeq
where
\beq
\nonumber
\widetilde{\Sigma}_{\rm n}( h q,  \xi_0) = \sum_{\nu=0}^{\infty}   D^x_{1,2\nu+1} \bar{D}^x_{2\nu+1,1}    \frac{R_{1,2\nu+1}^{(3)} (-i e^{i \pi/4} h q, i \xi_0 )}{d R_{1,2\nu+1}^{(3)} (-i e^{i \pi/4} h q, i \xi_0 )/d \xi_0}. 
\eeq
Then the polarization coefficient is extracted from $A^{1}_{1}$ and is given by
\beq
\label{eq:Px-2}
P^{\rm II}_{n} =  \frac{1}{3} \frac{\sigma^{\rm II}_{n} -\sigma_{w,n}^{\rm II} }{ L_{n} \sigma^{\rm II}_{n} +(1-L_{n}) \sigma_{w,n}^{\rm II}}.
\eeq
Here, we have used the depolarization factor in Eq.~\eqref{eq:L_n}. The effective particle and modified water conductivities for the $x$-direction ($y$-direction) are given by 
\beq
\label{eq:sigma-n}
\sigma^{\rm II}_{n} =  -\sigma_w  \frac{\widetilde{\Gamma}_0}{2 a N_0 } \frac{ \beta_{1}^{1} }{\sqrt{1+\xi_0^2}} \frac{P^1_1( i \xi_0)}{ d P^1_1( i \xi_0)/d\xi_0}  ,
\eeq
and
\beq
\label{eq:sigma-w-x-nonuniform}
\sigma_{w,n}^{\rm II} = \sigma_w \left( 1+  \frac{\widetilde{\Gamma}_0}{2 a N_0 }  \frac{ \beta_{1}^{1}  }{\sqrt{1+\xi_0^2}}  \widetilde{\Sigma}_{n}( h q, \xi_0)   \right).
\eeq

The Taylor expansion of $\widetilde{\Sigma}_{n}( h q, \xi_0)$ to the $(hq)^3$ order is the same as that of $\Sigma^{(0)}_{n}( h q, \xi_0)$ in Eq.~\eqref{eq:Sigma-x0_11-uniform}. Hence, the Pad\' e approximation, $\mathcal{P}[\widetilde{\Sigma}_{n}( h q, \xi_0)]^{(1,2)}$, is given by Eq.~\eqref{eq:Sigma_x0_11-uniform-Pade}. The Pad\' e approximated modified water conductivity then reads
\beq
\label{eq:modified-sigma_w-x-nonuniform}
\sigma_{\rm w,n}^{\rm II}  \sim \sigma_w \left( 1+  \frac{\widetilde{\Gamma}_0}{a N_0 }  \frac{ \beta_{1}^{1}(\xi_0)  }{2 \sqrt{1+\xi_0^2}}   \mathcal{P}[\Sigma^{(0)}_{n}( h q, \xi_0)]^{(1,2)}  \right).
\eeq
Again, Eq.~\eqref{eq:modified-sigma_w-x-nonuniform} gives a good approximation for the low-frequency polarization response with a smooth transition to higher frequencies without unphysical behavior. We will hence use this form for our discussions.

\end{appendix}

\bibliography{Ref-low-freq}

\end{document}